\documentclass[twocolumn,trackchanges]{aastex7}

\newcommand{\footnoterule}{%
    \vspace{-0.5em} 
    \hrule width 0.3\textwidth height 0.3pt 
    \vspace{1em} 
}

\usepackage{amsmath}
\usepackage{bm}
\DeclareUnicodeCharacter{02BC}{'}

\begin{document}

\title{B-fields And dust in interstelLar fiLAments using Dust POLarization (BALLAD-POL): VI. Grain alignment mechanisms in the massive quiescent filament G16.96+0.27 using dust polarization observations from JCMT/POL-2}

\author[0009-0002-6171-9740]{Saikhom Pravash}
\affiliation{Indian Institute of Astrophysics, II Block, Koramangala, 560034, India; \href{mailto:saikhom.singh@iiap.res.in}{saikhom.singh@iiap.res.in}, \href{mailto:spravash11@gmail.com}{spravash11@gmail.com}}
\affiliation{Pondicherry University, R.V. Nagar, Kalapet, Puducherry, 605014, India}
\email{saikhom.singh@iiap.res.in}

\author[0000-0003-2017-0982]{Thiem Hoang}
\affiliation{Korea Astronomy and Space Science Institute, 776 Daedeokdae-ro, Yuseong-gu, Daejeon 34055, Republic of Korea}
\affiliation{University of Science and Technology, Korea, 217 Gajeong-ro, Yuseong-gu, Daejeon 34113, Republic of Korea}
\email{thiemhoang@kasi.re.kr}

\author[0000-0002-6386-2906]{Archana Soam}
\affiliation{Indian Institute of Astrophysics, II Block, Koramangala, 560034, India; \href{mailto:saikhom.singh@iiap.res.in}{saikhom.singh@iiap.res.in}, \href{mailto:spravash11@gmail.com}{spravash11@gmail.com}}
\affiliation{Pondicherry University, R.V. Nagar, Kalapet, Puducherry, 605014, India}
\email{archana.soam@iiap.res.in}

\author[0000-0002-2826-1902]{Qi-Lao Gu}
\affiliation{Shanghai Astronomical Observatory, Chinese Academy of Sciences, No.80 Nandan Road, Xuhui, Shanghai 200030, Peopleʼs Republic of China}
\email{qlgu@shao.ac.cn}

\author[0000-0002-5286-2564]{Tie Liu}
\affiliation{Shanghai Astronomical Observatory, Chinese Academy of Sciences, No.80 Nandan Road, Xuhui, Shanghai 200030, Peopleʼs Republic of China}
\email{liutie@shao.ac.cn}

\author[0000-0002-2808-0888]{Pham Ngoc Diep}
\affiliation{Center for Astrophysics and Science Exploration, Vietnam National Space Center, Vietnam Academy of Science and Technology, 18 Hoang Quoc Viet, Hanoi, Vietnam}
\affiliation{Graduate University of Science and Technology, Vietnam Academy of Science and Technology, 18 Hoang Quoc Viet, Hanoi, Vietnam}
\email{pndiep@vnsc.org.vn}

\author[0000-0002-6488-8227]{Le Ngoc Tram}
\affiliation{Leiden Observatory, Leiden University, PO Box 9513, 2300 RA Leiden, The Netherlands}
\email{nle@mpifr-bonn.mpg.de}

\author[0000-0002-5913-5554]{Nguyen Bich Ngoc}
\affiliation{Center for Astrophysics and Science Exploration, Vietnam National Space Center, Vietnam Academy of Science and Technology, 18 Hoang Quoc Viet, Hanoi, Vietnam}
\affiliation{Graduate University of Science and Technology, Vietnam Academy of Science and Technology, 18 Hoang Quoc Viet, Hanoi, Vietnam}
\email{capi37capi@gmail.com}


\begin{abstract}
Dust polarization induced by aligned non-spherical grains acts as an important tool to trace the magnetic field (B-field) morphologies and strengths in molecular clouds and constrain grain properties and their alignment mechanisms. The widely accepted grain alignment theory is the alignment induced by RAdiative Torques (RATs). In this work, we investigate grain alignment mechanisms in a massive, quiescent and filamentary Infrared Dark Cloud G16.96+0.27 using thermal dust polarization observation with JCMT/POL-2 at 850 $\mu$m. We observe the so-called phenomenon of polarization hole attributed to the decrease in polarization fraction in denser regions of higher total intensity and gas density. Our study finds that B-field tangling effect is minimal to cause the polarization hole, and the dominant factor is the reduction in grain alignment efficiency in denser regions, consistent with RAT mechanism. To test RAT theory, we calculate various quantities describing grain alignment, including minimum size of aligned grains, magnetic and magnetic relaxation parameter, and show that RAT mechanism can explain observational data. Our study also reveals evidence for magnetically-enhanced RAT (M-RAT) mechanism required to explain the observed high polarization fractions of above 10\% in the outer regions of the filament. Finally, we perform detailed modeling of thermal dust polarization using \textsc{DustPOL\_py} based on M-RAT theory and find that the modeling could successfully reproduce the observational data when maximum grain size is around 0.45 $\mu$m accompanied by an increase in grain axial ratio, along with the consideration of variations in the magnetic field's inclination angle with the line of sight.

\end{abstract}



\section{Introduction} \label{section:Introduction}

Dust grains, although constituting only a minor fraction of the interstellar medium (ISM) by mass, are fundamental and indispensable components of the ISM. They play a central role in the physical, chemical, and dynamical evolution of the ISM, by significantly influencing in various astrophysical processes, including star and planet formation, thermal balancing of the ISM, and acting as catalytic surfaces for the formation of water and other complex molecules (for a review see \citealt{2003ARA&A..41..241D}). The landmark discovery of interstellar starlight polarization by \cite{1949Sci...109..166H} and \cite{1949ApJ...109..471H} revealed that interstellar dust grains are non-spherical and preferentially aligned with interstellar magnetic fields (B-fields), which induces starlight polarization through dichroic extinction. 


Aligned dust grains also emit polarized thermal radiation, known as thermal dust polarization, in far-IR/sub-mm/mm wavelengths \citep{1988QJRAS..29..327H}. Dust grains are aligned in such a way that their short axes are parallel and long axes are perpendicular to the ambient B-field (e.g., \citealt{2007JQSRT.106..225L, 2015ARA&A..53..501A, 2015psps.book...81L}). For starlight polarization, the observed polarization angle (PA) traces the plane-of-sky (POS) projected B-field orientation, while it needs to be rotated by 90$^\circ$ in the case of thermal dust polarization (e.g., \citealt{1988QJRAS..29..327H}). Hence, both the dust polarization, i.e., starlight (optical to NIR) and thermal dust polarization (far-IR to sub-mm/mm), have been widely used to trace and study the B-field morphologies and strengths in various environments, ranging from the diffused ISM, molecular clouds to star-forming regions, and also to study properties of dust grains like shape, size, and composition (e.g., \citealt{2021ApJ...919...65D}). 

Magnetic fields are thought to have a significant role in the formation and evolution of molecular clouds and star formation processes (e.g., \citealt{2012ARA&A..50...29C, 2019FrASS...6...15P}). The question that up to how much depth in dense star-forming regions (SFRs) grains can still be aligned is not yet clear, which affects the robustness of using dust polarization for tracing B-fields in dense SFRs \citep{2015psps.book...81L}. To use dust polarization as a reliable tool to study B-fields in SFRs, a detailed understanding of grain alignment mechanisms is of utmost importance. Also, recently \cite{2024ApJ...965..183H} and \cite{2024arXiv240714896T} suggest in their studies the potential of tracing three dimensional (3D) B-field with full dust polarization data including the polarization angle and the polarization fraction by relying on the comparison of the observed dust polarization with the accurate dust polarization model predicted from the grain alignment theory and dust properties like shape and composition. 


The inefficiency of different proposed grain alignment theories to explain and take into full account of various observational results (see \citealt{2003JQSRT..79..881L, 2007JQSRT.106..225L}) renders the grain alignment mechanism one of the most enduring and long-standing problems in astrophysics.
The advent of RAdiative Torque (RAT) alignment theory or RAT theory revolutionized the field. The RAT theory was first introduced by \cite{1976Ap&SS..43..291D}, then numerically demonstrated in \cite{1997ApJ...480..633D} and later developed analytically by \cite{2007MNRAS.378..910L} and \cite{2008MNRAS.388..117H}, positioning it as the leading theoretical framework for grain alignment. In accordance with the RAT theory, large non-spherical grains exposed to an anisotropic radiation field acquire RATs which can induce suprathermal rotation of the grains and align them with the ambient B-field (\citealt{1997ApJ...480..633D, 2007MNRAS.378..910L}). 


The magnetic properties of dust grains are also very important in the alignment efficiency of grains. The presence of iron atoms as clusters embedded inside the grains, expected in dense regions due to grain growth and evolution, can make the grains to behave as super-paramagnetic in nature (\citealt{2022AJ....164..248H}). This super-paramagnetic nature of the grains in the presence of external magnetic fields can enhance their magnetic relaxation strength, thereby enhancing the RAT alignment efficiency of grains, termed as Magnetically enhanced RAT (M-RAT) alignment mechanism \citep{2016ApJ...831..159H}, and could increase the dust polarization fraction. The M-RAT mechanism acts as the unified grain alignment theory.

\begin{figure*}
    \centering
    \begin{tabular}{ccc}
        \includegraphics[scale=0.4]{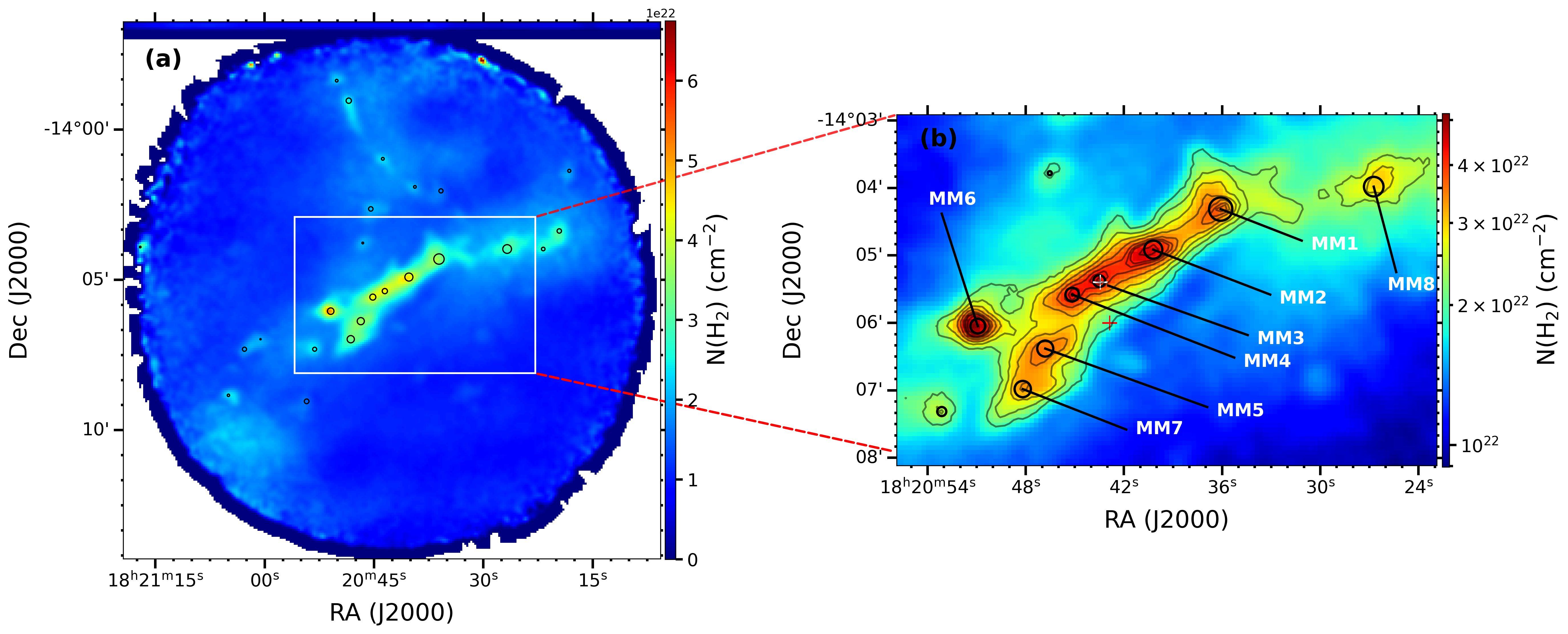} & \\ 
        \hspace{-265pt}
        \includegraphics[scale=0.338]{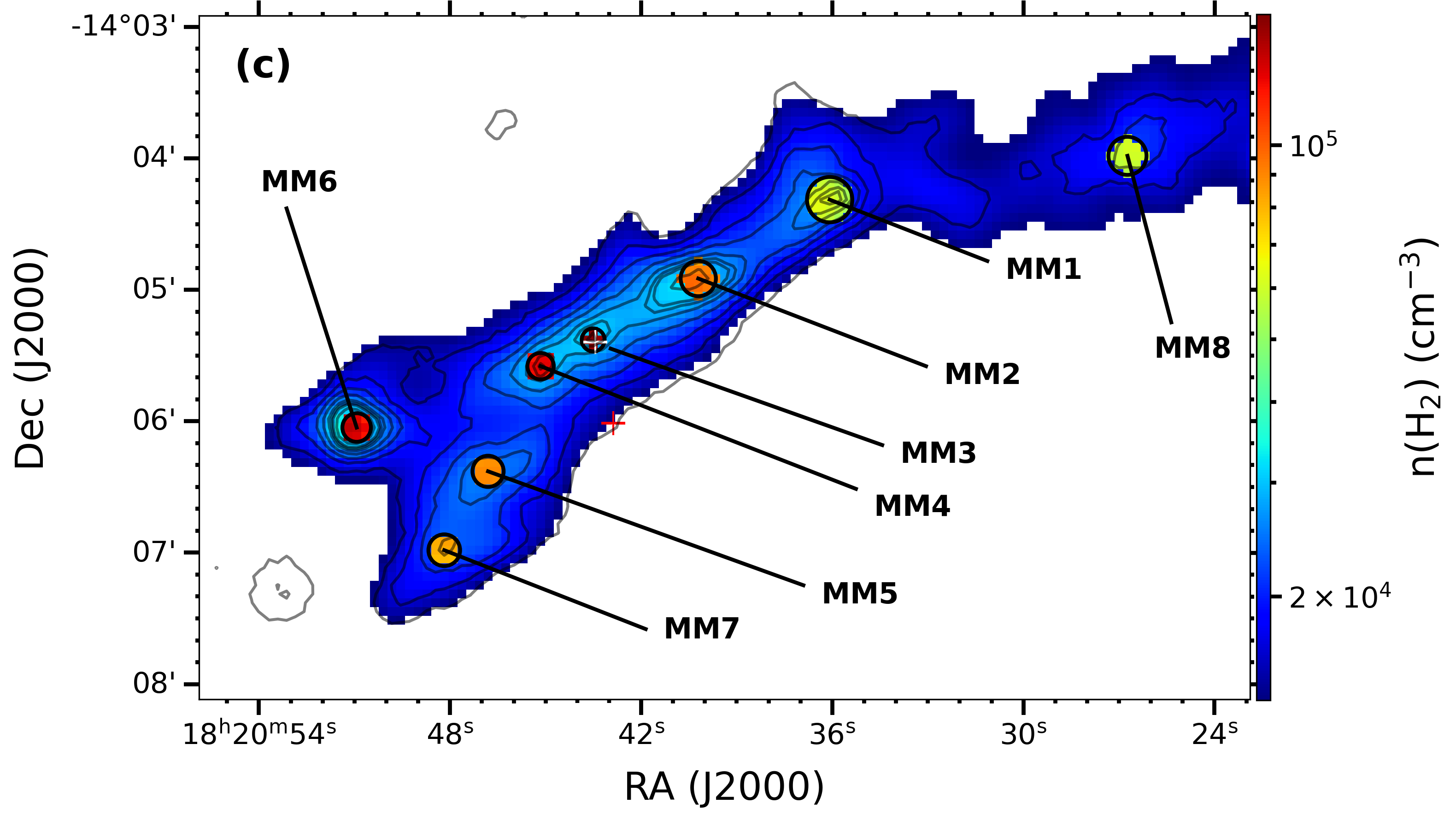} & 
        \hspace{-300pt}
        \includegraphics[scale=0.338]{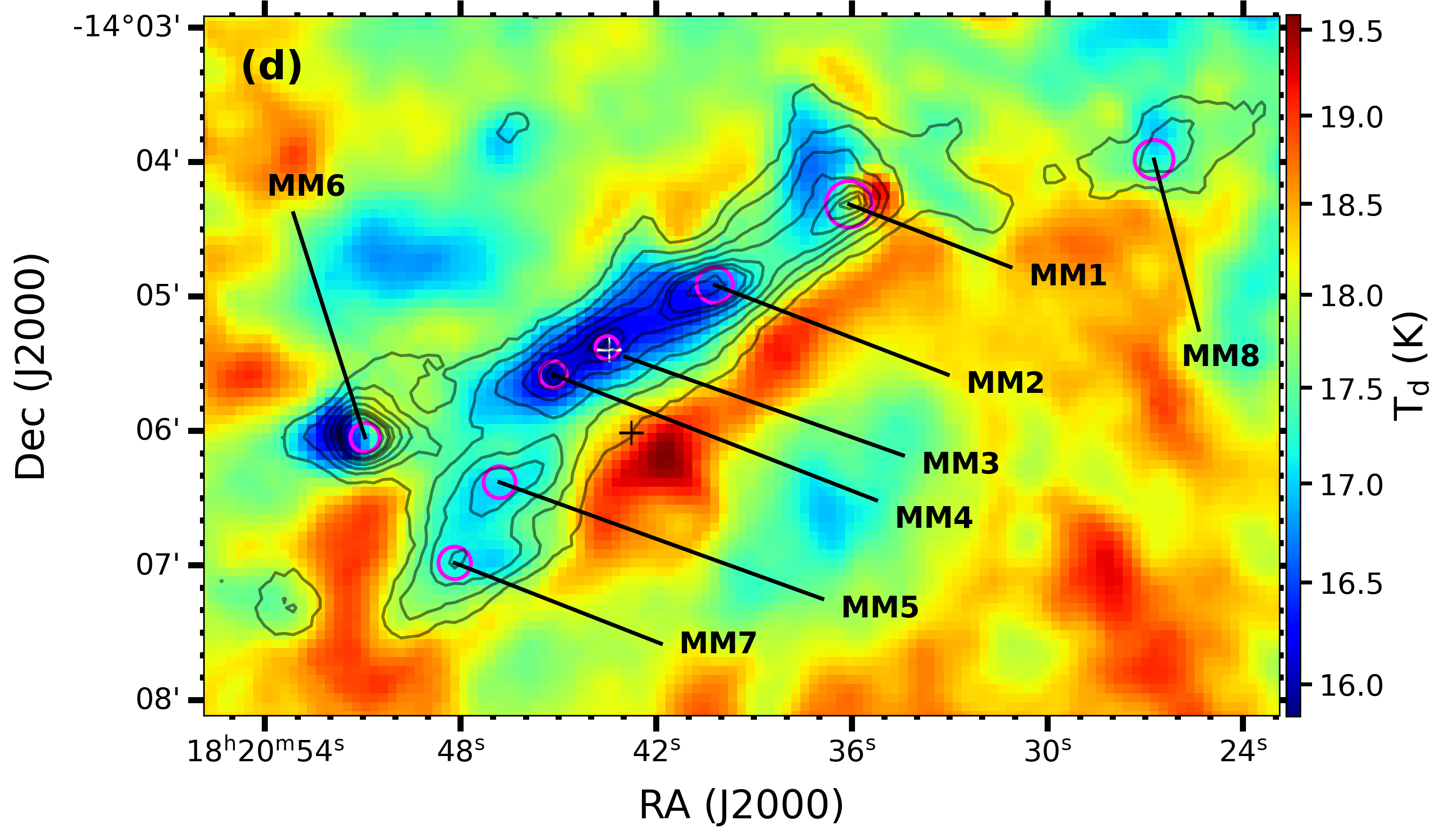} &      
    \end{tabular}
    \caption{(a) Map of gas column density $N(\mathrm{H_2})$ of the whole observed field of G16 region with the white rectangle indicating the region of the G16 filament; (b) Zoomed-in map of $N(\mathrm{H_2})$ for the rectangular region; (c) Map of gas volume density $n(\mathrm{H_2})$; and (d) Map of dust temperature $T_\mathrm{d}$. The black circles overlaid on (a) represent the dense cores identified on the whole observed field of G16 region, and those in (b) and (c) with black colors and (d) with magenta color represent the dense cores MM1, MM2, MM3, MM4, MM5, MM6, MM7 and MM8 identified on the G16 filament. The black contours are drawn at JCMT/POL-2 850 $\mu$m total emission intensity $I$ values of 50, 100, 150, 200, 250, 280, 300, 350 mJy/beam. The "+" symbols in (b), (c) and (d) indicate the pixels of outermost and innermost regions having polarization measurements.}
    \label{Figure:NH2_Td_map}
\end{figure*}

\begin{figure*}
    \centering
        \includegraphics[scale=0.48]{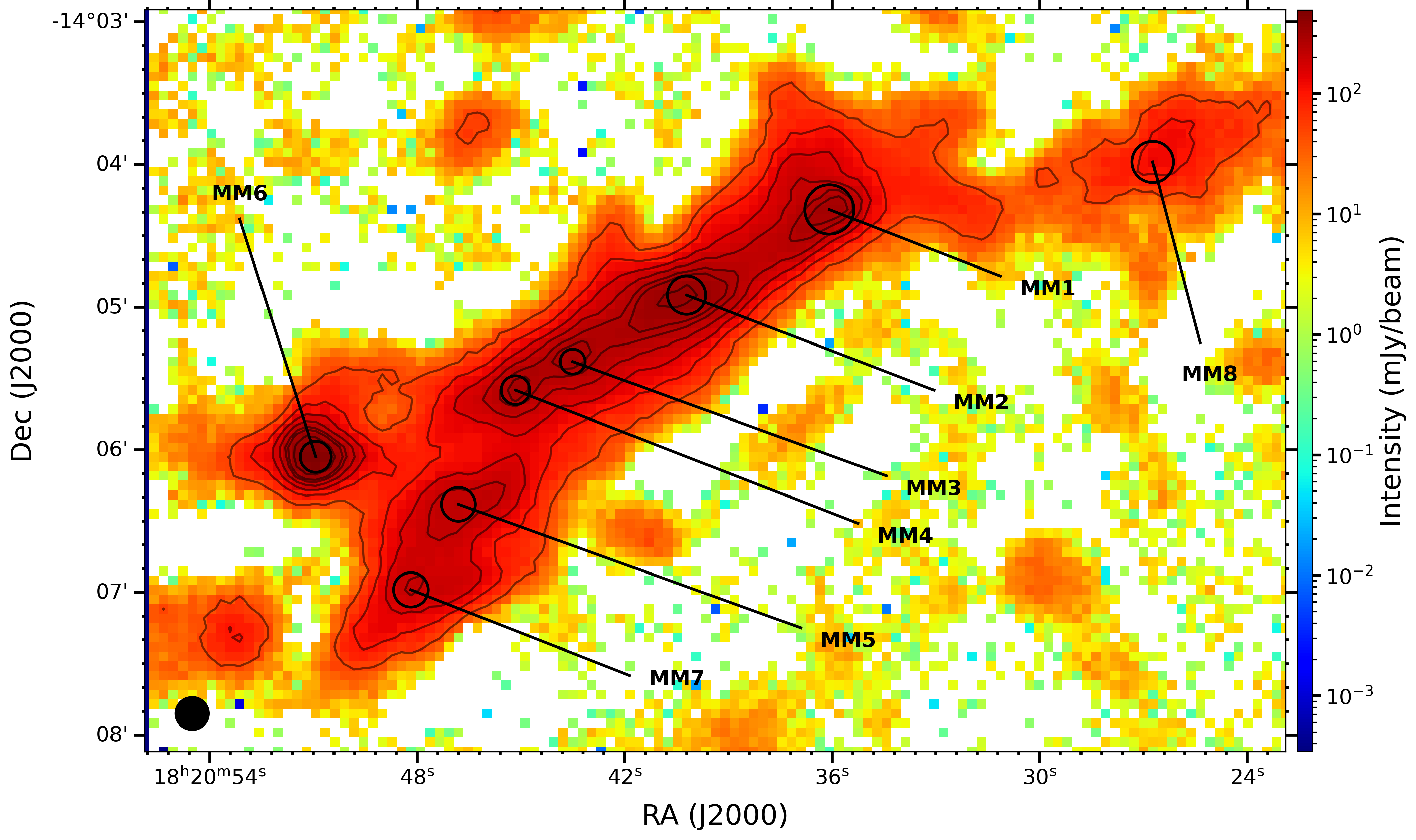} 
    \caption{Map of total emission intensity $I$ of the G16 filament observed by JCMT/POL-2 at 850 $\mu$m with a resolution of $14''.4$ indicated with a solid black circle. The black circles overlaid on the map represent the dense cores MM1, MM2, MM3, MM4, MM5, MM6, MM7 and MM8 identified on the filament and are labelled with black colors. The black contours are drawn at $I$ values of 50, 100, 150, 200, 250, 280, 300, 350 mJy/beam.} 
    \label{Figure:Intensity_map}
\end{figure*}

\begin{figure*}
    \centering
    \begin{tabular}{cc}
        \includegraphics[scale=0.4]{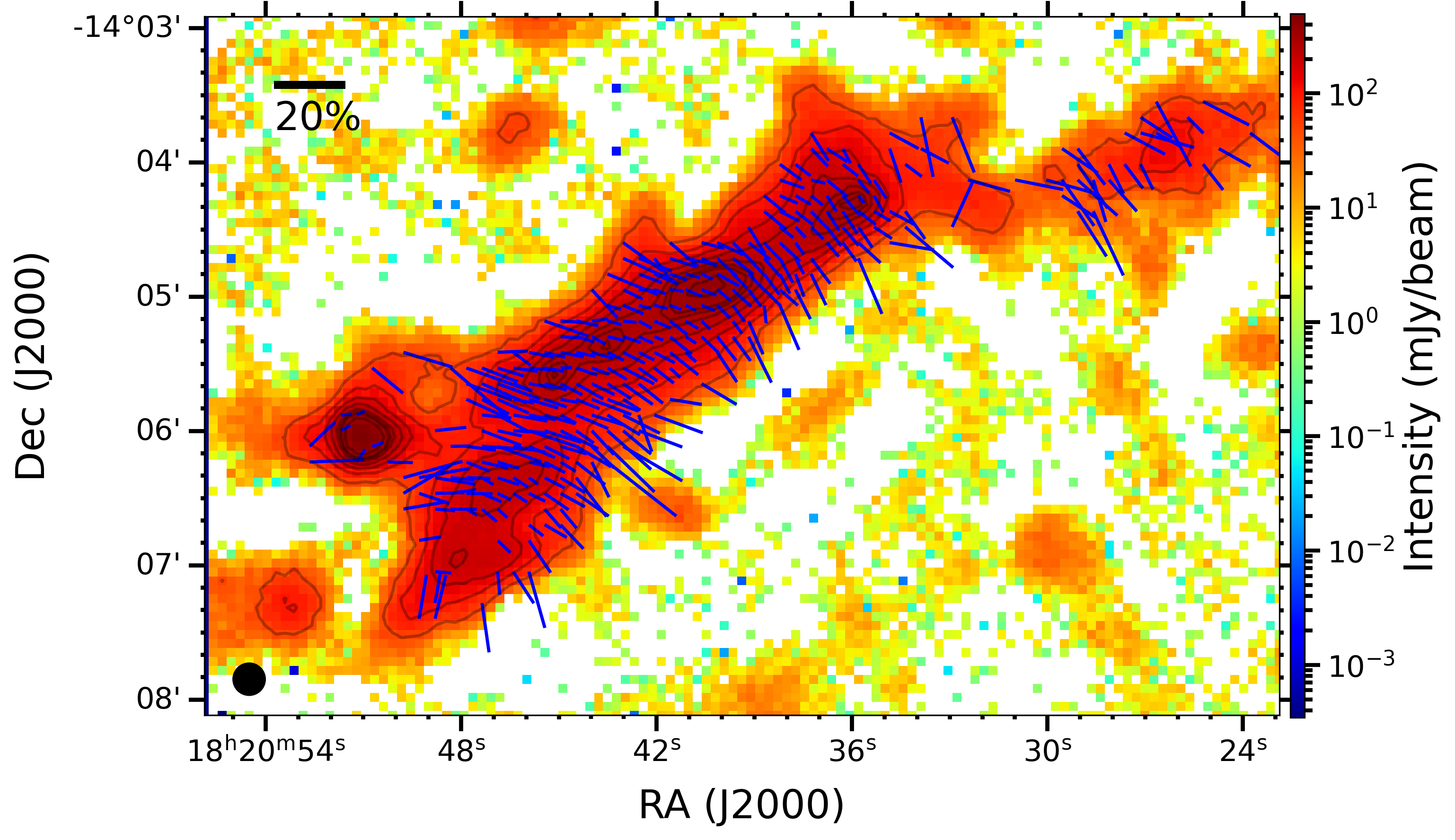} &  
        \hspace{5pt}
        \includegraphics[scale=0.33]{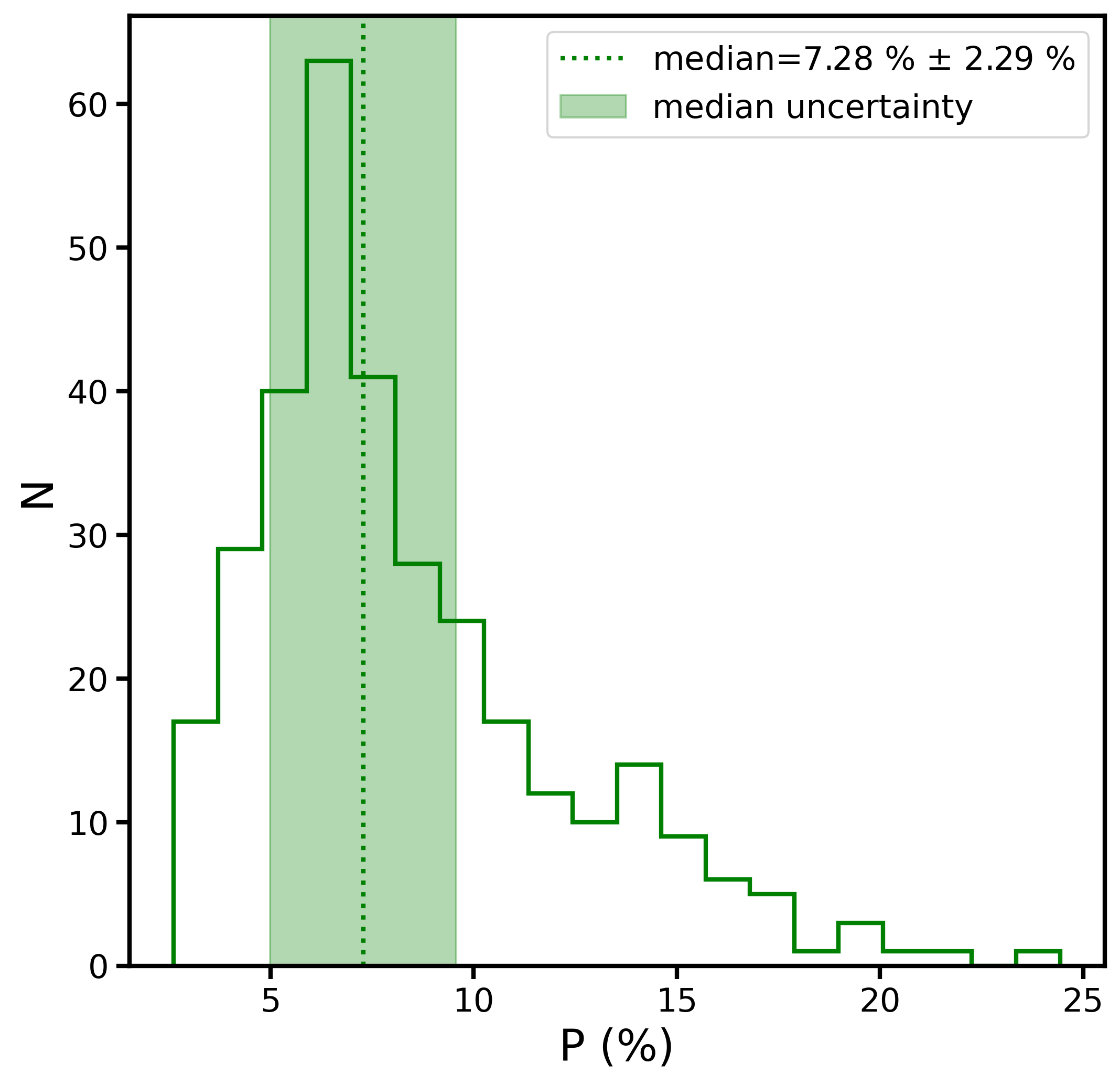}   
    \end{tabular}
    \caption{(Left) Same map as in Figure \ref{Figure:Intensity_map} but overlaid with polarization vectors shown with blue color. The lengths of the vectors are proportional to the polarization fractions $P$, and the directions of the vectors indicate the magnetic field orientations. A reference scale of $P$ of 20\% is also shown. (Right) Histogram of $P$ with the vertical dotted green line and shaded region denoting the median value of $P$ and its uncertainty.}
    \label{Figure:Intensity_P_map_Histogram_P}
\end{figure*}

The RAT theory is well supported by numerous observational studies in different molecular clouds using starlight polarization (for reviews, see \citealt{2015ARA&A..53..501A, 2015psps.book...81L}). However, the testing of RAT theory in the dense star-forming regions using thermal dust polarization observations is conducted only recently (e.g., \citealt{2022FrASS...9.3927T, 2023ApJ...953...66N, 2024ApJ...974..118N, 2025ApJ...981..128P}) with the advancement in far-IR/sub-mm/mm polarimetric instruments which facilitate the testing of grain alignment in these dense regions. However, there is still a lack of comprehensive testing of the grain alignment mechanisms in such dense regions. The synergy between various observations, theoretical advancements, and numerical modeling is essential to resolving the grain alignment problem, which remains a cornerstone issue in our understanding of interstellar dust and magnetic fields. In this paper, we aim to perform a comprehensive testing of the grain alignment mechanism in the context of the unified theory of the M-RAT mechanism, in a massive and dense star-forming Infrared Dark Cloud (IRDC).

\subsection{IRDC: An ideal environment for grain alignment studies}

IRDCs are considered as cradles for the formation of stars, especially high-mass stars and star clusters \citep{2006ApJ...641..389R}, and they are quiescent molecular clouds containing mostly cold and dense molecular gas. For testing grain alignment mechanisms, the gas column density or volume density and the radiation field strength or equivalently dust temperatures are the crucial physical parameters. A significant variation in these parameters over the clouds can favor the study of grain alignment mechanisms, which is shown by filamentary IRDCs making them ideal regions for grain alignment study in the context of the RAT alignment mechanism. Our study focuses on a massive quiescent filamentary IRDC G16.96+0.27, which is one of the brightest filaments in the JCMT SCOPE survey \citep{2018ApJS..234...28L}. It is located at a distance of 1.87 kpc and shows several fragmentations, which have been identified as cores MM1, MM2, MM3, MM4, MM5 and MM6 with MM1, MM2 and MM6 as protostellar cores and MM3, MM4 and MM5 as starless cores \citep{2020ApJS..249...33K, 2021A&A...654A.123M, 2021ApJS..256...25T}. This IRDC is associated with a simple filamentary structure and is dark at infrared wavelengths, with no prominent bright embedded sources. It is a quiescent filament at the very early stage of the star formation process. It shows significant variations in gas density and dust temperature from the outer region towards the inner region. All these environmental conditions make this filamentary IRDC an ideal region for our study. We aim to perform a comprehensive study using thermal dust polarization observations and detailed numerical modeling to constrain dust grain alignment mechanisms and dust properties in this filament. 


The rest of the paper is organised as: Section \ref{section:Observational data} provides the details on the observational data, Section \ref{section:Analysis and Results} presents the data analysis and results, Section \ref{section:Dust polarization modeling and results} presents the numerical modeling of dust polarization and the results, Section \ref{section:Discussions} discusses our results and Section \ref{section:Conclusions} summarizes our work.

\section{Observational data} \label{section:Observational data}
\subsection{Archival polarization data}
For our work, we use the archival thermal dust polarization data from \cite{2024ApJ...976..249G}. The polarimetric observations of G16.96+0.27 were done with the POL-2 instrument mounted on the James Clerk Maxwell Telescope (JCMT), from 2020 August to 2020 October using SCUBA-2/POL-2 DAISY mapping mode \citep{2013MNRAS.430.2513H, 2016SPIE.9914E..03F, 2018SPIE10708E..3MF} under Band 2 weather conditions. The JCMT/POL-2 has an effective beam size of $14''.4$ at 850 $\mu$m \citep{2021AJ....162..191M}, which corresponds to $\approx$ 0.13 pc at a distance of 1.87 kpc of the G16 filament. The data were reduced using the $pol2map$ routine in the Sub-Millimeter User Reduction Facility (SMURF) package \citep{2013MNRAS.430.2545C} of the STARLINK software \citep{2014ASPC..485..391C}. For the details on the polarimetric observations and data reduction processes, please refer to Section 2 of \cite{2024ApJ...976..249G}. For our analysis, we use debiased polarization data with the selection criteria of $I/\delta_I \geq 10$, $PI/\delta_{PI} \geq 3$ and $\delta_P \leq 5\%$, where $\delta_I$ is the uncertainty of the Stokes $I$, $PI$ is the debiased polarized intensity with its uncertainty $\delta_{PI}$ and $\delta_P$ is the uncertainty of the polarization fraction $P$.

\setlength{\tabcolsep}{0.4cm} 
\renewcommand{\arraystretch}{1.3}

\begin{table*}[ht]
    \centering
        \caption{Properties of cores identified on G16 filament}
    \begin{tabular}{ccccccc}
        \hline \hline
        Core ID & RA (J2000) & Dec (J2000) & Major size & Minor size & Effective size & Effective size \\
                & ($^\circ$) & ($^\circ$) & ($''$) & ($''$) & ($''$) & (pc) \\ \hline \hline
        MM1 & 275.150 & $-14.072$ & 25.16 & 17.00 & 20.69 & 0.19 \\
        MM2 & 275.168 & $-14.082$ & 17.39 & 14.91 & 16.10 & 0.15 \\
        MM3 & 275.181 & $-14.089$ & 12.44 & 8.82 & 10.48 & 0.09 \\
        MM4 & 275.188 & $-14.093$ & 15.63 & 9.31 & 12.06 & 0.11 \\
        MM5 & 275.195 & $-14.106$ & 18.41 & 11.03 & 14.25 & 0.13 \\
        MM6 & 275.212 & $-14.101$ & 15.41 & 11.04 & 13.04 & 0.12 \\
        MM7 & 275.201 & $-14.116$ & 18.85 & 11.11 & 14.47 & 0.13 \\
        MM8 & 275.111 & $-14.066$ & 23.57 & 12.82 & 17.38 & 0.16 \\ \hline    
    \end{tabular}

    \vspace{0.8em}
    
    \begin{minipage}{1\textwidth}
    Notes: (i) The major and minor sizes are deconvolved with the beam, i.e, $size = \sqrt{size^2_{uncorrected} - beam^2}$ \citep{2015A&C....10...22B}. \\
    (ii) The effective sizes both in $''$ and pc are the geometric means of the major and minor sizes, i.e, $Effective$ $size$ = $\sqrt{Major \hspace{0.1cm} size \times Minor \hspace{0.1cm} size}$. \\
    (iii) The physical effective size in pc is estimated considering a distance of 1.87 kpc of the G16 filament.
    \end{minipage}
    \label{Table:Core properties}
\end{table*}

\subsection{$H_2$ column density, volume density and dust temperature maps}
In this work, we use the $\mathrm{H_2}$ column density $N(\mathrm{H_2})$ and the dust temperature $T_\mathrm{d}$ maps of the G16 region from \cite{2024ApJ...976..249G}. The maps were obtained using level 2.5 processed archival Herschel PACS/SPIRE data at 70, 160, 250, 350, 500 $\mu$m and JCMT 850 $\mu$m through J-comb algorithm \citep{2022SCPMA..6599511J} and the derived maps have resolutions of $18''$. For the details of the derivation, please refer to Appendix B in \cite{2024ApJ...976..249G}.

We derive the $\mathrm{H_2}$ volume density map using the column density map. We assume a cylindrical geometry of the overall filament so that the depth of the filament can be considered to be equal to its width. However, since this filament hosts dense cores MM1, MM2, MM3, MM4, MM5, MM6, MM7 and MM8, we can not use the overall filament width value in the core regions. We assume the core regions to have spherical geometries so that their widths or diameters are equal to their depths. To detect the dense core regions and get the estimated size of the cores, we apply FellWalker algorithm \citep{2015A&C....10...22B} to the 850 $\mu$m total emission intensity map. In the process, we use pixels with intensities $> 3\sigma$, where $\sigma=5.3$ mJy/beam is the rms noise level of the background region in the intensity map. To identify real dense cores, we use the peak intensity threshold to be greater than 10$\sigma$ and the size greater than the beam size of $14''.4$. For the condition of having neighbouring peaks, these two peaks are considered to exist separately if the difference between the peak values and the minimum value (dip value) between the peaks is greater than 2$\sigma$. We find 24 cores over the whole observed field of G16 region. We find 8 dense cores in the filamentary region, named MM1, MM2, MM3, MM4, MM5, MM6, MM7 and MM8. The dense cores MM1, MM2, MM3, MM4, MM5 and MM6 were identified in earlier detections \citep{2020ApJS..249...33K, 2021A&A...654A.123M, 2021ApJS..256...25T}. In our detection, we also find two more clumps/cores, which we name as MM7 and MM8. The properties of the detected 8 dense cores are listed in Table \ref{Table:Core properties}. We use the core sizes given in the last column of Table \ref{Table:Core properties} to derive the volume density for these core regions. \cite{2024ApJ...976..249G} estimated the average width of the filament to be $\approx$ 0.46 $\pm$ 0.06 pc. We use this value of 0.46 pc as the depth value of the overall region of the filament except for the core regions to derive the volume density as follows:

\begin{equation}
{
n(\mathrm{H_2}) = \frac{N(\mathrm{H_2})}{d},
}
\end{equation}
where $d$ is the depth of the filament. We also note that the estimation of the width or depth of the filament is biased based on the angular resolutions of the observations \citep{2022A&A...657L..13P}. Figure \ref{Figure:NH2_Td_map}(a) shows the $\mathrm{H_2}$ column density map overlaid with circles that denote the dense core regions, and the region of the filament is marked with a solid rectangle; Figure \ref{Figure:NH2_Td_map}(b) shows the zoomed-in region of the rectangular region; Figure \ref{Figure:NH2_Td_map}(c) and (d) shows the derived volume density map and the dust temperature map, respectively. We find that the gas column density and volume density increase from the outer regions towards the inner regions, with the core regions showing higher values. MM6 core is the densest of the other cores. The core regions have a typical volume density value of the order of $10^5$ $\mathrm{cm^{-3}}$. The dust temperature significantly reduces from the outer regions towards the inner regions, with the lowest values of around 16 K observed in the MM2, MM3, MM4 and MM6 core regions. This implies the absence of bright embedded sources inside the filament. The only source of grain heating and thus grain temperature is the diffused interstellar radiation field (ISRF) which is strong in the outer regions but attenuates significantly towards the inner denser regions.



    

\begin{figure}
    \centering
        \includegraphics[scale=0.47]{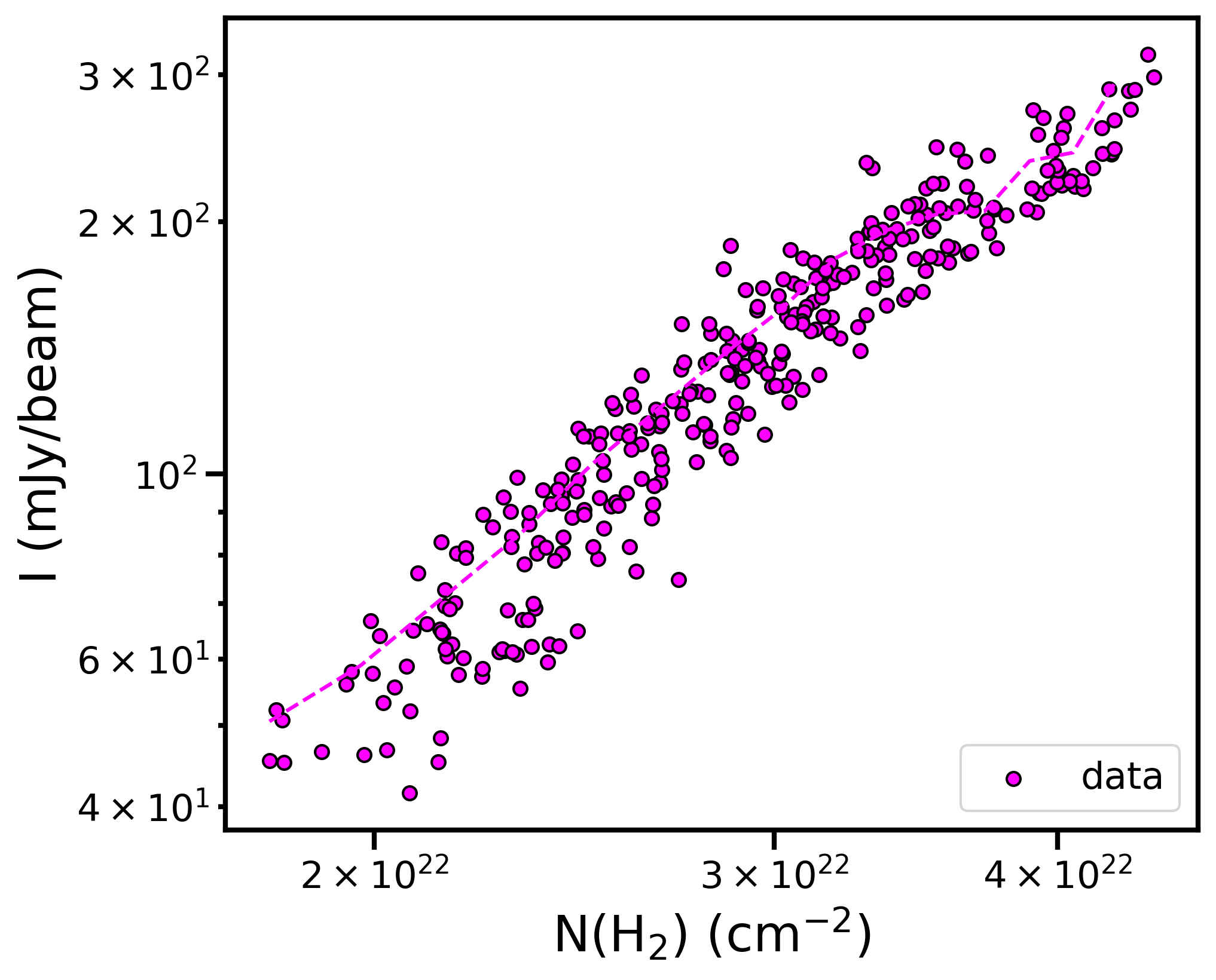} 
    \caption{Variation of total emission intensity with the gas column density for the pixels having polarization measurements. The magenta dashed line represents the running mean.} 
    \label{Figure:I_NH2}
\end{figure}

\begin{figure*}
    \centering
    \begin{tabular}{cc}
        \includegraphics[scale=0.49]{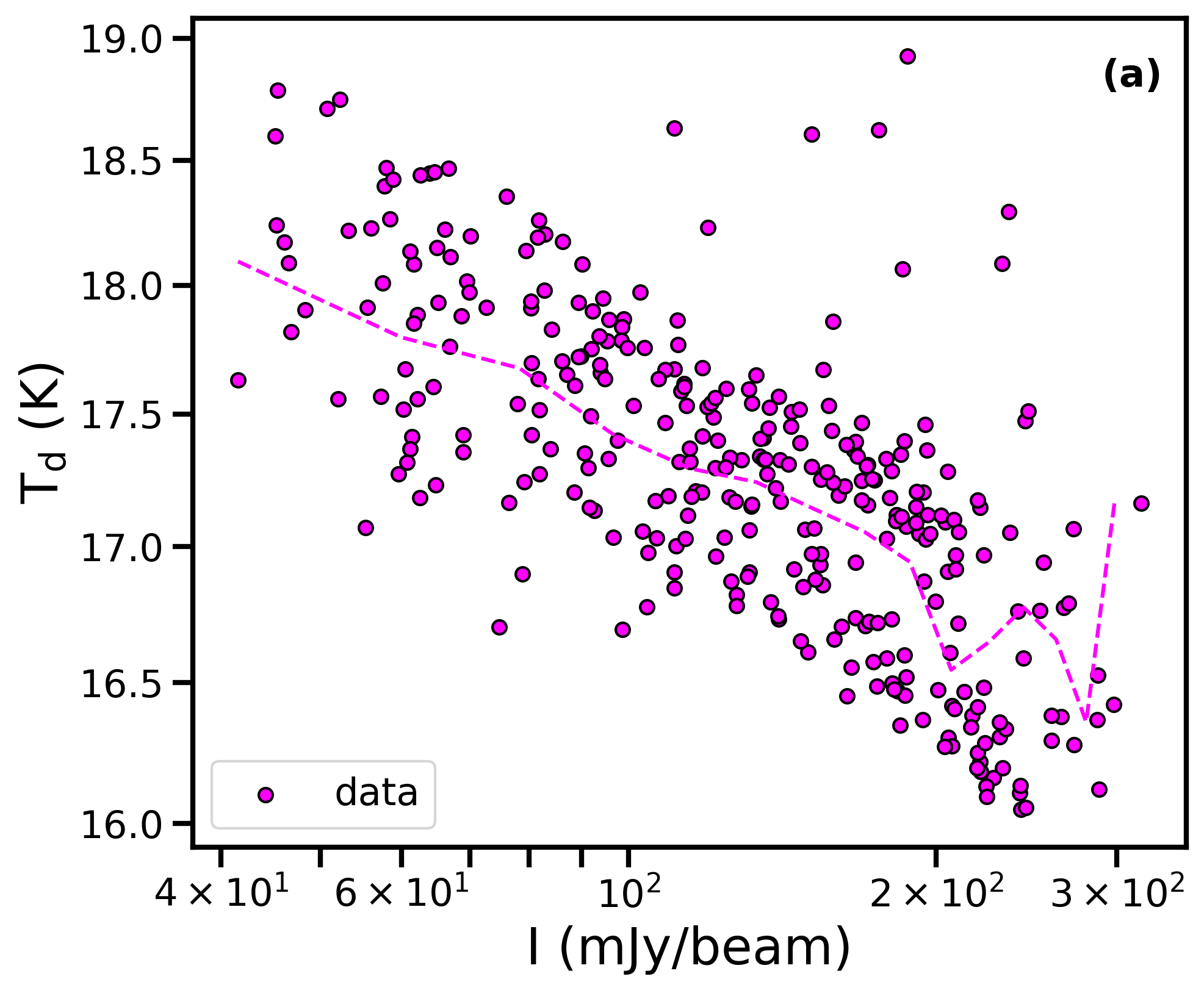} &  
        \hspace{5pt}
        \includegraphics[scale=0.49]{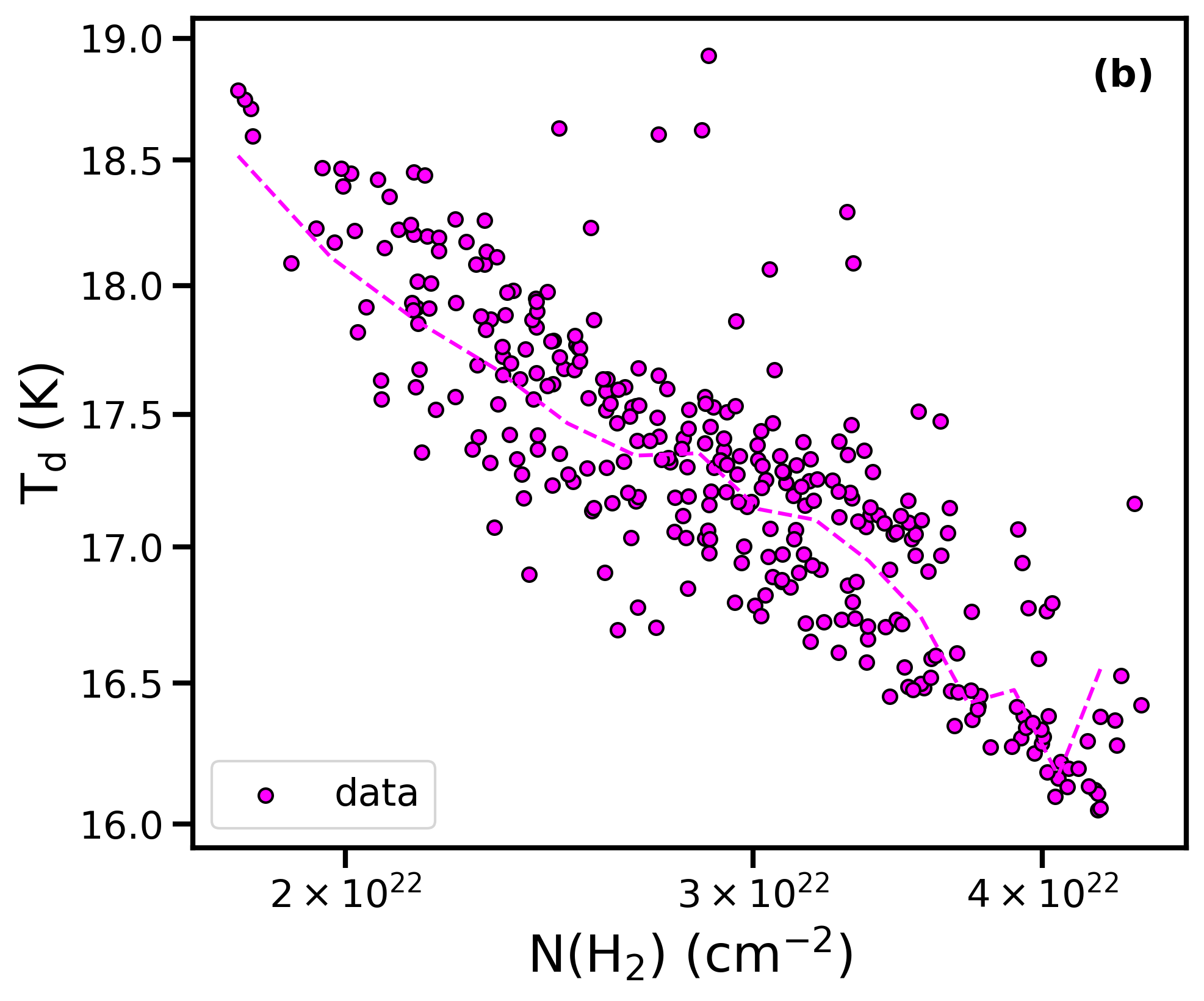}   
    \end{tabular}
    \caption{Variations of dust temperature with (a) the total emission intensity and (b) the gas column density for the pixels having polarization measurements. The magenta dashed lines represent the running means.}
    \label{Figure:Td_I_NH2}
\end{figure*}

\begin{figure*}
    \centering
    \begin{tabular}{cc}
        \includegraphics[scale=0.5]{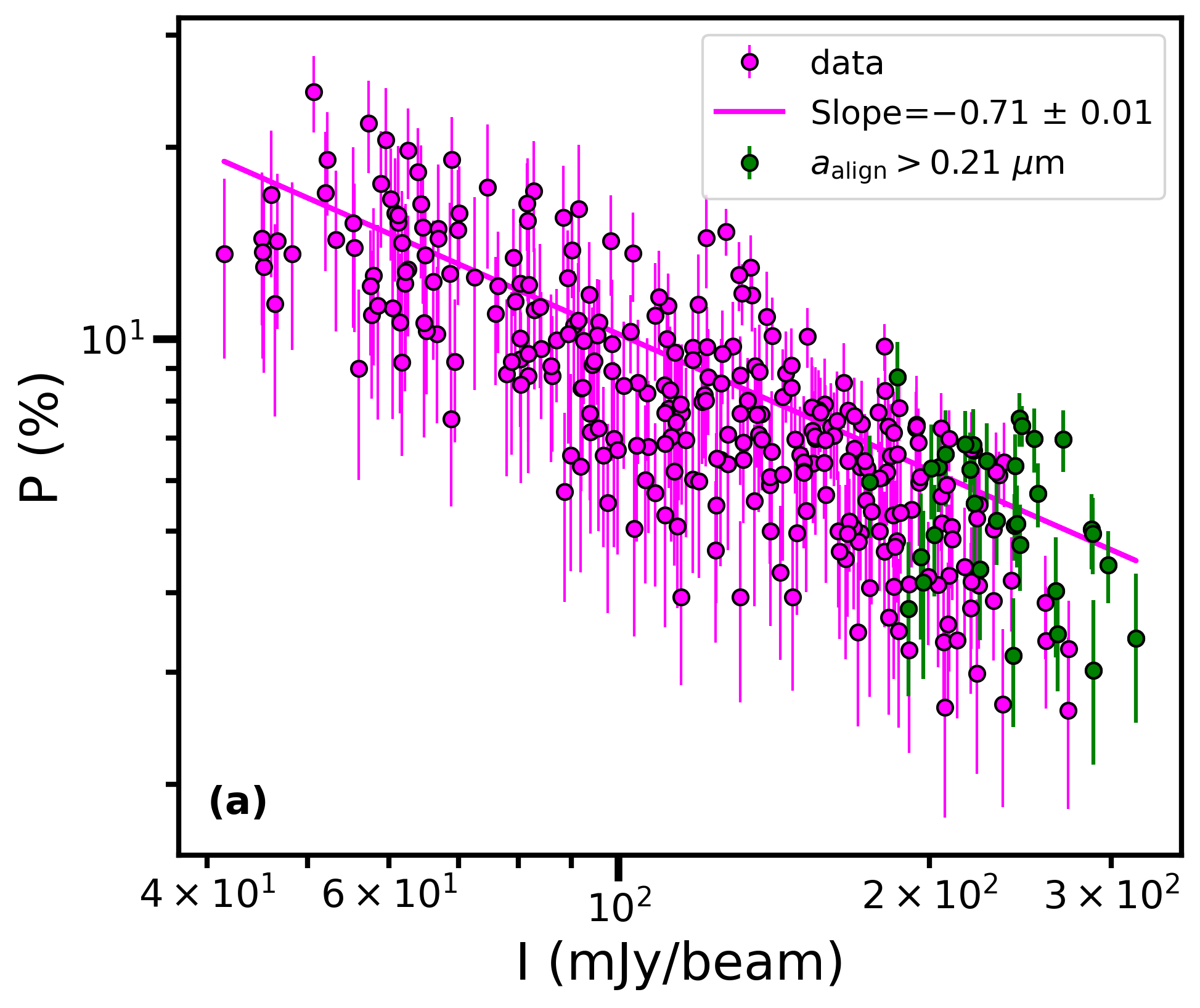} &  
        \hspace{5pt}
        \includegraphics[scale=0.5]{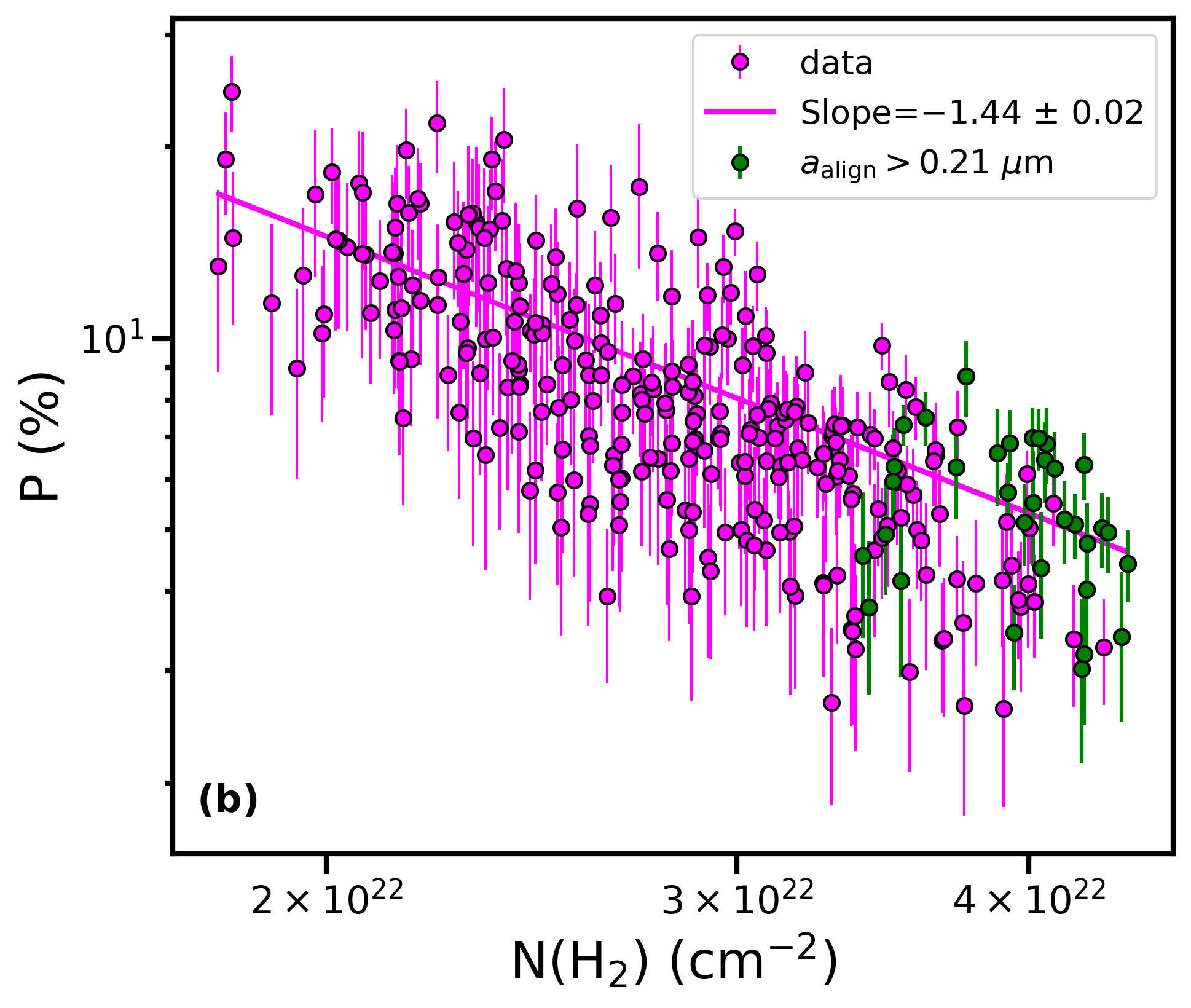}   
    \end{tabular}
    \caption{Variations of polarization fraction with (a) the total emission intensity and (b) the gas column density. The solid magenta lines represent the weighted best power-law fits. The data points shown with green color correspond to minimum grain alignment size, $a_\mathrm{align} > 0.21$ $\mu$m (for details on the parameter $a_\mathrm{align}$, please refer to Section \ref{Section: Minimum alignment size of grains}).}
    \label{Figure:P_I_NH2}
\end{figure*}

\begin{figure}
    \centering
        \includegraphics[scale=0.48]{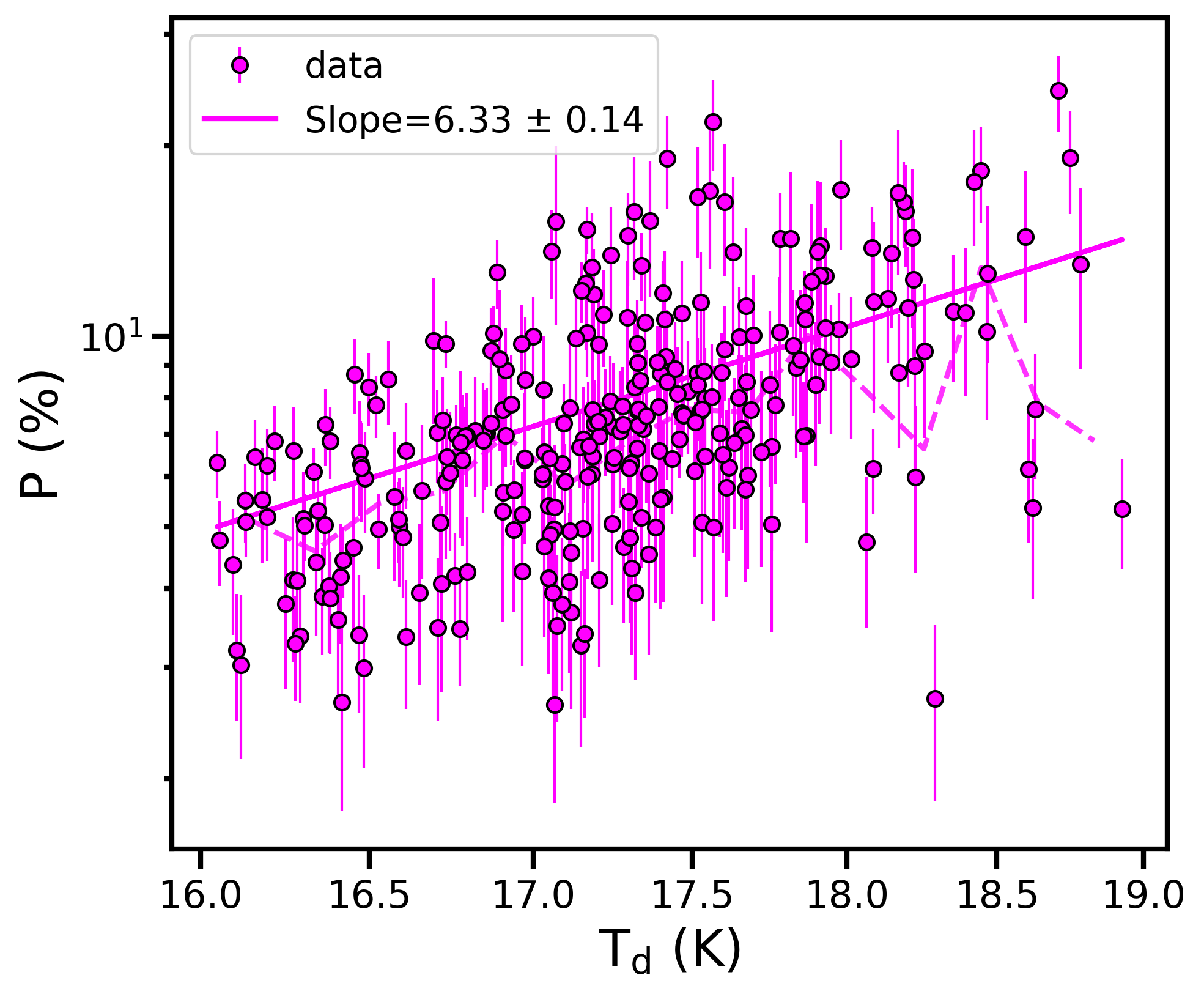} 
    \caption{Variation of polarization fraction with the dust temperature. The dashed magenta line represents the running means, whereas the solid magenta line represents the weighted best power-law fit.} 
    \label{Figure:P_Td}
\end{figure}

\section{Analysis and Results} \label{section:Analysis and Results}
\subsection{Data Analyses}
\subsubsection{Stokes 850 $\mu$m total intensity and magnetic field orientation maps}
Figure \ref{Figure:Intensity_map} shows the 850 $\mu$m total emission intensity Stokes $I$ map of the G16 filamentary region observed by JCMT/POL-2. We see that the intensity increases from the outer regions towards the inner regions of the filament. The dense cores MM1, MM2, MM3, MM4, MM5, MM6, MM7 and MM8 indicated with black circles on the figure, show higher values of total intensity, with the highest being found in the densest core MM6.

Figure \ref{Figure:Intensity_P_map_Histogram_P} (left) shows the same intensity map of Figure \ref{Figure:Intensity_map} but overlaid with the polarization half-vectors (hereafter, we use the term vectors to denote half-vectors for simplicity) with $PI/\delta_{PI} \geq 3$. The orientations of the vectors determine the B-field orientations, and the lengths are proportional to the polarization fraction. A reference length scale of 20\% polarization fraction is also indicated in the figure. We observe that the polarization fraction is higher in the outer regions associated with lower $I$ values and gets decreased towards the inner regions of the filament associated with higher $I$ values. Figure \ref{Figure:Intensity_P_map_Histogram_P} (right) shows the histogram distribution of the polarization fraction, which shows a right-skewed distribution with a median value of $7.3 \pm 2.3\%$. We observe high $P$ values of above 10\% up to around 23\%.   

\subsubsection{Relations among total intensity, gas column density and dust temperature}
We see the relations among $I$, $N(\mathrm{H_2})$ and $T_\mathrm{d}$ for the pixel coordinate regions where polarization is detected. Figure \ref{Figure:I_NH2} shows the variation of $I$ with $N(\mathrm{H_2})$, which shows a strong correlation, implying that higher dust emission intensity or denser regions are associated with higher gas column densities.

Figure \ref{Figure:Td_I_NH2} shows that $T_\mathrm{d}$ decreases with increasing $I$ and $N(\mathrm{H_2})$, implying the decrease in dust temperatures in the denser regions. This could be due to the absence of bright embedded sources inside the filament and the dust heating, and hence the dust temperature relies only on the diffused interstellar radiation field (ISRF), which is strong in the outer, less dense regions but attenuates significantly in the denser regions.

After analyzing the relations among $I$, $N(\mathrm{H_2})$ and $T_\mathrm{d}$, we analyze the variations of polarization fraction $P$ with the increase of the different parameters $I$, $N(\mathrm{H_2})$ and $T_\mathrm{d}$ in the following section.

\subsubsection{Variations of polarization fraction with total intensity, gas column density and dust temperature}
Figure \ref{Figure:P_I_NH2} shows the variations of polarization fraction $P$ with increasing total intensity $I$ (left) and $N(\mathrm{H_2})$ (right). We find that $P$ is very high reaching up to around 23\% in the outer regions with $I < 80$ mJy/beam and $N(\mathrm{H_2}) < 2.3 \times 10^{22}$ $\mathrm{cm^{-2}}$ but decreases with the increase in both $I$ and $N(\mathrm{H_2})$ towards the filament's spine. Best power-law fits of the forms $P=k_1I^{a_1}$ and $P=k_2[N(\mathrm{H_2})]^{a_2}$ are plotted with solid lines for the $P-I$ and $P-N(\mathrm{H_2})$ plots, respectively, where $k_1$ and $k_2$ are some proportionality constants and $a_1$ and $a_2$ are the respective slopes. The power-law fits yield $a_1$ value of $-0.71 \pm 0.01$ and $a_2$ value of $-1.44 \pm 0.02$. We mark some data points with green colors to indicate the data points that correspond to the minimum size of aligned grains, $a_\mathrm{align} > 0.21$ $\mu$m. For more details on the $a_\mathrm{align}$ parameter and the reason for marking these data points, please refer to Section \ref{Section: Minimum alignment size of grains} and \ref{section: Implication for anisotropic grain growth of aligned grains}.

Figure \ref{Figure:P_Td} shows the variation of $P$ with $T_\mathrm{d}$. We find that $P$ increases overall with the increase in $T_\mathrm{d}$. Best power-law fit of the form $P=k_3T_\mathrm{d}^{a_3}$, where $k_3$ is the proportionality constant and $a_3$ is the slope, is plotted with a solid line, thereby yielding $a_3$ value of $6.33 \pm 0.14$. An increase in $P$ with increasing $T_\mathrm{d}$ or, equivalently, radiation field strength is an expectation of the RAT alignment mechanism of grains.

The decreases in $P$ with increasing $I$ and $N(\mathrm{H_2})$ as observed in Figure \ref{Figure:P_I_NH2} are termed as polarization hole or depolarization. The exact cause of depolarization is not yet fully clear. However, popular explanations include the decrease in grain alignment efficiency in the denser regions or magnetic field tangling due to turbulence or a combination of both. To investigate whether there is any significant role of magnetic field tangling to cause the observed depolarization, we estimate the polarization angle dispersion function and analyze the effect of magnetic field tangling in the following section.

\begin{figure*}
    \centering
    \begin{tabular}{ccc}
        \includegraphics[scale=0.5]{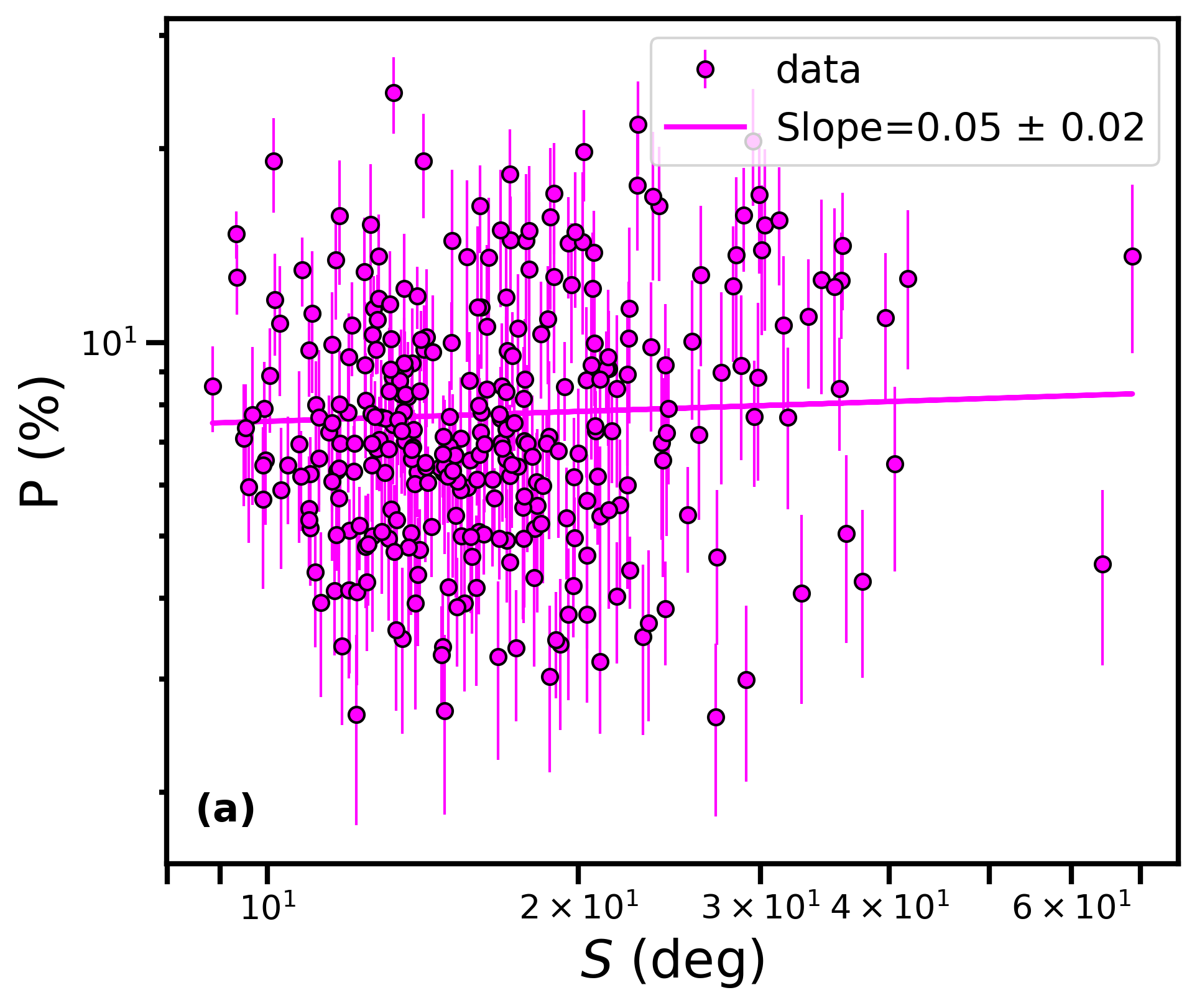} & 
        \hspace{5pt}
        \includegraphics[scale=0.5]{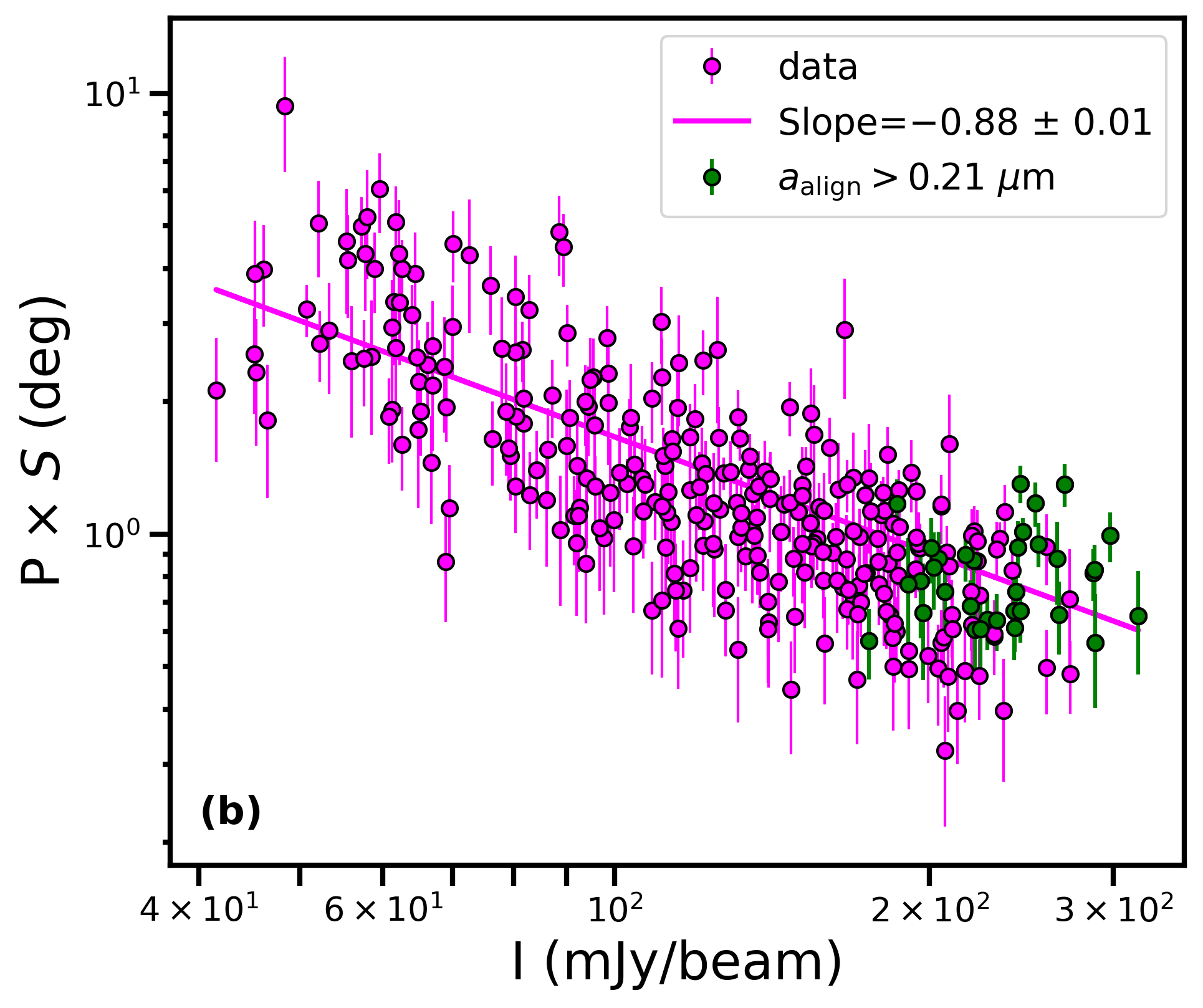} & \\ 
        \includegraphics[scale=0.5]{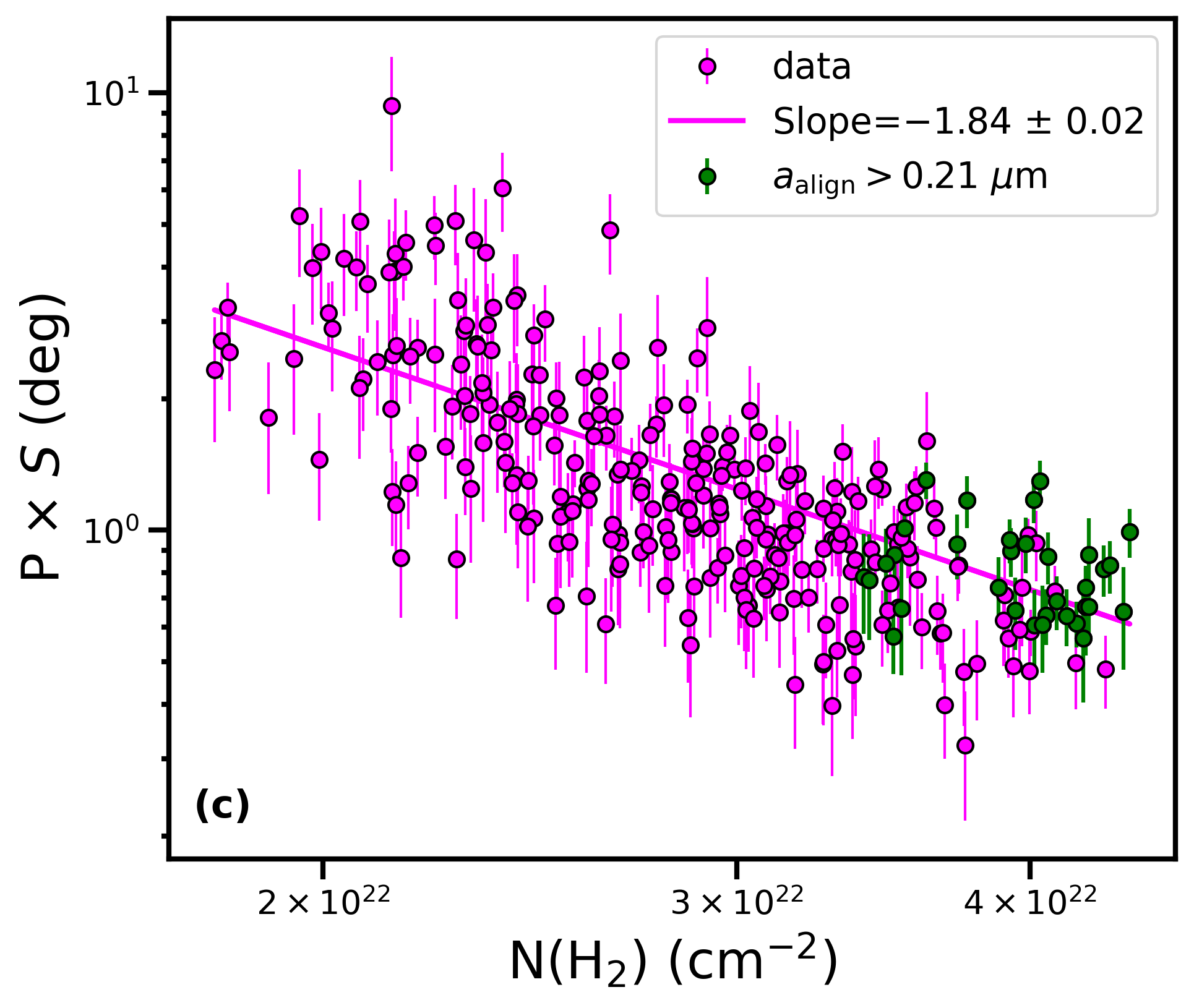} &
        \hspace{5pt}
        \includegraphics[scale=0.5]{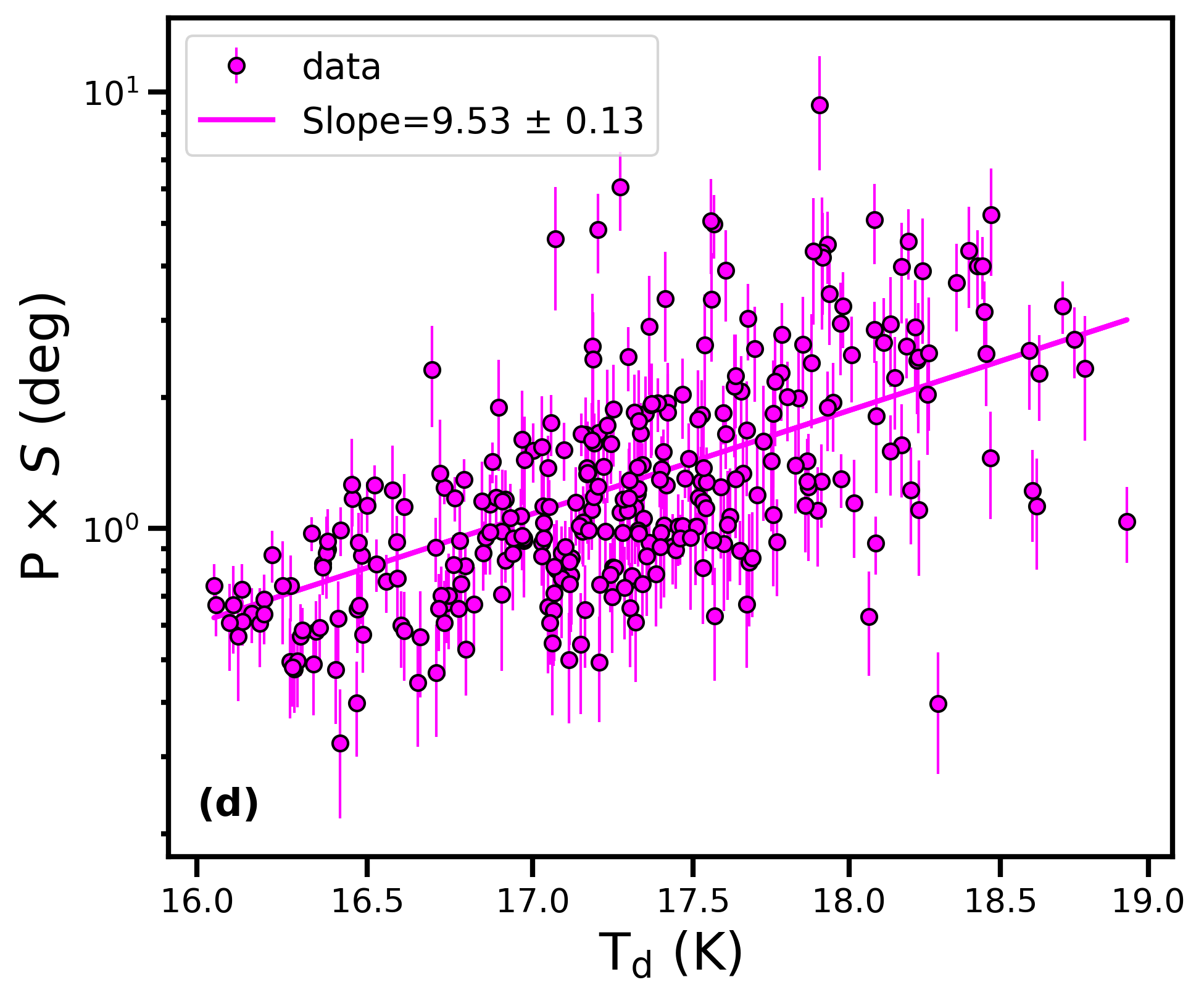}     
    \end{tabular}
    \caption{Variations of (a) $P$ with $S$; (b) $P \times S$ with $I$; (c) $P \times S$ with $N(\mathrm{H_2})$; and (d) $P \times S$ with $T_\mathrm{d}$. The solid magenta lines represent the weighted best power-law fits. The data points with green color correspond to $a_\mathrm{align} > 0.21$ $\mu$m, same as mentioned in Figure \ref{Figure:P_I_NH2}.}
    \label{Figure:P_S}
\end{figure*}

\begin{figure*}
    \centering
    \begin{tabular}{cc}
        \includegraphics[scale=0.41]{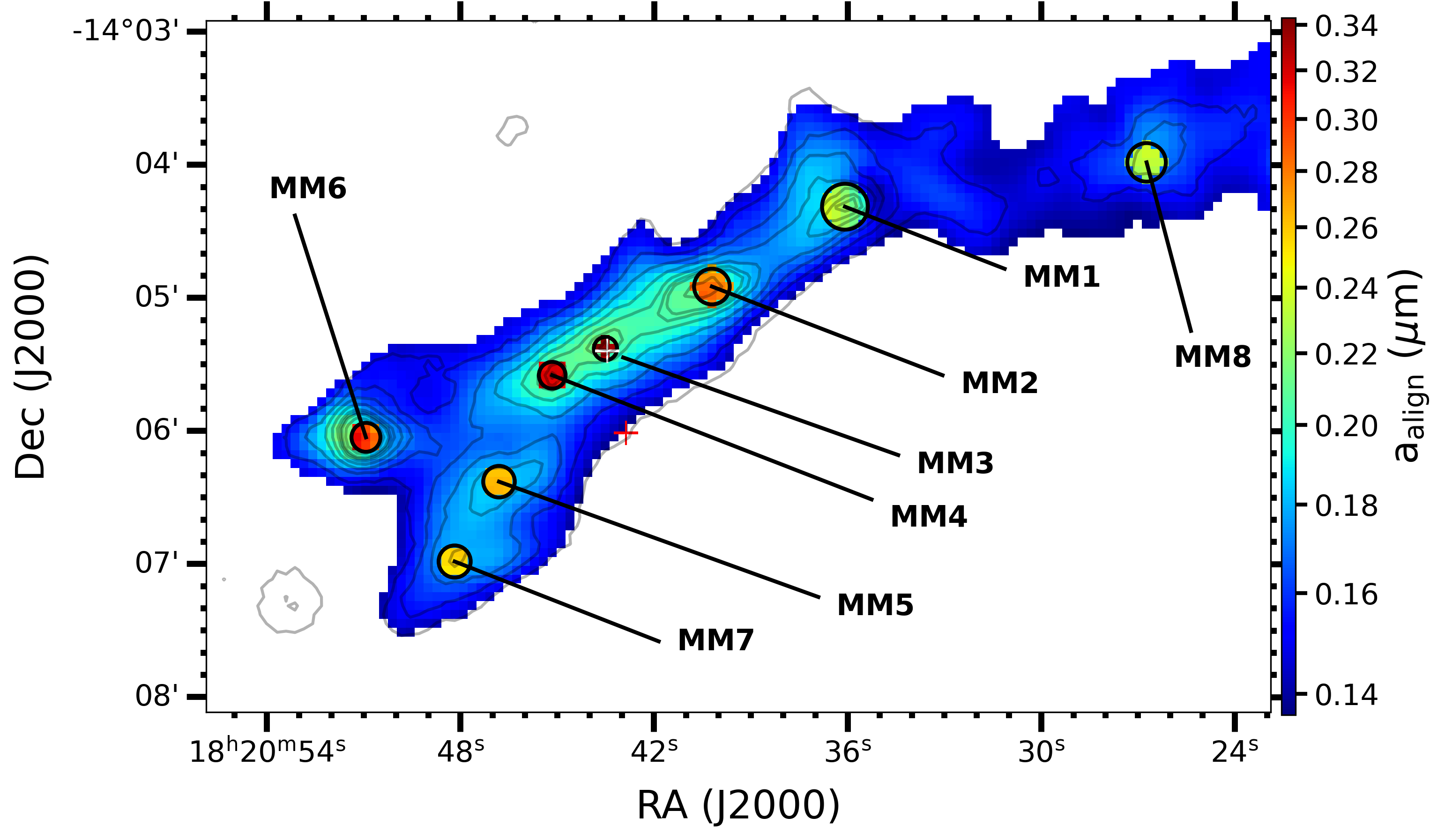} &  
        \hspace{5pt}
        \includegraphics[scale=0.35]{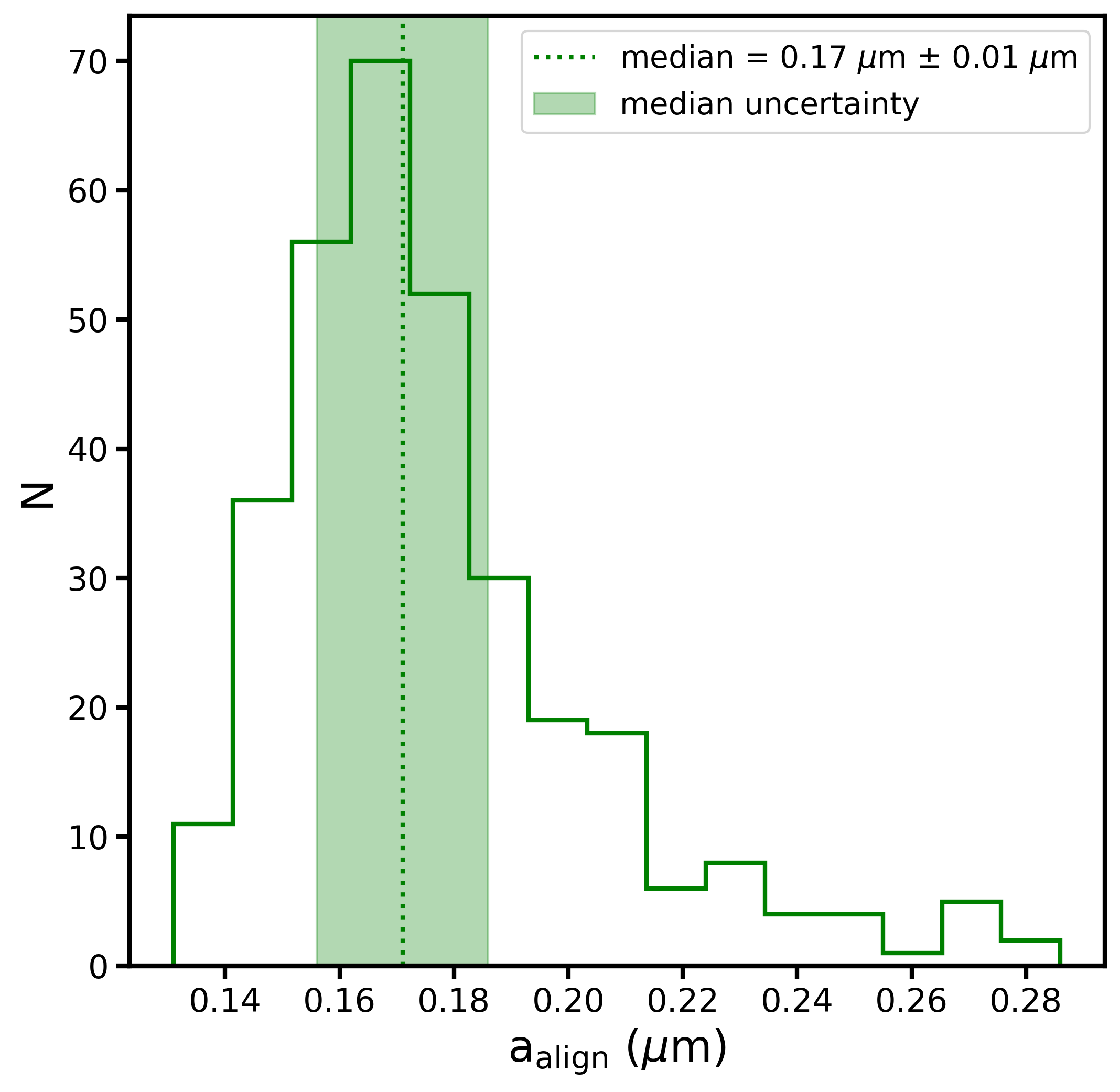}   
    \end{tabular}
    \caption{(Left) Map of minimum alignment size of grains, $a_\mathrm{align}$, with the circles representing the dense cores. The "+" symbols indicate the same locations as in Figure \ref{Figure:NH2_Td_map}. (Right) Histogram of $a_\mathrm{align}$ with the vertical green dotted line and the shaded region representing the median value of $a_\mathrm{align}$ and its uncertainty, respectively.}
    \label{Figure:a_align_map_Histogram_a_align}
\end{figure*}

\begin{figure}
    \centering
        \includegraphics[scale=0.47]{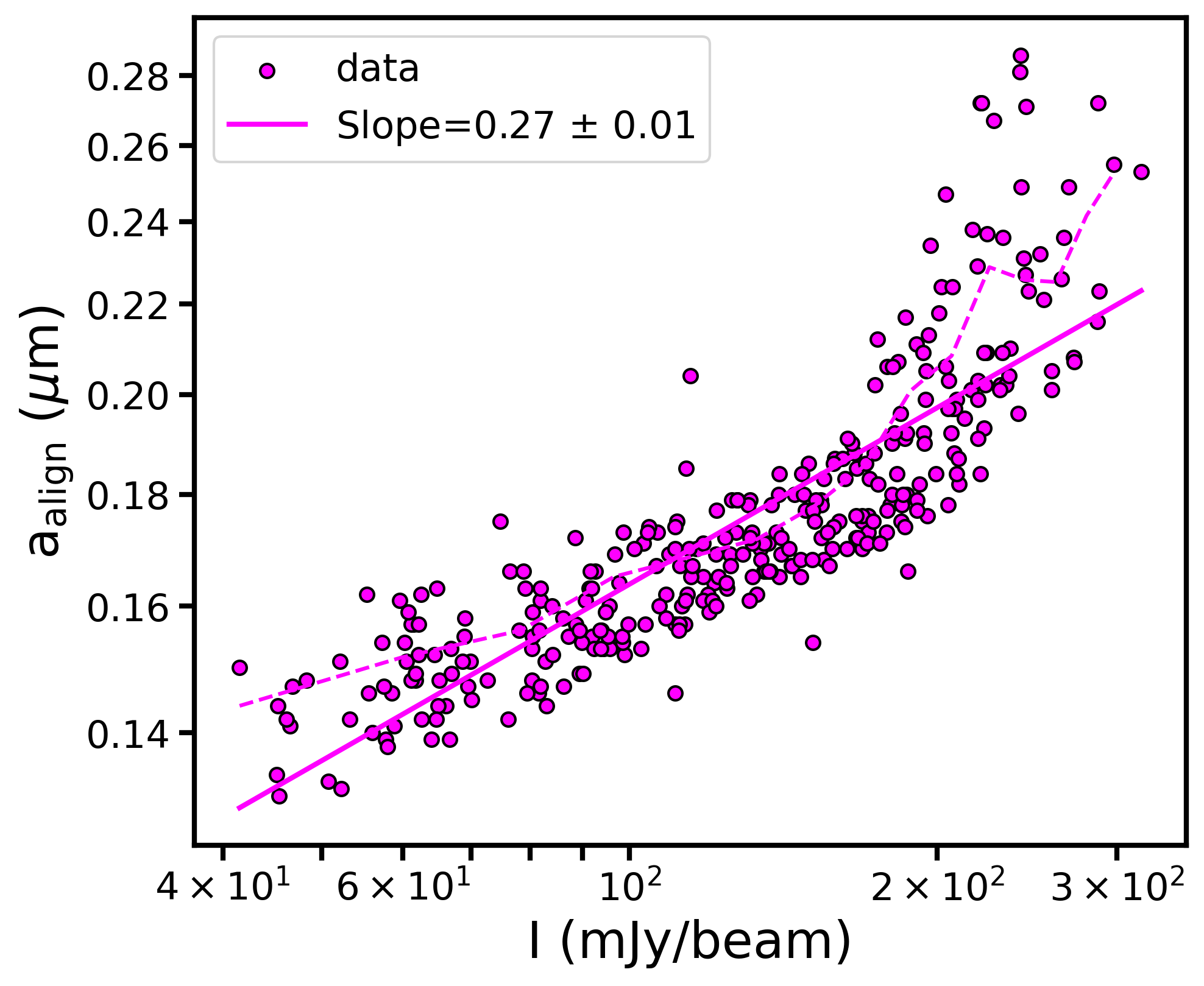} 
    \caption{Variation of minimum alignment size of grains with the total emission intensity. The magenta dashed and solid lines represent the running means and the best power-law fit, respectively.} 
    \label{Figure:a_align_I}
\end{figure}

\begin{figure*}
    \centering
    \begin{tabular}{cc}
        \includegraphics[scale=0.5]{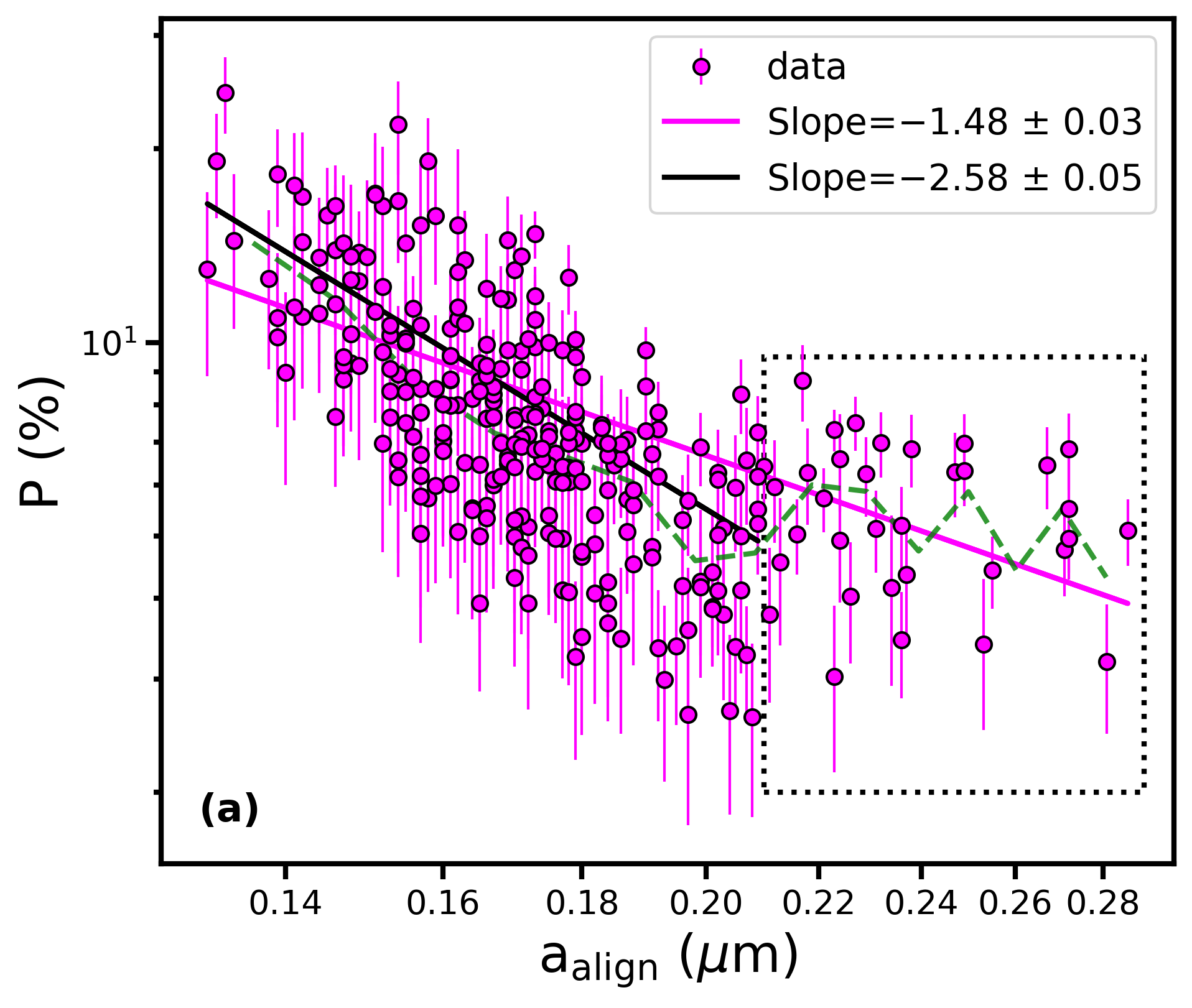} &  
        \hspace{5pt}
        \includegraphics[scale=0.5]{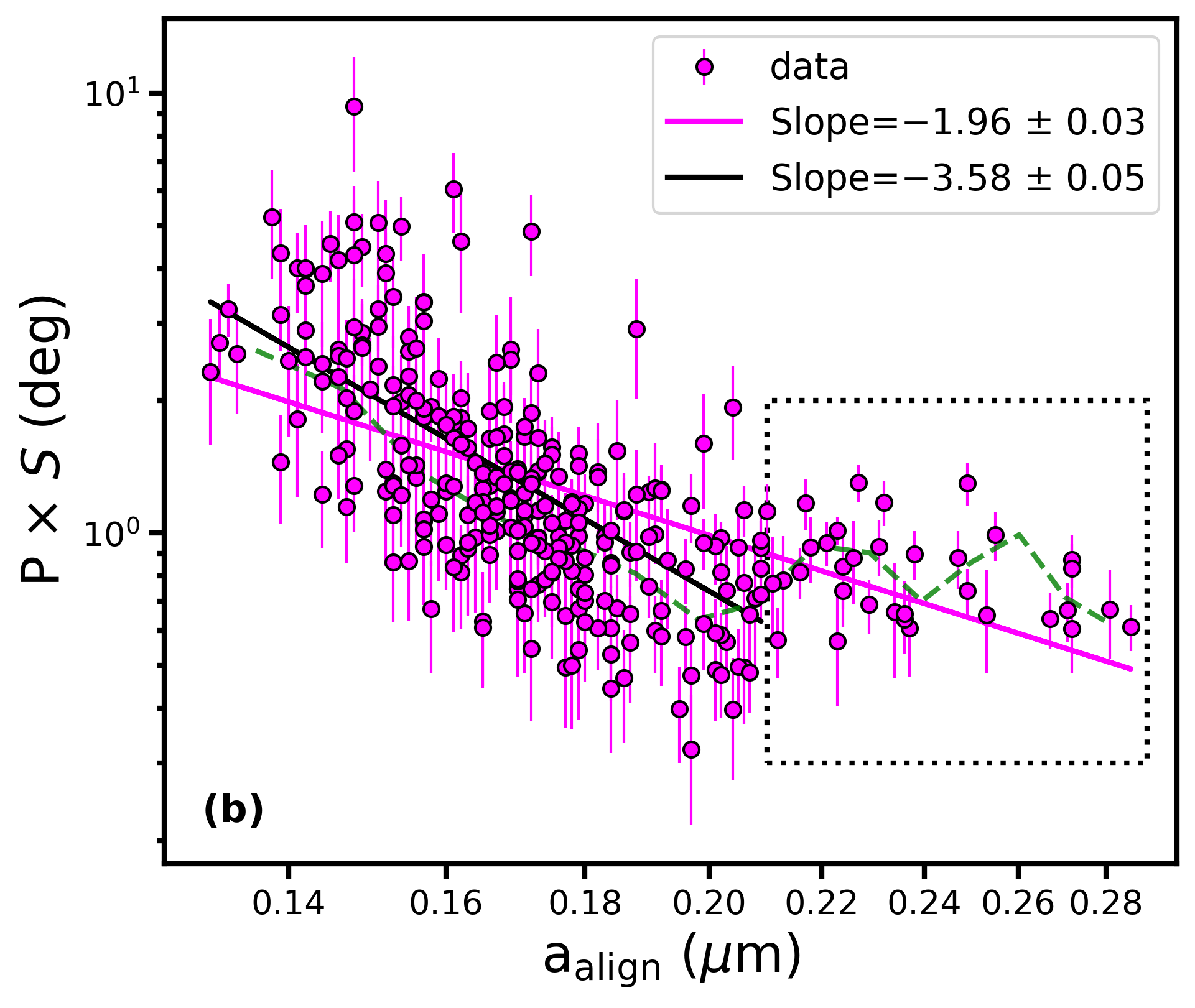}   
    \end{tabular}
    \caption{Variations of (a) $P$ with $a_\mathrm{align}$ and (b) $P \times S$ with $a_\mathrm{align}$. The green dashed and the magenta solid lines represent the weighted running means and the weighted best power-law fits, respectively, for the whole data. The black solid line is the weighted best power-law fit for the data with $a_\mathrm{align} < 0.21$ $\mu$m. The data points within the dotted rectangles correspond to $a_\mathrm{align} > 0.21$ $\mu$m.}
    \label{Figure:P_P_S_a_align}
\end{figure*}


\subsubsection{Polarization angle dispersion function; magnetic field tangling effect}

To assess for any significant contribution of magnetic field tangling to cause the observed polarization hole, we derive the polarization angle dispersion function denoted by $S$ and calculate the product $P \times S$. The value of $S$ provides an insight into the local non-uniformity in the distributions of the magnetic field morphology, and $P \times S$ provides information on the averaged grain alignment efficiency along the line of sight \citep{2020A&A...641A..12P}. A higher value of $S$ implies stronger magnetic field tangling, which can decrease the polarization fraction, and a lower $S$ value implies weaker magnetic field tangling, which can result in higher polarization fraction, considering a constant grain alignment efficiency along the line-of-sight.

For calculating $S$, we refer to the definition described in Section 3.3 of \cite{2020A&A...641A..12P} and the relation is given below:

\begin{equation}
{
S^2(r,\delta) = \frac{1}{N}\sum\limits_{i=1}^{N} {\left[\psi(r+\delta_i) - \psi(r)\right]}^2,
}
\end{equation}
 where the sum extends over the $N$ pixels, indexed by $i$ and located at positions $r+\delta_i$, within a circle centered on $r$ and having radius of $\delta$ taken as two times the beam size of JCMT/POL-2. The term $[\psi(r+\delta_i) - \psi(r)]$ is the difference in the polarization angles at positions $r+\delta_i$ and $r$.

As the Stokes parameters $Q$ and $U$ are associated with noise, $S$ becomes biased. This bias of $S$ can be positive or negative depending on whether the true value is smaller or larger than the random polarization angle of $52^\circ$ \citep{2016A&A...595A..57A}. An estimation of the variance of $S$ ($\sigma_S$) resulting from noise along with the debiased values of $S$ ($S_\mathrm{db}$) is described in Section 3.5 of \cite{2020A&A...641A..12P} and are given by the following relations 

\begin{equation}
\begin{split}
&\sigma_S^2(r,\delta)=\frac{\sigma_{\psi}^2(r)}{N^2S^2}\left[\sum\limits_{i=1}^{N} {\psi(r+\delta_i) - \psi(r)}\right]^2 + \\ &\hspace{1.5cm} \frac{1}{N^2S^2}\sum\limits_{i=1}^{N} \sigma_{\mathrm{\psi}}^2(r+\delta_i)\left[\psi(r+\delta_i) - \psi(r)\right]^2
\end{split}
\end{equation}


and
\begin{equation}
{
S_\mathrm{db}^2(r,\delta) = S^2 - \sigma_S^2 \hspace{0.5cm} \rm{if} \hspace{0.5cm} \it{S > \sigma_S}
}
\end{equation}

We take only those values of $S_\mathrm{db}$ with $S > \sigma_S$ and other values not meeting this criterion (around 3\%) are discarded. Hereafter, we will denote $S_\mathrm{db}$ by $S$ for convenience. Then, we analyse the variation of $P$ with the increase in $S$ as shown in Figure \ref{Figure:P_S}(a). We plot the weighted running means with a black dashed line. We do not find a significant correlation between $P$ and $S$, and the data shows a large spread. Most of the values of $S$ are concentrated below 25$^\circ$, signifying weak magnetic field tangling and a more ordered magnetic field.

We analyze the variations of the averaged grain alignment efficiency $P \times S$ with increasing $I$ and $N(\mathrm{H_2})$ as shown in Figure \ref{Figure:P_S}(b) and (c). We find that $P \times S$ decreases with $I$ and $N(\mathrm{H_2})$ almost similar to the decreases in $P$ with $I$ and $N(\mathrm{H_2})$ shown in Figure \ref{Figure:P_I_NH2}(a) and (b). We mark those data points that correspond to $a_\mathrm{align} > 0.21$ $\mu$m with green colors similar to Figure \ref{Figure:P_I_NH2}. Again, we analyze the variation of $P \times S$ with $T_\mathrm{d}$ as shown in Figure \ref{Figure:P_S}(d) and find that $P \times S$ increases with $T_\mathrm{d}$ almost similar to the increase of $P$ with $T_\mathrm{d}$ shown in Figure \ref{Figure:P_Td}. These imply that the effect of magnetic field tangling in causing the depolarization is minor and not significant. The observed depolarization is mainly due to the decrease in grain alignment efficiency in the denser regions, in good agreement with the RAT alignment mechanism of grains.

\subsection{Grain alignment mechanisms}\label{section:Grain alignment mechanisms}
\subsubsection{Minimum alignment size of grains} \label{Section: Minimum alignment size of grains}
The study of grain sizes is of great importance in the context of RAT-A theory to explain the grain alignment mechanism. According to RAT-A theory, efficient alignment of the grains can be achieved only when they rotate suprathermally with a rate exceeding about 3 times the thermal angular velocity \citep{2008MNRAS.388..117H, 2016ApJ...831..159H}. Under these conditions, the randomization of grains by gas-grain collisions can be ignored. The size distribution of aligned grains, spanning from the minimum alignment size, $a_{\mathrm{align}}$ to the maximum size, $a_{\mathrm{max}}$ \citep{2014MNRAS.438..680H, 2020ApJ...896...44L} determines the polarization fraction. In order to get an insight into the variations in the polarization fractions in different regions of the filaments in the context of grain sizes, we estimate the values of $a_\mathrm{align}$ over all the filaments using the analytical relation given below as established in \cite{2021ApJ...908..218H} 

\begin{equation}
\begin{split}
a_{\mathrm{align}}\simeq0.055\hat{\rho}^{-1/7}\left(\frac{\gamma U}{0.1}\right)^{-2/7}\left(\frac{n_\mathrm{H}}{10^3 \: \mathrm{cm^{-3}}}\right)^{2/7} \\ \times \left(\frac{T_\mathrm{gas}}{10 \: \mathrm{K}}\right)^{2/7} \left(\frac{\bar{\lambda}}{1.2 \: \mu \text{m}}\right)^{4/7} \left(1 + F_\mathrm{IR}\right)^{2/7}, 
\end{split}
\label{equation:a_align} 
\end{equation}
where $\hat{\rho} = \rho_\mathrm{d}/(3$ $\mathrm{gcm^{-3}})$ with $\rho_\mathrm{d}$ being the dust mass density; $\gamma$ is the anisotropy degree of the radiation field; $\bar{\lambda}$ represents the mean wavelength of the radiation; $U$ is the radiation field strength; $n_\mathrm{H}$ is the number density of hydrogen atoms; $T_\mathrm{gas}$ is the gas temperature and $F_\mathrm{IR}$ is the ratio of the IR damping to the collisional damping rate. For the diffused interstellar radiation field, $\gamma = 0.1$ \citep{1997ApJ...480..633D, 2007ApJ...663.1055B}. We use $\rho_\mathrm{d}=3$ $\mathrm{gcm^{-3}}$, $\bar{\lambda}=1.2$ $\mu$m, $n_\mathrm{H} = 2n(\mathrm{H_2})$ where $n(\mathrm{H_2})$ is the volume density of molecular hydrogen gas and $T_\mathrm{gas}=T_\mathrm{d}$ is considered as this thermal equilibrium between gas and dust is valid for dense and cold environments. For dense molecular clouds, $F_\mathrm{IR} \ll 1$. 



To calculate $U$, we use the relation between dust temperature and the radiation strength for silicate grains having sizes in the range of 0.01-1$\mu$m with dust heating and cooling balance and radiation strength $U < 10^4$ $(\approx 75$ K) i.e $U$ $\approx$ $(T_\mathrm{d}/16.4$ $\mathrm{K})^6$ \citep{2011piim.book.....D}. The map of the alignment size and the histogram distributions are shown in the left and the right panels of Figure \ref{Figure:a_align_map_Histogram_a_align}. The regions of the dense cores show higher $a_\mathrm{align}$ values compared to the other regions and $a_\mathrm{align}$ increases from the outer regions towards the inner regions of the filament. The histogram of $a_\mathrm{align}$ shows a right-skewed distribution with a median value of $0.17 \pm 0.01$ $\mu$m.

Within the framework of the RAT-A theory, the size distributions of aligned grains range from $a_\mathrm{align}$ to $a_\mathrm{max}$, and the polarization fraction is determined by this range of grain size distributions. The processes of grain growth and destruction determine the value of $a_\mathrm{max}$. When $a_{\mathrm{max}}$ is fixed, an increase in the value of $a_{\mathrm{align}}$ produces a narrower size distribution of aligned grains, which decreases the polarization fraction $P$. Again, a decrease in $a_{\mathrm{align}}$ results in a wider size distribution of aligned grains, which increases $P$ (see Figure 7 in \citealt{2022FrASS...9.3927T}). Therefore, an anti-correlation is expected between $a_{\mathrm{align}}$ and $P$. Also, RAT-A theory expects $a_\mathrm{align}$ to increase with the increase in total intensity $I$ or in denser regions in starless clouds.

\setlength{\tabcolsep}{1cm} 
\renewcommand{\arraystretch}{1.3}

\begin{table}
    \centering
        \caption{Slope values of different analyses}
    \begin{tabular}{cc}
        \hline \hline
        Relation between & Slope \\ \hline \hline
        $P$ vs. $I$ & $-0.71 \pm 0.01$ \\ \hline
        $P$ vs. $N(\mathrm{H_2})$ & $-1.44 \pm 0.02$ \\ \hline
        $P$ vs. $T_\mathrm{d}$ & $6.33 \pm 0.14$ \\ \hline
        $P \times S$ vs. $I$ & $-0.88 \pm 0.01$ \\ \hline
        $P \times S$ vs. $N(\mathrm{H_2})$ & $-1.84 \pm 0.02$ \\ \hline
        $P \times S$ vs. $T_\mathrm{d}$ & $9.53 \pm 0.13$ \\ \hline
        $a_\mathrm{align}$ vs. $I$ & $0.27 \pm 0.01$ \\ \hline

        $P$ vs. $a_\mathrm{align}$ & \parbox{2cm}{$-1.48 \pm 0.03$ (overall) \\ $-2.58 \pm 0.05$ ($a_\mathrm{align} < 0.21$ $\mu$m)}   \\ \hline
        $P \times S$ vs. $a_\mathrm{align}$ & \parbox{2cm}{$-1.96 \pm 0.03$ (overall) \\ $-3.58 \pm 0.05$ ($a_\mathrm{align} < 0.21$ $\mu$m)}  \\ \hline    
    \end{tabular}
    \label{Table:Slope values of different analyses}
\end{table}



Figure \ref{Figure:a_align_I} shows the variation of $a_\mathrm{align}$ with $I$. We find that $a_\mathrm{align}$ increases with the increase in $I$ as expected by the RAT-A mechanism. Again, we analyze the variations of $P$ and $P \times S$ with $a_\mathrm{align}$ as shown in Figure \ref{Figure:P_P_S_a_align} and find that both $P$ and $P \times S$ overall decrease with increasing $a_\mathrm{align}$, which implies that the effect of magnetic field tangling on the decrease in polarization fraction is minor and the major cause of the decrease in $P$ is the decrease in the grain alignment efficiency with the increase in $a_\mathrm{align}$. The decrease in the polarization fraction as $a_\mathrm{align}$ increases is due to the reduction in the fraction of aligned grains as shown in the numerical modeling in \cite{2020ApJ...896...44L} and \cite{2021ApJ...908..218H}. Equivalently, the decrease in the polarization fraction is due to the decrease in the RAT alignment efficiency of grains in the denser regions, which are associated with higher total dust emission intensity, higher gas column density, lower dust temperature, and hence higher $a_\mathrm{align}$ values. However, if we closely observe the variations of $P$ and $P \times S$ with $a_\mathrm{align}$, we see that both $P$ and $P \times S$ decrease significantly up to $a_\mathrm{align}$ value of 0.21 $\mu$m and then after 0.21 $\mu$m we see an increment and become nearly flat. Therefore, we see a tendency of the larger aligned grains with $a_\mathrm{align} > 0.21$ $\mu$m to produce higher polarization fraction. For a discussion on these observed features, please refer to Section \ref{section: Implication for anisotropic grain growth of aligned grains}.

Table \ref{Table:Slope values of different analyses} summarizes the slope values of all the different analyses of relations among different parameters as described above.

\begin{figure*}
    \centering
    \begin{tabular}{cccc}
        \includegraphics[scale=0.24]{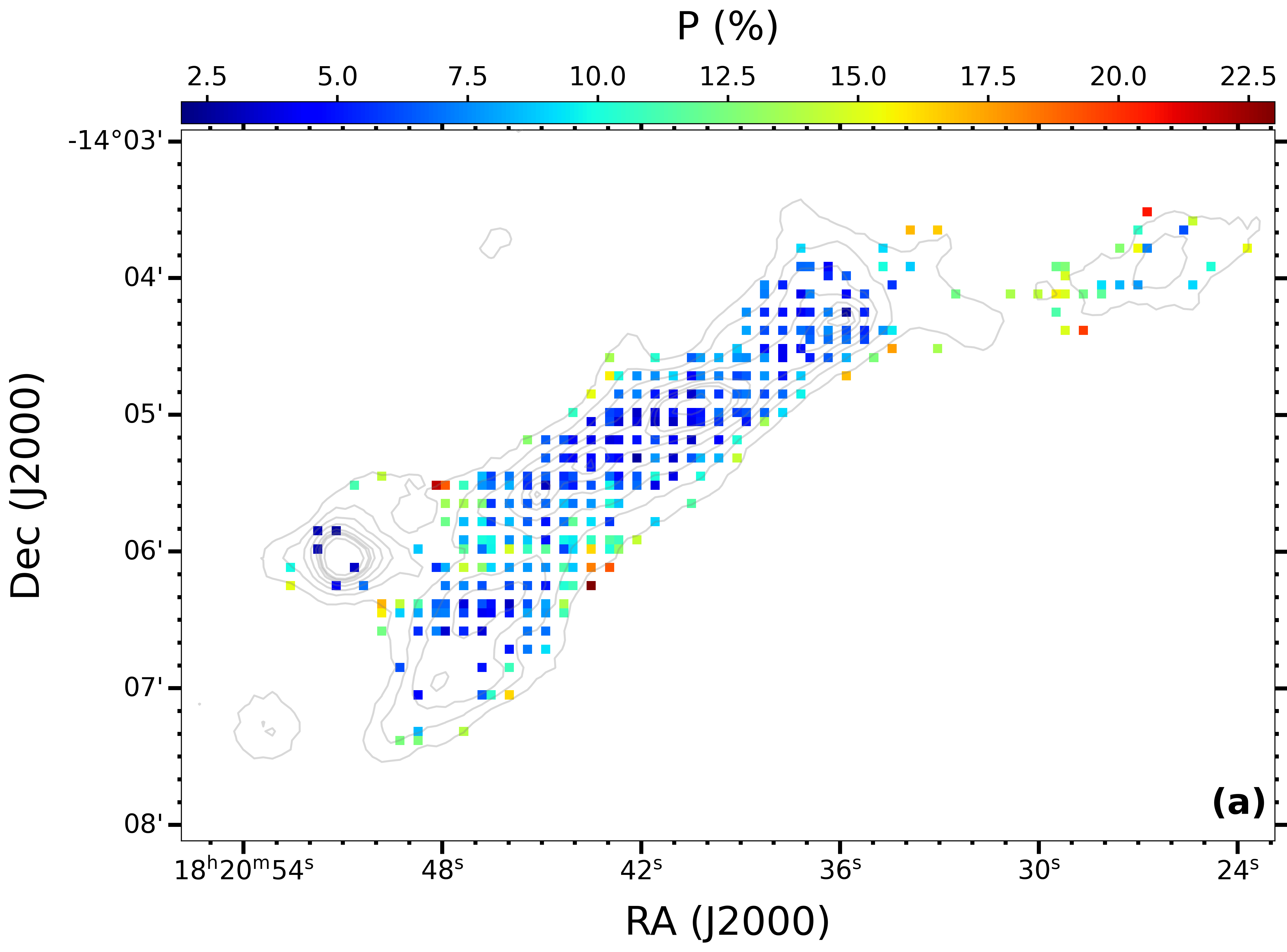} & 
        \hspace{-50pt}
        \includegraphics[scale=0.24]{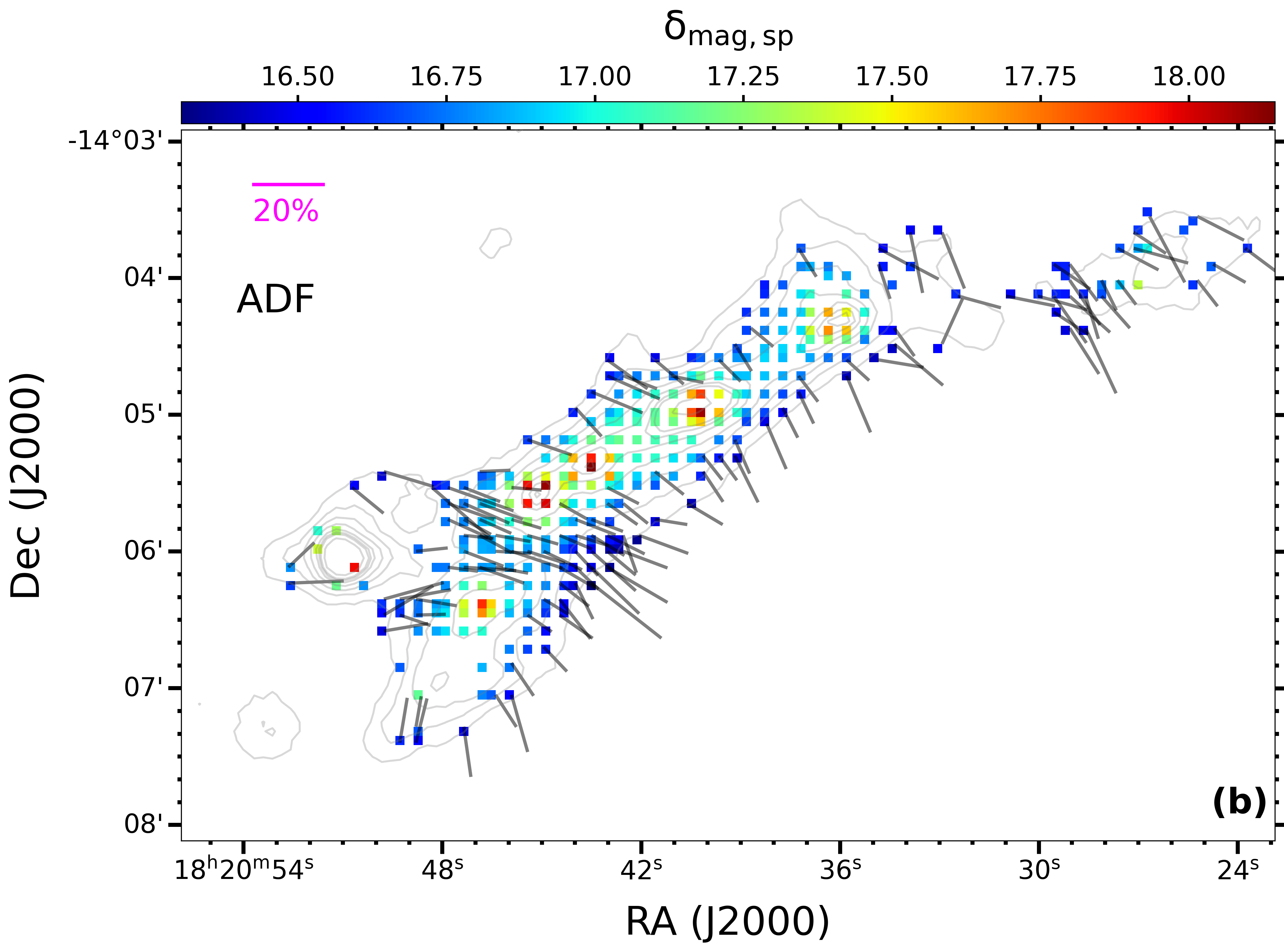} & \\ 
        \includegraphics[scale=0.24]{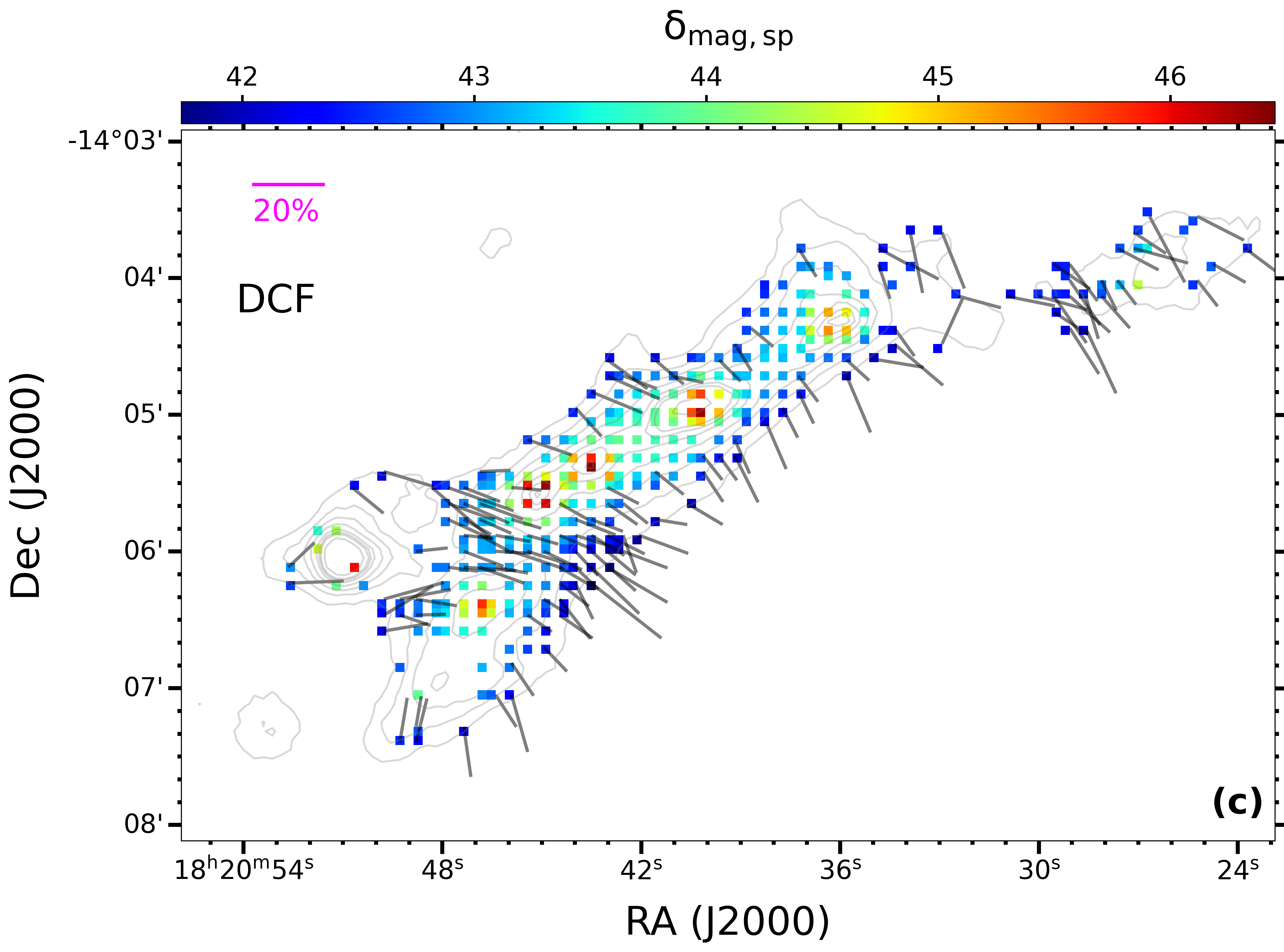} &
        \hspace{-50pt}
        \includegraphics[scale=0.24]{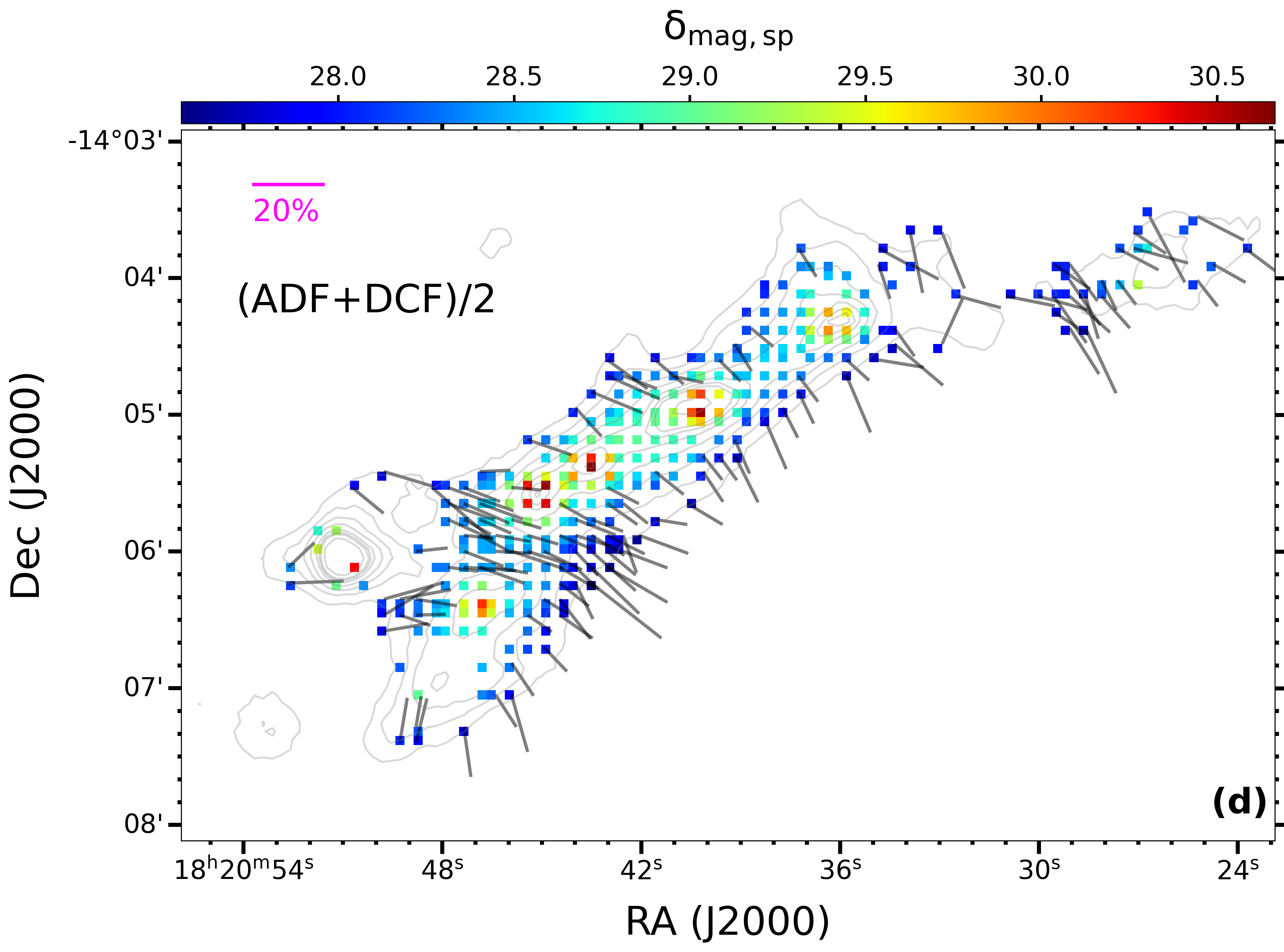}    
    \end{tabular}
    \caption{(a) Map of observed polarization fraction $P$; (b) Map of magnetic relaxation strength $\delta_\mathrm{mag,sp}$ estimated using the magnetic field strength values derived from ADF method, and overlaid with polarization vectors of $P \geq 8$\% shown in gray color; (c) same as (b) but for DCF method; (d) same as (b) but for the average of the ADF and DCF methods.}
    \label{Figure:mr_ADF_DCF}
\end{figure*}

\setlength{\tabcolsep}{1.2cm} 
\renewcommand{\arraystretch}{1.3}

\subsection{Effect of magnetic relaxation on RAT Alignment} \label{section:Effect of magnetic relaxation on RAT Alignment}

In the study of grain alignment mechanisms, the magnetic properties of dust grains are crucial as these properties enable the grains to interact with the external magnetic field. When there is diffuse distribution of iron atoms within a silicate grain, the grain behaves as an ordinary paramagnetic material. However, when iron atoms are distributed as clusters, the grain becomes super-paramagnetic \citep{2022AJ....164..248H}. A paramagnetic grain that rotates with an angular velocity $\omega$ in the presence of an external magnetic field $B$ undergoes paramagnetic relaxation \citep{1951ApJ...114..206D} that induces dissipation of the rotational energy of the grains into heat, resulting in the gradual alignment of angular velocity and angular momentum with $B$, known as the classical Davis-Greenstein mechanism which is applicable to any magnetic material. However, the efficient alignment of grains can not be achieved by the paramagnetic relaxation alone due to the randomization of grains by gas-grain collisions. Again, only RATs may not be able to produce perfect alignment of grains as the RAT alignment efficiency depends on various factors such as the angle between the radiation and the magnetic field directions, grain properties like shape and compositions \citep{2016ApJ...831..159H, 2021ApJ...913...63H}. In the filament of our study, it is observed that there are high $P$ values of more than 10\% and reaching up to around 23\%. Therefore, we explore the effect of the magnetic relaxation on the RAT alignment efficiency by considering super-paramagnetic grains having embedded iron atoms as clusters, to explain the observed high $P$ values, especially in the outer regions of the filament. This consideration of super-paramagnetic nature of the grains is expected in denser regions due to the evolution of grains.


\begin{figure*}
    \centering
    \begin{tabular}{ccc}
        \includegraphics[scale=0.4]{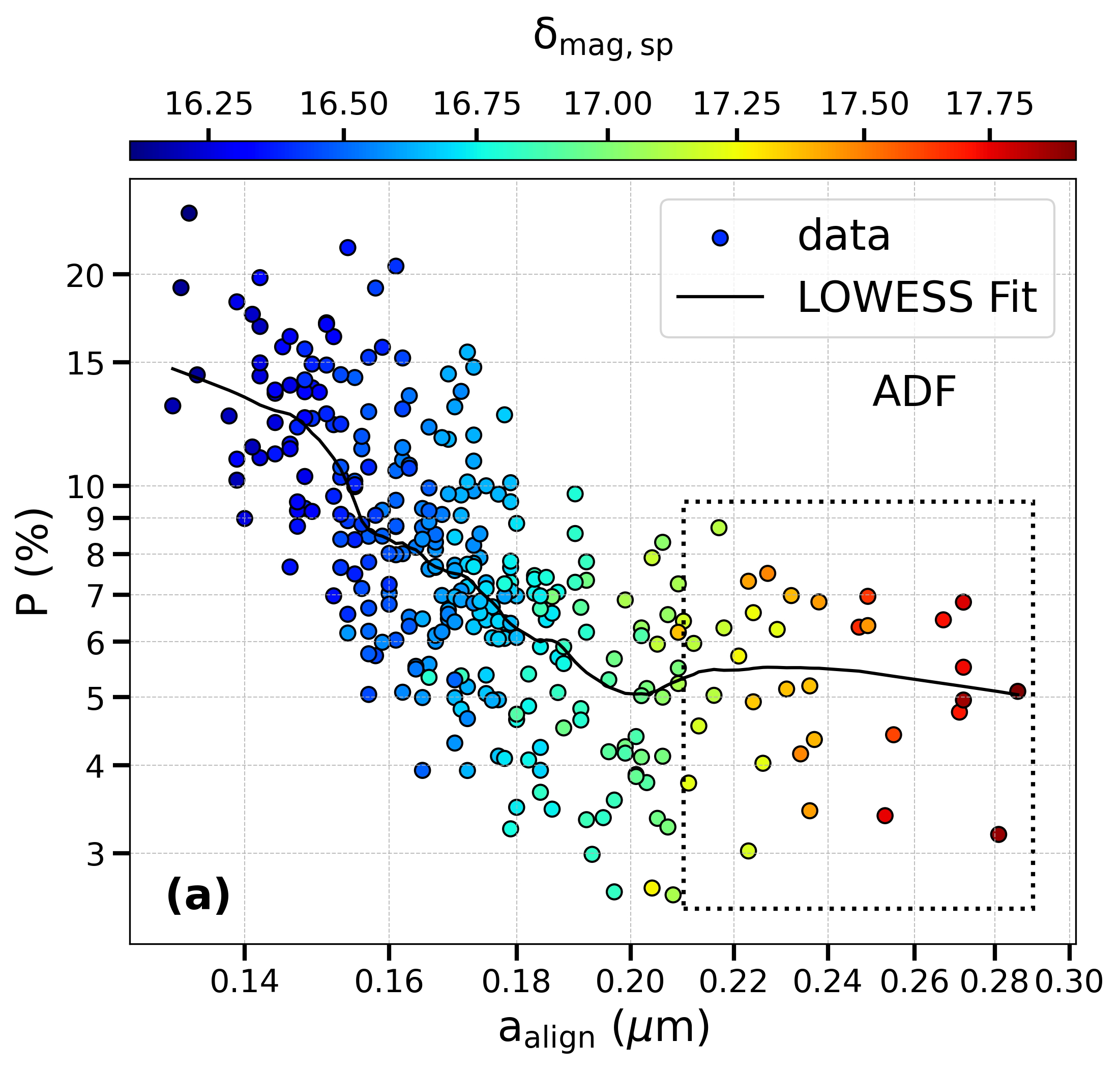} & 
        \hspace{-30pt}
        \includegraphics[scale=0.4]{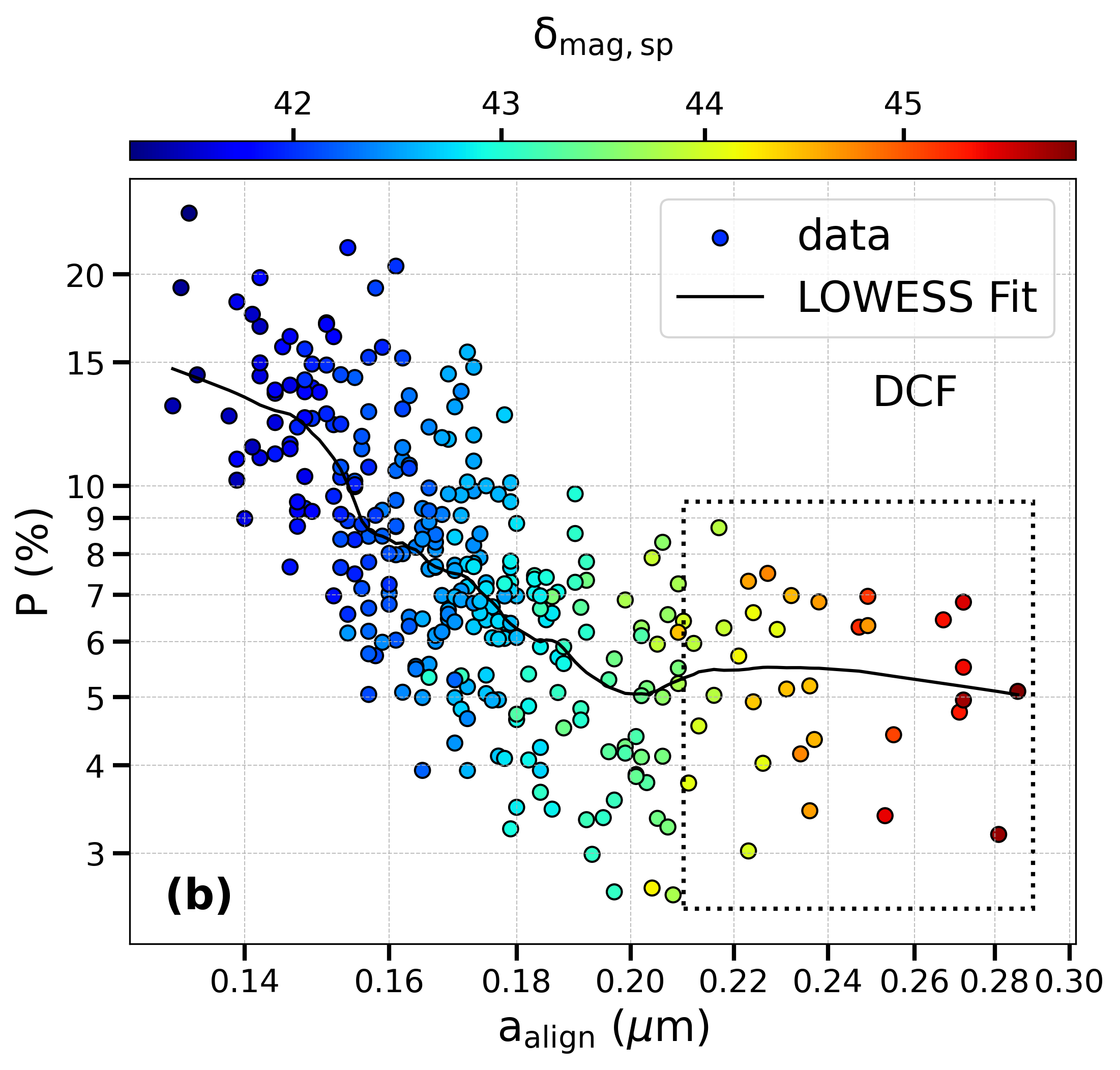} & \\ 
        \includegraphics[scale=0.4]{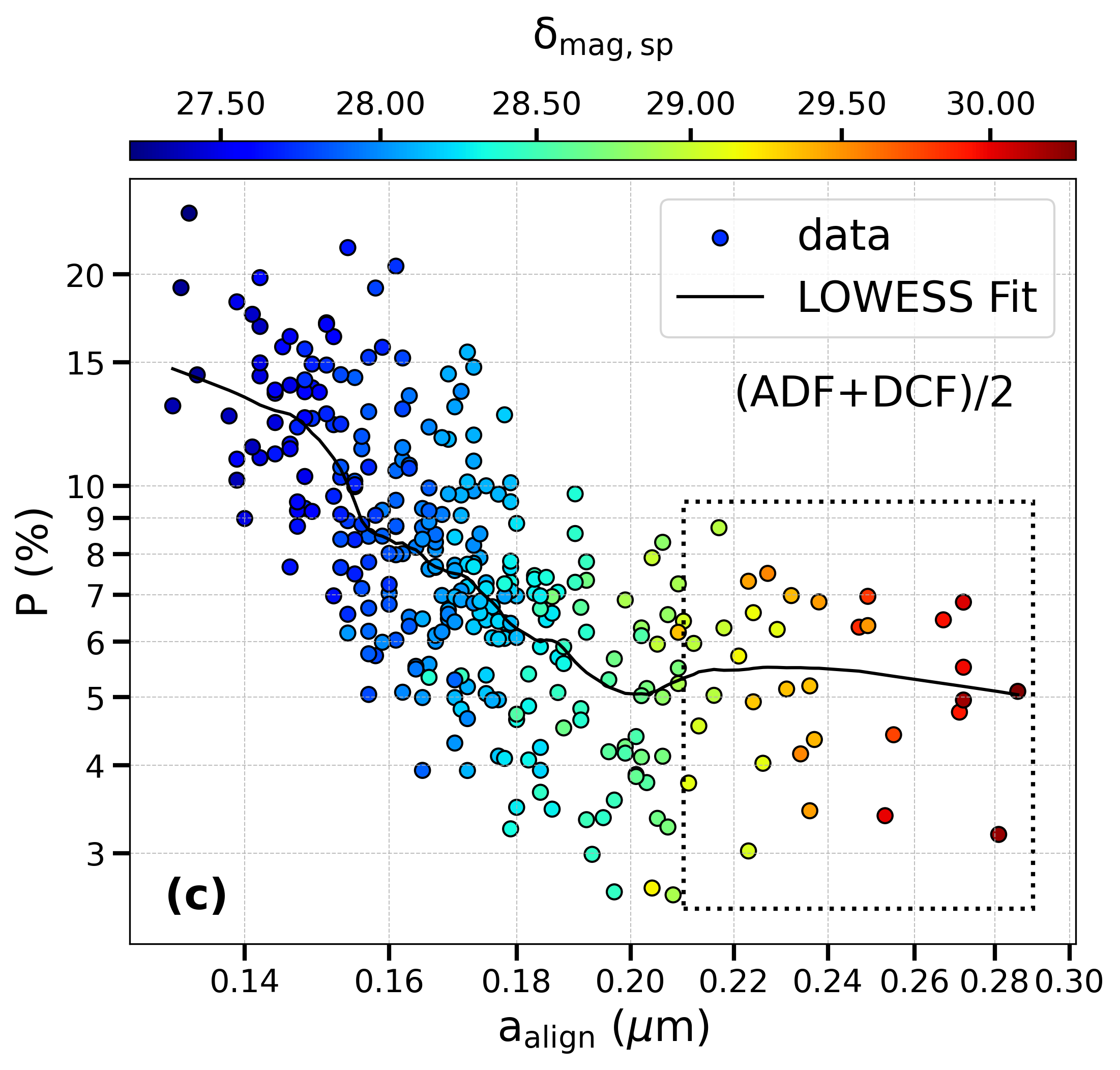}   
    \end{tabular}
    \caption{Same as Figure \ref{Figure:P_P_S_a_align}(a), but the data points are shown with colors that denote the values of magnetic relaxation strengths derived using the magnetic field strengths from (a) ADF, (b) DCF and (c) average of ADF and DCF methods.}
    \label{Figure:P_a_align_mr}
\end{figure*}

\setlength{\tabcolsep}{1.2cm} 
\renewcommand{\arraystretch}{1.3}

\begin{table*}[ht]
    \centering
        \caption{Magnetic fields derived from ADF, DCF and average of ADF and DCF methods; and magnetic relaxation strength values derived using the magnetic field values from each method.}
    \begin{tabular}{cccc}
        \hline \hline
        Parameters & ADF & DCF & (ADF+DCF)/2\\ \hline \hline
        Mean $B_\mathrm{POS}^*$ & $60 \pm 10$ $\mu$G & $96 \pm 17$ $\mu$G & $78 \pm 20$ $\mu$G\\ \hline
        Mean $B_\mathrm{tot}^*$ & $78 \pm 13$ $\mu$G & $124.8 \pm 22.1$ $\mu$G & $101.4 \pm 26$ $\mu$G\\ \hline
        Median $\delta_\mathrm{mag,sp}$ & 16.6 & 42.6 & 28.1\\ \hline
        Range of $\delta_\mathrm{mag,sp}$ & $16.1-17.9$ & $41.2-45.9$ & $27.2-30.3$ \\ \hline    
    \end{tabular}

    \vspace{0.8em}
    
    \begin{minipage}{1\textwidth}
    Note: $^*$ $B_\mathrm{POS}$ values are taken from \cite{2024ApJ...976..249G} and $B_\mathrm{tot}$ values are derived using the relation given in \cite{2004ApJ...600..279C}.
    \end{minipage}

    \label{Table:Magnetic relaxation strength values}
\end{table*}

A dimensionless parameter $\delta_\mathrm{mag}$ was introduced in \cite{2016ApJ...831..159H} to describe the aligning effect of magnetic relaxation relative to the disalignment caused by gas collisions. This parameter $\delta_\mathrm{mag}$ provides the strength of the magnetic relaxation and is defined as the ratio of the gas collision damping timescale, $\tau_\mathrm{gas}$, to the magnetic relaxation time, $\tau_\mathrm{mag,sp}$. For super-paramagnetic grains having embedded iron atoms as clusters, the strength of the magnetic relaxation is given by the following relation

\begin{equation}
{
\delta_\mathrm{mag,sp} = \frac{\tau_\mathrm{gas}}{\tau_\mathrm{mag,sp}} = 56a^{-1}_{-5} \frac{N_\mathrm{cl} \phi_\mathrm{sp,-2} \hat{p}^2 B_3^2}{\hat{\rho} n_4 T_\mathrm{gas,1}^{1/2}} \frac{k_\mathrm{sp}(\Omega)}{T_\mathrm{d,1}}, \label{equation:magnetic relaxation}
}
\end{equation} 
where $a_{-5}=a/(10^{-5}$ cm) with $a$ being the grain size taken as $a=a_\mathrm{align}$, $B_3=B_\mathrm{tot}/(10^3$ $\mu$G), $n_4=n_\mathrm{H}/(10^4$ $\mathrm{cm^{-3}}$) with $n_\mathrm{H} \approx 2n(\mathrm{H_2})$ for molecular gas, $T_\mathrm{gas,1}=T_\mathrm{gas}/(10$ K), $T_\mathrm{d,1}=T_\mathrm{d}/(10$ K), $\hat{p}=p/5.5$ with $p \approx 5.5$ the coefficient describing the magnetic moment of an iron atom, $N_\mathrm{cl}$ gives the number of iron atoms per cluster, $\phi_\mathrm{sp}$ is the volume filling factor of iron clusters with $\phi_\mathrm{sp,-2}=\phi_\mathrm{sp}/0.01$ and $k_\mathrm{sp}(\Omega)$ is the function of the grain rotation frequency $\Omega$ which is of order unity \citep{2022AJ....164..248H}.

When the magnetic relaxation occurs much faster than the gas collision damping, the magnetic relaxation strength is considered to be effective for the alignment of grains. The degree of RAT alignment of grains can be significantly enhanced by the combined effect of both the suprathermal rotation of grains by RATs and the strong magnetic relaxation strength, which is termed as Magnetically-enhanced RAdiative Torque (M-RAT) mechanism of grain alignment.

To study the effect of magnetic relaxation strength on the RAT alignment efficiency of grains, we estimate the $\delta_\mathrm{mag,sp}$ values in all regions of the filament using Equation \ref{equation:magnetic relaxation}. For the plane-of-sky magnetic field strength $B_\mathrm{POS}$, we use the estimated results from \cite{2024ApJ...976..249G} using different methods. The mean $B_\mathrm{POS}$ values using Angular Dispersion Function (ADF) and classical Davis-Chandrasekhar-Fermi (DCF) methods are estimated to be $60 \pm 10$ $\mu$G and $96 \pm 17$ $\mu$G with an average value of $78 \pm 20$ $\mu$G for both the methods. The total magnetic field strength $B_\mathrm{tot}$ is 1.3 times the $B_\mathrm{POS}$ values as shown in \cite{2004ApJ...600..279C}. We calculate $B_\mathrm{tot}$ for each of the methods. We use $N_\mathrm{cl}=100$ and $\phi_\mathrm{sp}=0.1$ (about 30\% of iron abundance as iron clusters, \citealt{2016ApJ...831..159H}). We estimate the $\delta_\mathrm{mag,sp}$ values from Equation \ref{equation:magnetic relaxation} using each of the $B_\mathrm{tot}$ values derived from ADF, DCF methods and the average of these two methods, while keeping the values of other parameters same. The magnetic field strength and the magnetic relaxation strength values for each of the $B_\mathrm{tot}$ values derived from ADF, DCF and average of ADF and DCF methods are given in Table \ref{Table:Magnetic relaxation strength values}.

Since there is lack of information on the pixel-by-pixel values of magnetic field strength using ADF or DCF method and we have only the mean $B_\mathrm{tot}$ values, we use the relation $B_\mathrm{tot} \sim n_\mathrm{H}^{2/3}$ from \cite{2010ApJ...725..466C} to estimate the pixel-by-pixel values of $B_\mathrm{tot}$ as a function of $n_\mathrm{H}$. We estimate the scaled $B_\mathrm{tot}$ values using the scaling relation given below:

\begin{equation}
{
B_\mathrm{tot}(n_\text{H})=B_\text{tot}(\bar{n}_\text{H}) \, \left(\frac{n_\text{H}}{\bar{n}_\text{H}}\right)^{2/3},
}
\end{equation} 
where $B_\text{tot}(\bar{n}_\text{H})$ is the mean $B_\text{tot}$ value taken from the ADF, DCF and average of ADF and DCF methods.

The maps of the polarization fraction and the magnetic relaxation strengths for each of ADF, DCF and the average of ADF and DCF methods are shown in Figure \ref{Figure:mr_ADF_DCF}. We see that the $\delta_\mathrm{mag,sp} > 10$ overall and becomes stronger in the denser regions than in the outer regions. The polarization vectors with $P \geq 8$\% are shown in gray color in the figure and these high $P$ values are associated with $\delta_\mathrm{mag,sp} > 10$, implying the potential significance of magnetic relaxation strength in enhancing the RAT alignment efficiency of grains to produce the high $P$ values in the outer regions and hence it provides an implication for M-RAT mechanism. The value of $\delta_\mathrm{mag,sp} > 10$ and increases slightly towards the denser regions. However, the $P$ values show lower values in these denser regions compared to the outer, less dense regions, which can be due to the saturation of M-RAT and an increase in $a_\mathrm{align}$ values.

Figure \ref{Figure:P_a_align_mr} shows the variation of $P$ with $a_\mathrm{align}$ with the colors representing the $\delta_\mathrm{mag,sp}$ values for each data point. We see that the data points for $a_\mathrm{align} > 0.21$ $\mu$m show a significant increase in $P$ and are also associated with higher values of $\delta_\mathrm{mag,sp}$, which may imply the significance of the magnetic relaxation strength to produce perfect alignment of the available large aligned grains by M-RATs. However, due to the increase in the $a_\mathrm{align}$ values, the fraction of available aligned grains becomes less considering a constant $a_\mathrm{max}$ and here we expect that there could be significant influence by the changes in the grain properties like increase in size and more elongation to produce the higher observed $P$ values of $5-8$\% in the denser regions. The perfect alignment of the large and more elongated aligned grains in these denser regions by M-RATs could result in a significant increase in $P$ values. In the subsequent section, we try to constrain the grain properties like the grain size and the axial ratio by performing numerical dust polarization modeling.

\subsection{Grain alignment behavior: fast or slow} \label{section:Grain alignment behavior}
The unified model of grain alignment by M-RAT mechanism can explain the efficient alignment of grains. For an ensemble of grains having different shapes and magnetic properties, only a fraction of grains can be aligned at an attractor with angular momentum above the thermal value, termed as high-J attractors. For those grains at high-J attractors, if they have initially random orientations, a fraction of the grains can be rapidly aligned at high-J attractors, known as fast alignment. The other grains are driven at an attractor at thermal angular momentum (low-J attractors). The collisional and magnetic excitations by gaseous random collisions with these grains can pump the grains out of low-J attractors and slowly transport them to high-J attractors, which can result in the perfect alignment of grains within several gaseous damping time scales due to M-RATs, and this is termed as perfect slow alignment \citep{2021ApJ...908...12L}.

\begin{figure}
    \centering
        \includegraphics[scale=0.35]{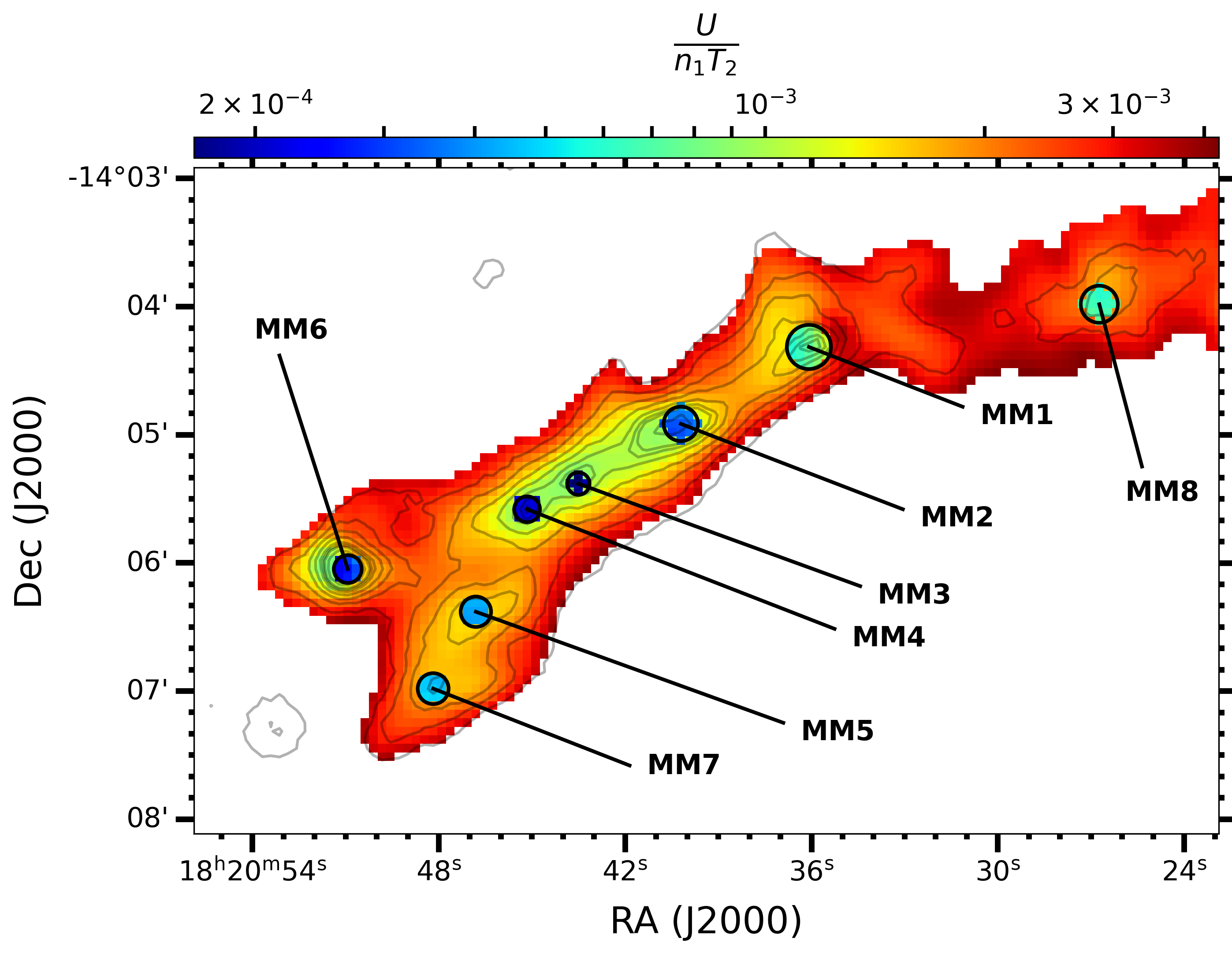} 
    \caption{Map of the parameter $U/n_1T_2$ over the G16 filament, showing values $\ll 1$.} 
    \label{Figure: U_n1T2_map}
\end{figure}

The local radiation field strength ($U$), gas number density ($n_\mathrm{H}$) and the gas temperature ($T_\mathrm{gas}$) are the very crucial parameters that can determine the nature of general astrophysical environments. Usually, the dimensionless parameter $U/(n_1 T_2)$, where $\mbox{$U \approx (T_\mathrm{d}/16.4$ K)$^6$}$, $n_1 = n_\mathrm{H}/(10\ \mathrm{cm^{-3}})$ and $T_2 = T_\mathrm{gas}/100$ K, describes the general astrophysical environments in two regimes based on the ratio $U/(n_1 T_2)$. If $U/(n_1T_2) \leq 1$, then the region is in Collision-Dominated (CD) regime, whereas for $U/(n_1T_2) > 1$, the region is in Radiation-Dominated (RD) regime. For the case of CD regime, the grains undergo perfect slow alignment, whereas for the RD regime, only a fraction of grains can undergo fast alignment at high-J attractors by M-RATs, and the majority of grains are trapped at low-J rotation due to strong RATs, a new effect known as RAT Trapping (RATT) introduced in \cite{2025arXiv250502157H}. The astrophysical environments under the CD regime include the standard ISM and star-forming regions, including molecular clouds and filaments, to dense cores and protostellar environments.

We estimate the value of $U/(n_1T_2)$ for our cold filament G16 and the map is shown in Figure \ref{Figure: U_n1T2_map}. We find that $U/(n_1T_2) \ll 1$ over the filament, implying that G16 filamentary region is in the CD regime. Therefore, the grains can achieve perfect slow alignment due to M-RATs. We note this observed feature to consider the maximum alignment degree of grains $f_\mathrm{max}$ to be 1 in our polarization modeling as described in the below Section \ref{section:Dust polarization modeling and results}.

\section{Dust polarization modeling and results} \label{section:Dust polarization modeling and results}
In this section, we will present the theoretical approaches and methods of thermal dust polarization modeling using RAT theory to test the RAT paradigm and constrain the grain properties in our G16 filament. For the modeling, we use the latest version of $\textsc{DustPOL\_py}$ \footnote{Github: \url{https://github.com/lengoctram/DustPOL_py}} code, first presented in \cite{2020ApJ...896...44L} and then in \cite{2021ApJ...906..115T,2024A&A...689A.290T,2025arXiv250116079T} with subsequent improvements in the code. In the initial, we first compute the model dust polarization fraction at 850 $\mu$m wavelength for an ideal condition, assuming that the B-fields lie in the plane-of-sky (the inclination angle of $\psi=90^\circ$) and are uniform with negligible effects of B-field tangling on depolarization. 

However, this ideal condition may not reflect correctly the actual condition of the polarization, as other parameters like the inclination angle of B-field with the line of sight, and the B-field tangling along the line of sight can contribute to the depolarization effect. For our modeling, we estimate the model dust polarization, taking into account the B-field tangling effect and considering the plane-of-sky B-field, and then compare with the observational data. The effect of variation in the B-field's inclination angle $\psi$ is discussed later in Section \ref{section: Evidence for varying 3D magnetic field and its effect}.

We briefly describe the approaches and methods of the modeling using \textsc{DustPOL\_py} along with the important physical input parameters listed in Table \ref{Table:Physical input parameters for the model}. Our modeling using \textsc{DustPOL\_py} is based on the dust alignment physics by RATs and well-suited for pixel-by-pixel polarization modeling at single wavelength data.

\subsection{Approaches and Methods of Modeling}
\subsubsection{Gas Density, Temperature and Radiation Fields}
The properties of gas like the gas number density $n_\mathrm{H} = 2n(\mathrm{H_2})$ and the temperature $T_\mathrm{gas}$ determine the randomization and alignment of dust grains in the framework of RAT alignment theory (e.g., \citealt{2021ApJ...908..218H}) and are crucial physical parameters in our model. We use the values of these parameters from the observational data and provide as input to the \textsc{DustPOL\_py} for the modeling.

Again, other key physical parameters in the realm of RAT-A theory are the radiation field strength ($U$), which is equivalent to the dust temperature $T_\mathrm{d}$, and the anisotropy degree of the radiation $\gamma$ (see \citealt{2021ApJ...908..218H}). The value of $\gamma$ can range from 0 to 1, with $\gamma=1$ representing a unidirectional radiation field from a nearby star and $\gamma=0.1$ for the case of diffused ISM (see \citealt{1997ApJ...480..633D}). The mean wavelength of the radiation $\bar{\lambda}$ depends on regions and the typical value for the diffuse ISRF is 1.2 $\mu$m. For the dense and cold environments, the gas and dust can be considered to be in thermal equilibrium, and hence for our G16 filament, we consider $T_\mathrm{gas} = T_\mathrm{d}$.

\subsubsection{Distribution of grain size}
We make use of a power-law grain size distribution, $dn/da \sim a^{-\beta}$, with $\beta$ in the range of $3.5-4.5$. For our ISM, the grain size distribution follows the Mathis-Rumpl-Nordsieck (MRN) distribution with $\beta = 3.5$ \citep{1977ApJ...217..425M}. We take the minimum grain size, $a_\mathrm{min}$ to be $10^{-3}$ $\mu$m, whereas the maximum grain size, $a_\mathrm{max}$ is taken up to 1 $\mu$m. 

\subsubsection{RAT Alignment Function of Grains}
Two crucial aspects of RAT alignment are the smallest grain size that RATs can align and how effectively this alignment is achieved. The minimum grain size that is aligned by RAT is known as the alignment size ($a_{\rm align}$), which is determined when the grain rotates suprathermally (that is, the angular velocity is three times the thermal angular velocity; see Equation \ref{equation:a_align}). For a typical diffuse interstellar medium, $a_{\rm align}\simeq 0.055\,\mu$m. Within the submillimeter wavelength observations used in this work, the alignment of small grains is neglected, and thus the efficiency is close to zero for $a\ll a_{\rm align}$ and achieves a certain value of maximum alignment efficiency $f_{\rm max}$ for $a\gg a_{\rm align}$. This transition is known as the alignment function, defined as
\begin{equation}
    f(a) = f_{\rm max} \, \left[1 \, - \text{e}^{-(0.5a/a_{\rm align})^{3}}\right]
\end{equation}

\cite{2016ApJ...831..159H} showed that super-paramagnetic grains can achieve perfect alignment due to M-RAT mechanism. We find that the G16 filament is effective for the M-RAT mechanism as described in Section \ref{section:Effect of magnetic relaxation on RAT Alignment}, and also the filament is found to be in the CD regime as shown in Section \ref{section:Grain alignment behavior}, and hence large grains can achieve perfect slow alignment due to M-RATs. Therefore, we consider $f_\mathrm{max} = 1$ (referred to as a perfect alignment of grains) in our modeling.

\subsubsection{Thermal Dust Polarization Model Using \textsc{DustPOL\_py} for Idealistic Case}
The dust models developed over the past decades assumed the two-component model which attributes to the presence of two separate populations of dust as silicate and carbon grains (e.g., \citealt{1977ApJ...217..425M, 1984ApJ...285...89D, 2001ApJ...548..296W}). 
However, recent studies favored a one-component model or composite dust model (a single grain containing a mixture of silicate and non-silicate materials) that could explain observational results. For example, \cite{2021ApJ...909...94D} found that the observations of far-infrared polarization fraction from 250 $\mu$m $-$ 3 mm favored the composite model, \textsc{Astrodust}. This composite model is likely expected in the diffuse, and cold and dense regions because of many competing processes such as photoprocessing, coagulation, accretion, erosion and gas-grain collisions (see e.g., \citealt{1990ASPC...12..193D, 2009ASPC..414..453D, 2013A&A...558A..62J}). 

For the case of two-component model, only the silicate dust grains could be aligned with the magnetic field because of their paramagnetic nature, whereas the ideal carbonaceous grains (e.g, pure graphite) could not be efficiently aligned with the magnetic field because of their diamagnetic nature (see \citealt{2016ApJ...831..159H}). The non-ideal carbonaceous grains, e.g., hydrogenated amorphous carbons (HACs) and polycyclic aromatic hydrocarbons (PAHs), develop some degree of paramagnetic nature and could be aligned with the magnetic field (e.g., see \citealt{2023ApJ...954..216H}). The separate populations of silicate and carbonaceous grains are expected in the environment of newly formed dust, e.g., the envelopes of Asymptotic Giant Branch (AGB) stars. Mostly carbon-rich grains are expected in the envelopes of carbon-rich AGB stars. For the case of the composite model, the bulk of the grain could be aligned with the magnetic field.

Considering the cold and dense environment of our study, where separate populations of silicate and carbonaceous grains are less likely expected and composite dust grains most likely dominate, we use the single composite dust model \textsc{Astrodust} \citep{2021ApJ...919...65D, 2021ApJ...909...94D, 2023ApJ...948...55H} for our dust polarization modeling. The \textsc{Astrodust} grain is a mixture of silicate ($\approx$ 50\% of the grain mass), carbonaceous and hydrocarbon materials in a single grain. The bulk of the \textsc{Astrodust} grain could be aligned with the magnetic field due to the presence of silicate composition and could emit polarized radiation. \cite{2024A&A...692A..60R} showed that the centrifugal force resulting from the fast rotation of grains can make them to be oblate in shape. For our modeling, we consider the oblate \textsc{Astrodust} grains.

The total and the polarized intensity for the thermal dust polarization can be derived analytically as given in \cite{2020ApJ...896...44L, 2021ApJ...906..115T, 2024ApJ...965..183H, 2024A&A...689A.290T}.
For the composite \textsc{Astrodust} model, the total emission intensity is given by the following equation:


\begin{align}
\frac{I_\text{em}(\lambda)}{N_\text{H}} \, &= \int_{a_\text{min}}^{a_\text{max}} Q_\text{abs}\pi a^2 \times B_{\lambda}(T_\text{d})\frac{1}{n_\text{H}}\frac{dn}{da} \, da, 
\end{align}
where $Q_\mathrm{abs}$ is the absorption efficiency (taken from \textsc{Astrodust} database), a function of $a$ and $\lambda$ for a given grain axial ratio, and $B_\lambda(T_\mathrm{d})$ is the Planck function.

The intensity of polarized emission with $\psi=90^\circ$ is given by the following equation:

\begin{align}
\frac{I_\text{pol}(\lambda)}{N_\text{H}} \, &= \int_{a_\text{align}}^{a_\text{max}} f(a) Q_\text{abs}^{\text{pol}}\pi a^2 
\times B_{\lambda}(T_\text{d})\frac{1}{n_\text{H}}\frac{dn}{da} \, da, 
\end{align}
where $Q_\mathrm{abs}^{\mathrm{pol}}$ is the absorption polarization efficiency, determined by the residual of the absorption efficiencies in the two components, where the electric field $\vec{\bm{E}}$ is perpendicular and parallel to the grain symmetry axis ($\vec{\bm{a}}_1$) as $0.5 \, \left[Q_\mathrm{abs}(\vec{\bm{E}} \perp \vec{\bm{a}}_1) - Q_\mathrm{abs}(\vec{\bm{E}} \parallel \vec{\bm{a}}_1)\right]$ \citep{2020ApJ...896...44L}. These components were taken from the \textsc{Astrodust} database.

The degree of polarization of thermal dust emission is then given by the following equation:

\begin{equation}
P_\text{mod}^\text{ideal} (\%) = 100 \% \times \frac{I_\text{pol}}{I_\text{em}}, 
\label{equation:P_mod_ideal}
\end{equation}
where the subscript "mod" denotes the model polarization and the superscript "ideal" denotes the ideal condition of estimating the polarization degree assuming uniform B-fields in the plane-of-sky.

\subsubsection{Thermal Dust Polarization Model for Realistic Case}
When we consider the ideal case in computing the thermal dust polarization degree in our modeling, the polarization degree could achieve the maximum value (Equation \ref{equation:P_mod_ideal}). However, for a realistic case, there are depolarization effects contributed by the inclination angle $\psi$ of B-fields with the line-of-sight due to the projection effect, and also by the fluctuations of B-fields along the line-of-sight due to turbulence. The depolarization due to B-field fluctuations can be described by a parameter $F_\mathrm{turb}$ which is a function of the angle between the local B-field and the mean B-field and is anticorrelated with the polarization angle dispersion function $S$ \citep{2024ApJ...965..183H}. The realistic degree of dust polarization can be described by the following equation:

\begin{equation}
P_\text{mod} = P_\text{mod}^\text{ideal} \, \text{sin}^2\psi \, F_\text{turb}  
\end{equation}
The parameter $F_\mathrm{turb}$ decreases with $S$ as $F_\mathrm{turb} \sim S^{-\eta_1}$, as shown in the numerical simulations in \cite{2024ApJ...965..183H}, where $\eta_1$ is dependent on the inclination angle. Again, \cite{2015A&A...576A.104P} showed the correlation between the average of the B-field inclination angle and $S$, described as $\langle sin^2\psi \rangle \sim S^{-\eta_2}$. Moreover, there can be another depolarization factor caused by the B-field tangling within the beam size or B-field fluctuations in the plane-of-sky and described by $F_\mathrm{beam}$ \citep{2024ApJ...965..183H}. Taking into account all the depolarization effects, the effective net degree of polarization of thermal dust emission can be described by the following equation:

\begin{equation}
P_\text{mod} = \phi \, P_\text{mod}^\text{ideal} \left(\frac{S}{1^\circ}\right)^{-\eta},  
\label{equation:P_mod}
\end{equation}
where $\phi$ is a coefficient that describes the depolarization effect due to the inclination angle of B-field, and $\eta > 0$ is a power index that describes the depolarization effect due to B-field fluctuations along the line of sight and in the plane-of-sky. We obtain the value of $\eta$ from the slope in the variation of $P$ with $S$. In our modeling, we consider the plane-of-sky magnetic field or $\phi=1$ and discuss the depolarization effect due to the variation in the magnetic field's inclination angle with the line of sight in Section \ref{section: Evidence for varying 3D magnetic field and its effect}.


\setlength{\tabcolsep}{0.25cm} 
\renewcommand{\arraystretch}{1.3}

\begin{table}[ht]
    \centering
        \caption{Physical input parameters for the model}
    \begin{tabular}{cc}
        \hline \hline
        Parameter & Value \\ \hline \hline
        Anisotropic degree, $\gamma$ & 0.1 \\ \hline
        Mean wavelength, $\bar{\lambda}$ ($\mu$m) & 1.2 \\ \hline
        Gas temperature, $T_\mathrm{gas}$ (K) & $T_\mathrm{d}$ \\ \hline
        Grain composition & \textsc{Astrodust} \\ \hline
        Dust mass density, $\rho$ (g $\mathrm{cm^{-3}}$) & 3 \\ \hline
        Minimum grain size, $a_\mathrm{min}$ ($\mu$m) & $10^{-3}$ \\ \hline
        Maximum grain size, $a_\mathrm{max}$ ($\mu$m) & $0.3-1$ \\ \hline
        Grain size distribution power index, $\beta$ & 3.5 \\ \hline
        Maximum alignment efficiency, $f_\mathrm{max}$ & 1 \\ \hline
        B-field inclination angle, $\psi$ & $90^\circ$ \\ \hline
    \end{tabular}
    \label{Table:Physical input parameters for the model}
\end{table}


\begin{figure*}
    \centering
    \begin{tabular}{cc}
        \includegraphics[scale=0.37]{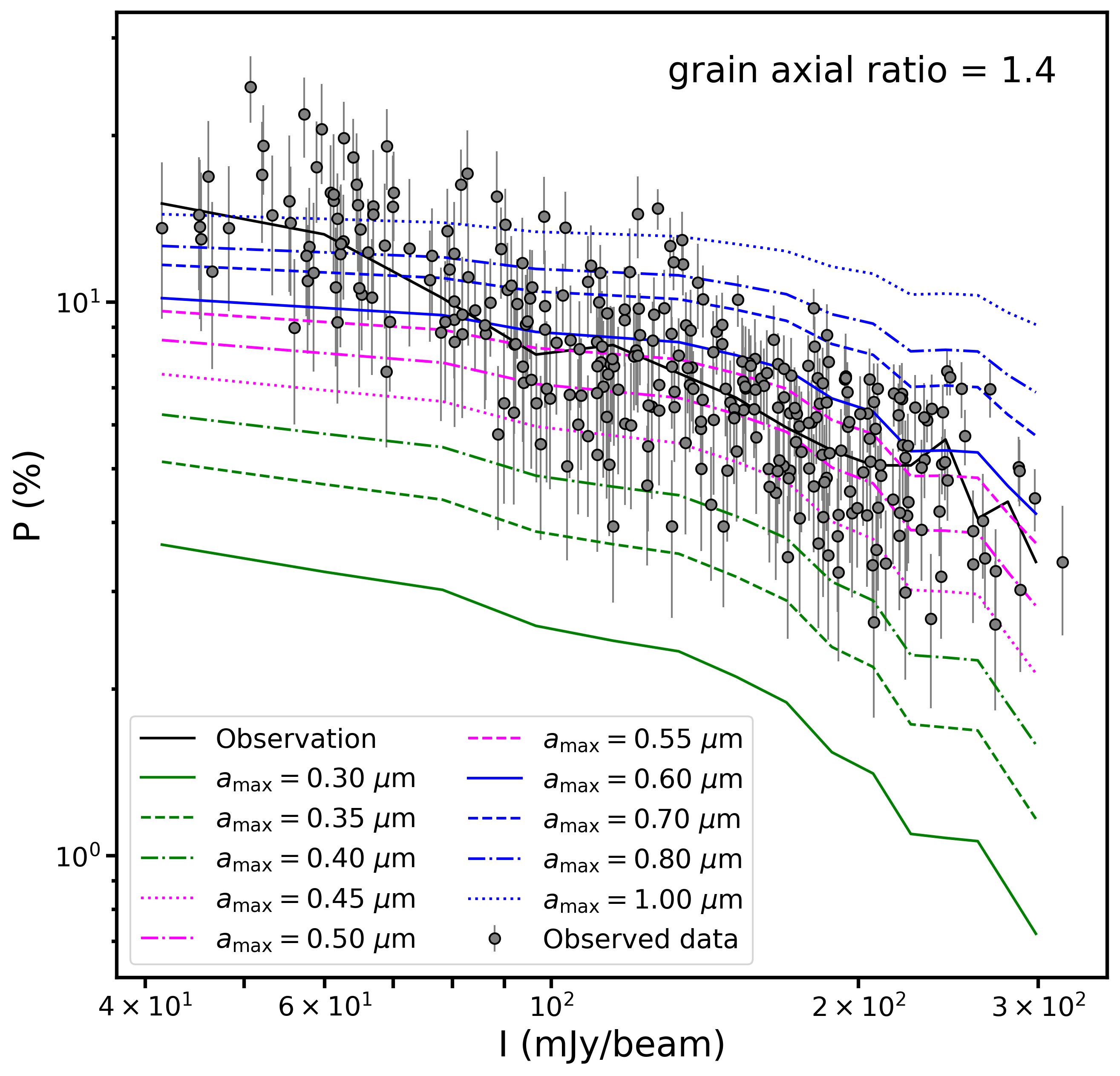} &  
        \hspace{5pt}
        \includegraphics[scale=0.37]{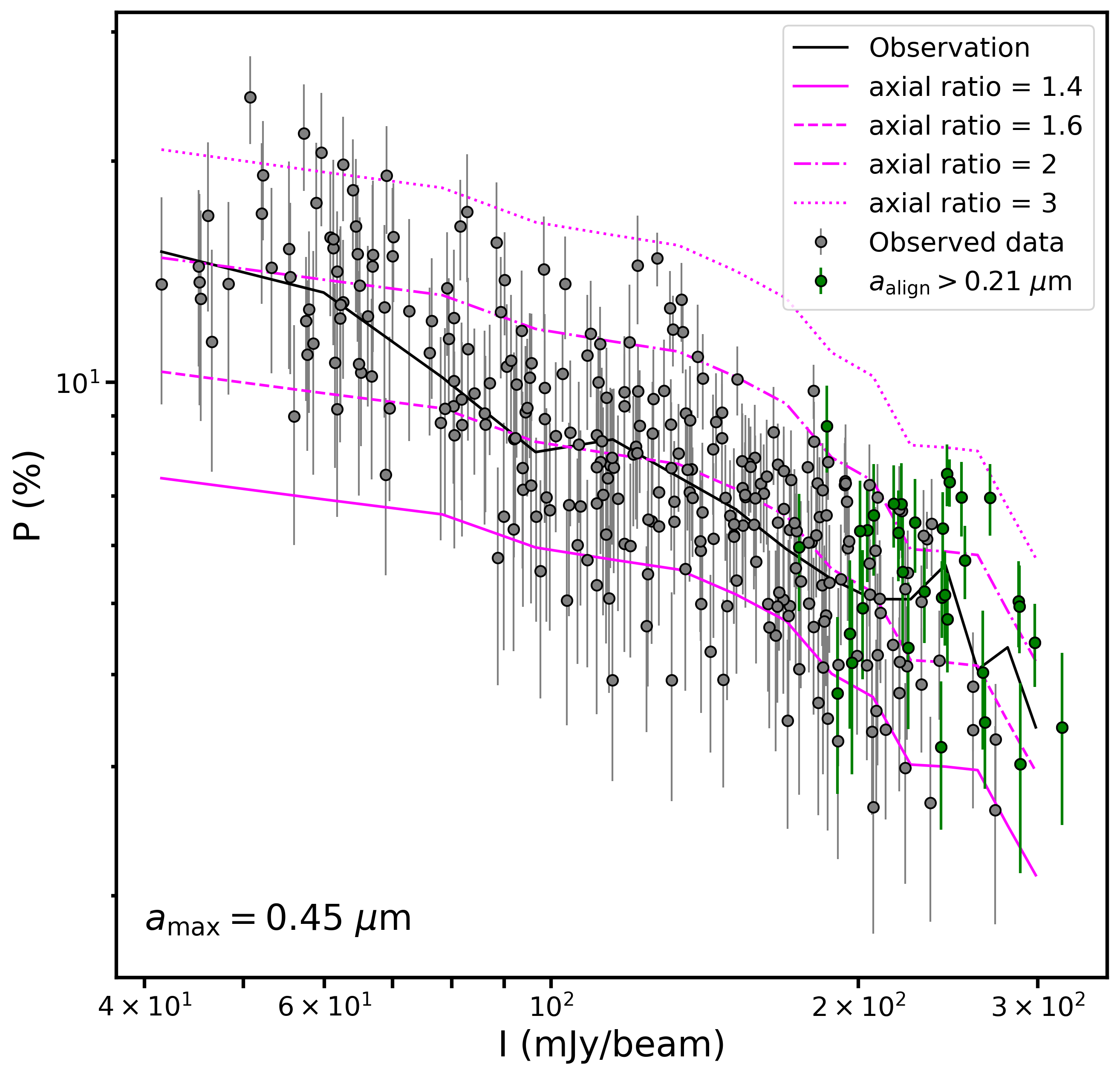}   
    \end{tabular}
    \caption{Left panel shows the comparison of our model polarizations for different $a_\mathrm{max}$ but fixed grain axial ratio of 1.4, with the observations in their variations with the total intensity. The model polarization predicted for $a_\mathrm{max} = 0.45$ $\mu$m seems to follow the observational trend more closely. The right panel depicts the comparison of the model polarizations for fixed $a_\mathrm{max}$ value of 0.45 $\mu$m but variable grain axial ratio or grain elongation, with the observation. The green data points correspond to $a_\mathrm{align} > 0.21$ $\mu$m. Models with an increase in grain elongation could reproduce the observational data.}
    \label{Figure:P_I_model_2_3}
\end{figure*}

\begin{figure*}
    \centering
    \begin{tabular}{cc}
        \includegraphics[scale=0.37]{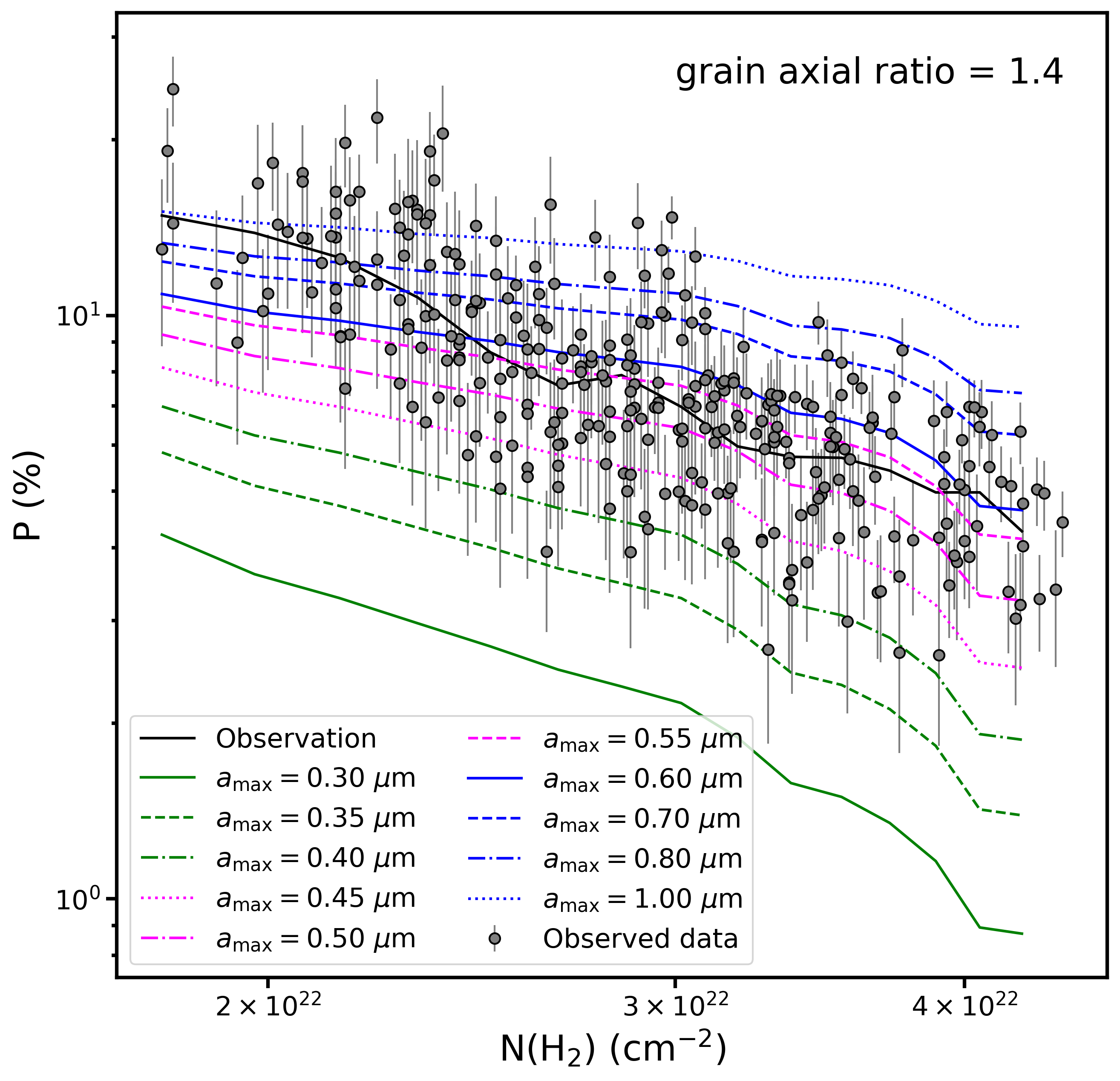} &  
        \hspace{5pt}
        \includegraphics[scale=0.37]{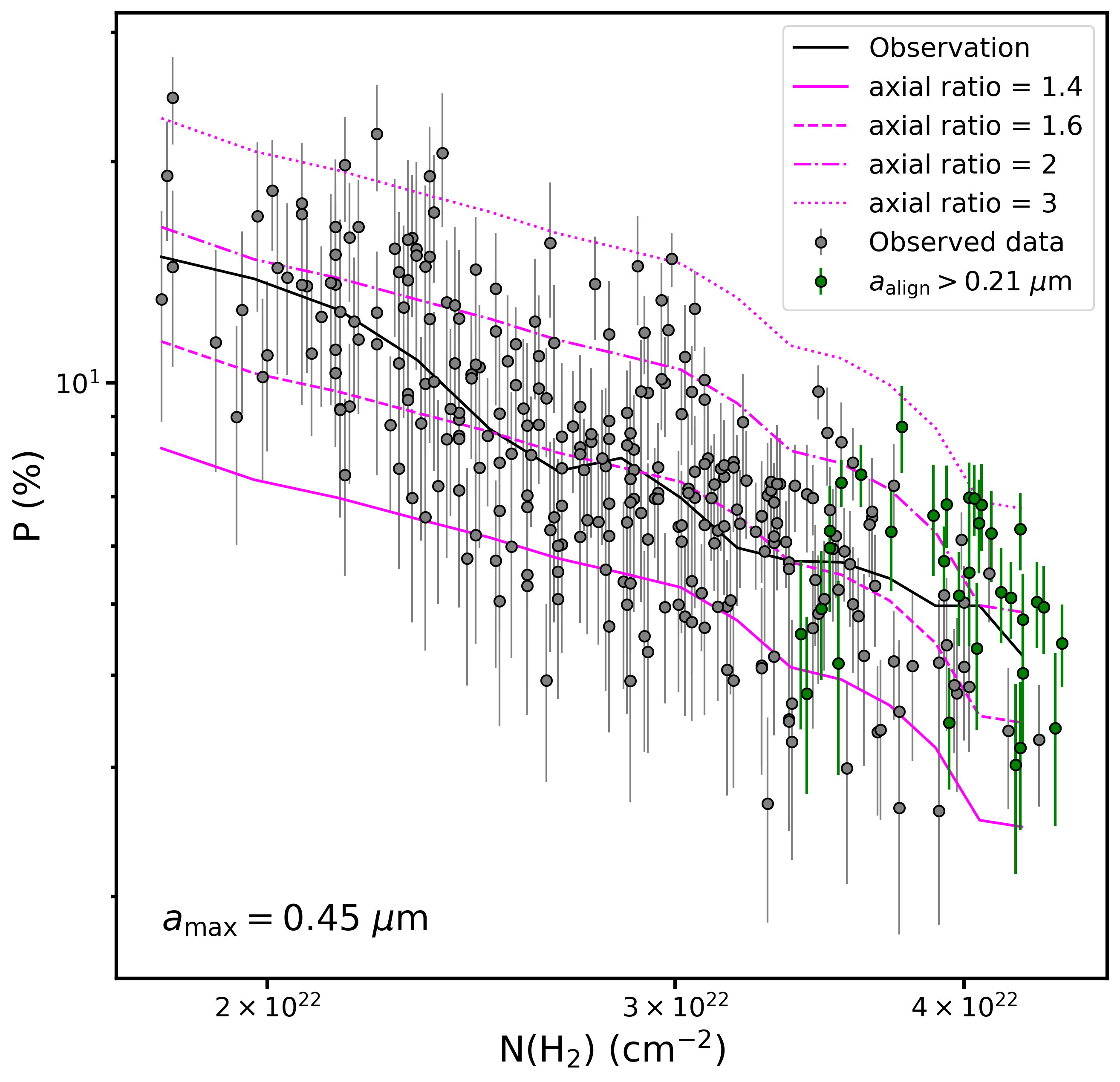}   
    \end{tabular}
    \caption{Same as Figure \ref{Figure:P_I_model_2_3}, but for the variations of the polarizations with the gas column density.}
    \label{Figure:P_NH2_model_1_2}
\end{figure*}

\begin{figure*}
    \centering
    \begin{tabular}{cc}
        \includegraphics[scale=0.37]{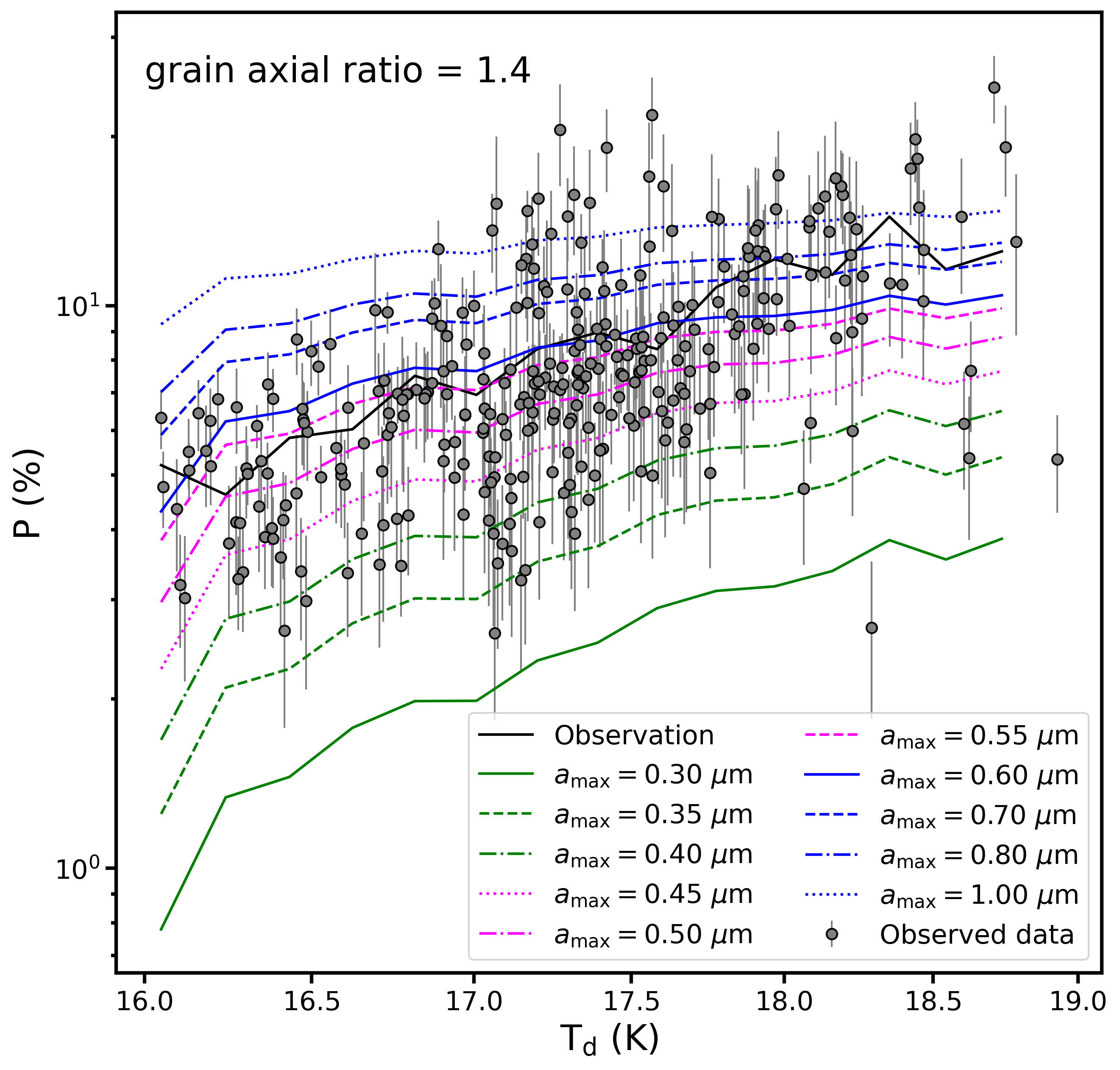} &  
        \hspace{5pt}
        \includegraphics[scale=0.37]{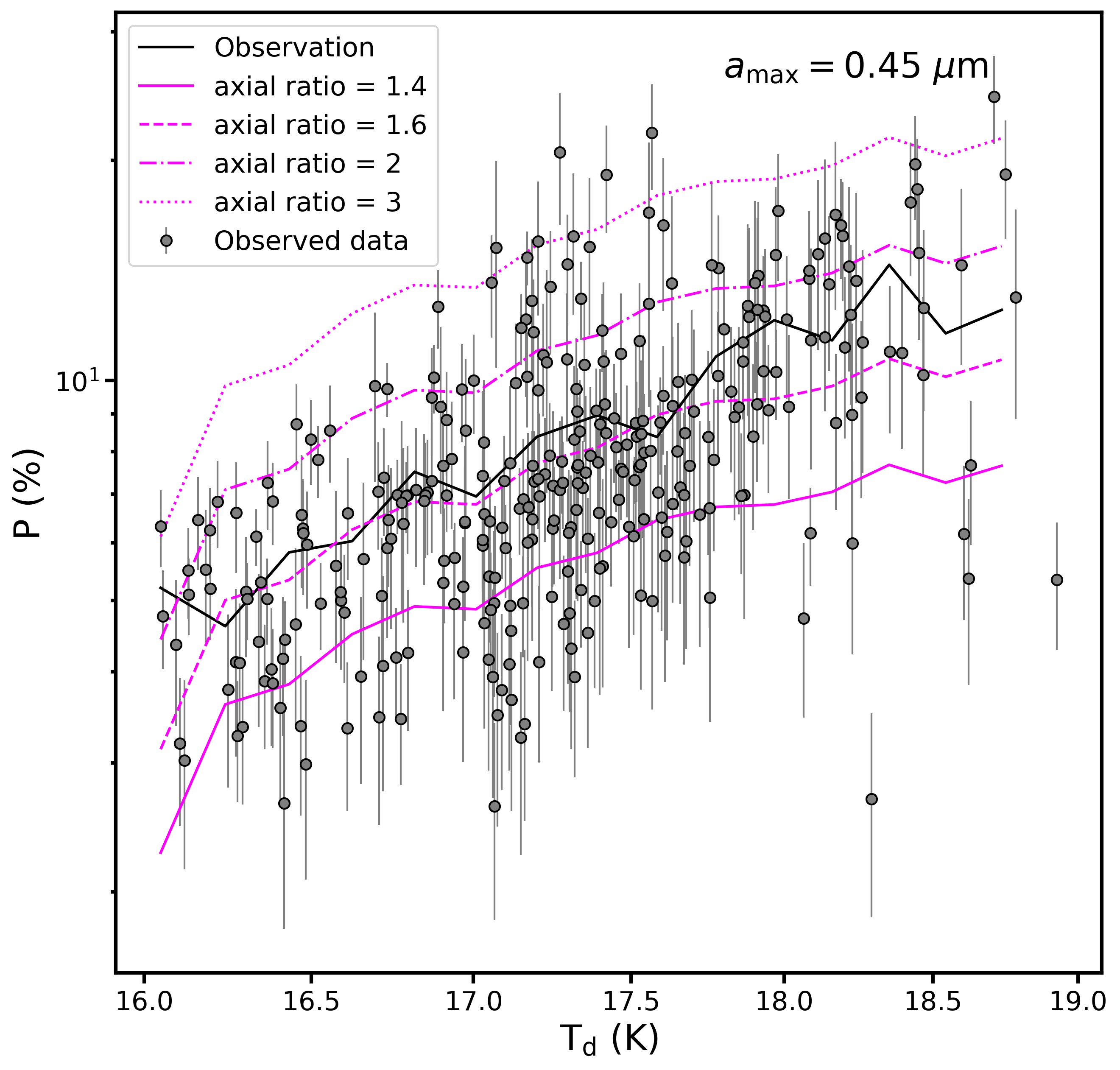}   
    \end{tabular}
    \caption{Same as Figure \ref{Figure:P_I_model_2_3}, but for the variations of the polarizations with the dust temperature.}
    \label{Figure:P_Td_model_1_2}
\end{figure*}

\subsection{Results of Modeling}
In this section, we will present the results of our modeling of thermal dust polarization in the G16 filament. 

Our model assumes that the magnetic fields lie on the plane-of-sky. We use the polarization fraction of the model, $P_\mathrm{mod}$ as given by Equation \ref{equation:P_mod} considering $\phi=1$ to compare and analyse with the trends in the observational data. Figure \ref{Figure:P_I_model_2_3} (left panel) shows the variations of polarization fraction (from both observations and our models with different $a_\mathrm{max}$ values) with the total intensity. We see that the different $P_\mathrm{mod}$ values for different $a_\mathrm{max}$ in the range from 0.4 to 0.5 $\mu$m seem to follow the trends that could reproduce the depolarization in the denser regions found in the observational data, whereas $P_\mathrm{mod}$ for higher values of $a_\mathrm{max} > 0.6$ $\mu$m becomes larger and the slopes become nearly flat in the denser regions due to increase in polarization fraction, which could not reproduce the observational trend. The $P_\mathrm{mod}$ for the value of $a_\mathrm{max} = 0.45$ $\mu$m more closely follows the variation trend of the observational data. 

However, it shows lower values than the observational data and the values would be decreased further, if the realistic situation of the variation in the magnetic field's inclination angle with the line of sight is considered, because of the reduction factor $\phi < 1$ from the magnetic field's inclination angle and the model with a fixed grain elongation of 1.4 with $a_{max}=0.45$ $\mu$m could not fully reproduce the observational data, suggesting that the grain elongations should be increased to reproduce the observational data. We fix $a_\mathrm{max} = 0.45$ $\mu$m and vary the grain axial ratio from 1.4 to 3 as shown in the right panel of Figure \ref{Figure:P_I_model_2_3}. We see that the model could better reproduce the observational data as the grain elongation increases, most effectively at axial ratios in the range [$1.6-2$]. A further elongation with axial ratio $\sim 3$ is required to reproduce the upper bound of the observed higher polarization fraction of the thermal dust emission. The observed data points shown in green color correspond to $a_\mathrm{align} > 0.21$ $\mu$m (we see an increase in $P$ after $a_\mathrm{align}$ value of 0.21 $\mu$m and then becomes nearly constant as shown in Figure \ref{Figure:P_P_S_a_align}) and some of these data points with observed higher polarization fractions could be reproduced by higher grain elongation with axial ratio of [$2-3$].

We also plot the variations of polarization (from both observation and model) with the gas column density as shown in Figure \ref{Figure:P_NH2_model_1_2}. This plot provides information on the gas damping effect. We find similar results in this variation also. Again, we analyse the variations of polarization (from both observation and model) with the dust temperature or equivalently the radiation field strength as shown in Figure \ref{Figure:P_Td_model_1_2}. In this variation, we get similar results. Hence, our modeling of the polarization implies the requirement of grain growth, with $a_\mathrm{max}$ value typically of around 0.45 $\mu$m, accompanied by an increase in the grain elongation to reproduce the observational data. We also see from Figures \ref{Figure:P_I_model_2_3}, \ref{Figure:P_NH2_model_1_2} and \ref{Figure:P_Td_model_1_2} that to reproduce the observed polarization, micron-sized grains of $a_\mathrm{max}=1$ $\mu$m are required in the outermost regions if the typical axial ratio of the ISM grains of 1.4 is adopted. However, these large grains are not expected in the outer, less dense regions. Hence, the model with $a_\mathrm{max}=0.45$ $\mu$m with grain axial ratio = 2 could well reproduce the observational data.

\begin{figure*}
    \centering
    \begin{tabular}{cc}
        \includegraphics[scale=0.43]{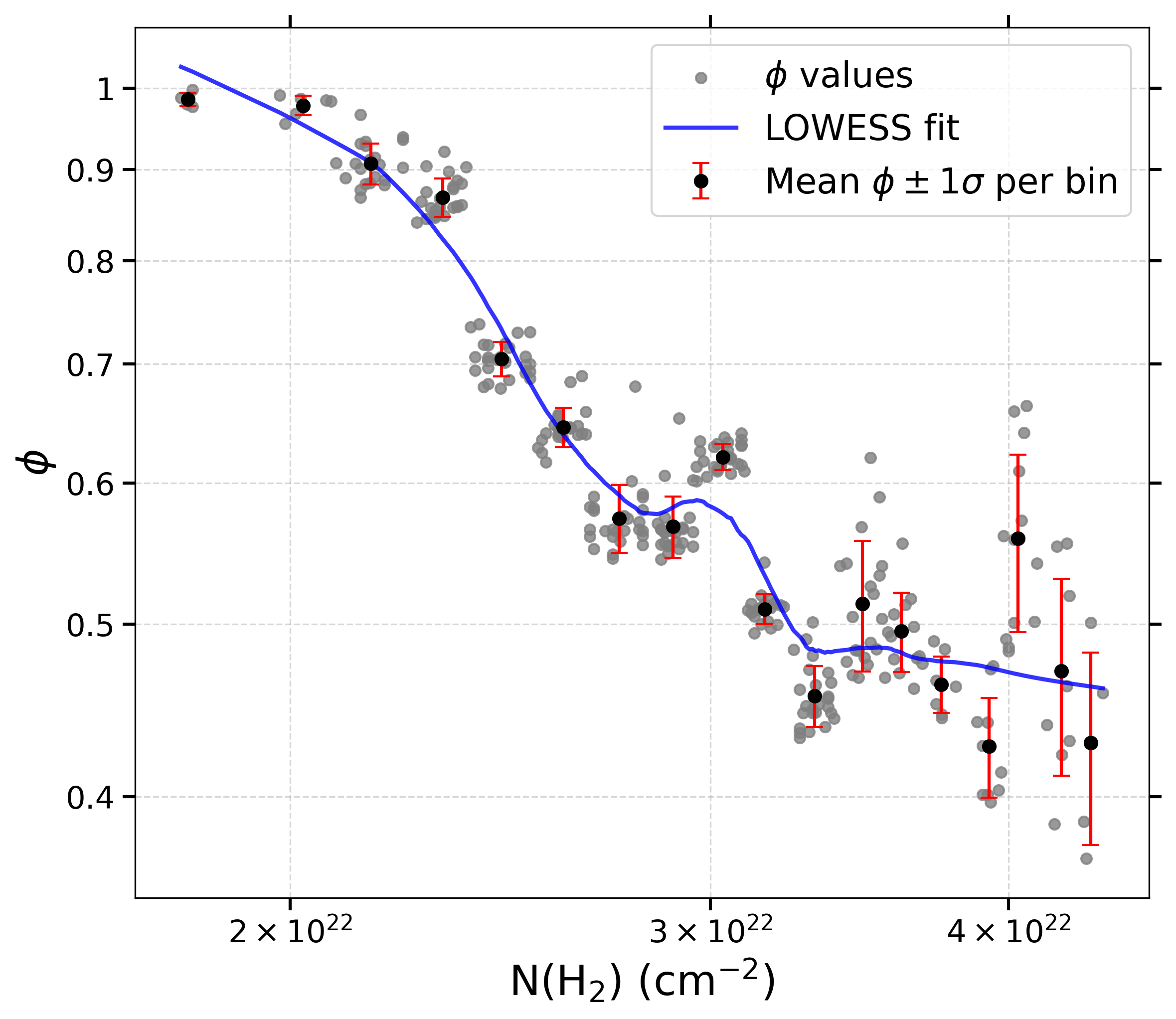} &  
        \includegraphics[scale=0.43]{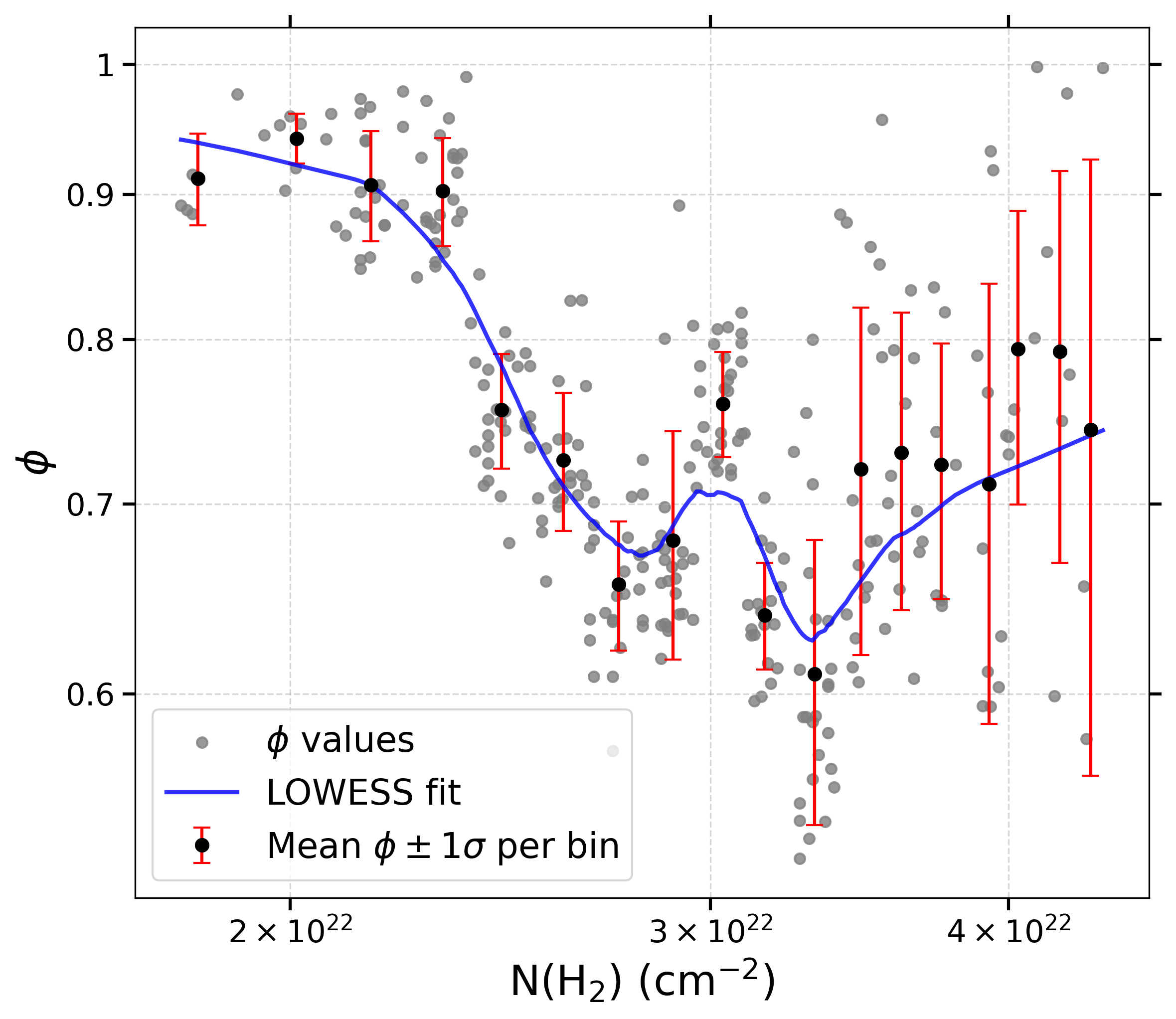}   
    \end{tabular}
    \caption{Variations of the factor $\phi$ with the gas column density $N(\mathrm{H_2})$ for the $\phi$ values estimated from the model polarizations with parameters of $a_{max}=1$ $\mu$m and grain axial ratio = 1.4 (left panel), and $a_\mathrm{max}=0.45$ $\mu$m and grain axial ratio = 2 (right panel). The black data points are the mean values of $\phi$ in each bin of $N(\mathrm{H_2})$ with bin size of $1.45 \times 10^{21}$ $\mathrm{cm^{-2}}$.}
    \label{Figure:phi_NH2_model_1_2}
\end{figure*}


In our modeling, we assume a constant $a_\mathrm{max}$ value along an individual line of sight, however, it could vary due to the local variations in the physical parameters like gas density, temperature, radiation field, etc. A constant value of $a_\mathrm{max}=0.45$ $\mu$m is more likely valid for the denser regions compared to the outer, less dense regions. Hence, the results of our modeling are in particular more suitable towards the denser regions, and we get implications for grain growth and elongation in these denser regions.

Our model assumes the plane-of-sky magnetic field with $\phi=1$. However, in reality $\phi$ could vary from the outer to the inner regions of the filament and this variation implies the evolution of the 3D magnetic fields. We can derive quantitatively the magnetic field's inclination angle $\psi$ or the factor $\phi$ using the technique as described in \cite{2024ApJ...965..183H}. First, we see the models that could reproduce the polarization fraction at the outermost region associated with the lowest gas column density. From Figure \ref{Figure:P_NH2_model_1_2}, we find that the models with parameters of $a_\mathrm{max}=1$ $\mu$m with axial ratio = 1.4, and $a_\mathrm{max}=0.45$ $\mu$m with axial ratio = 2 could reproduce the polarization fraction at the outermost region associated with the lowest gas density. Then, for each of these models, assuming that the model parameters are constant from the outer to the inner regions, we can infer the mean $\phi$ which gives the regular component of the magnetic field, as $\phi=\bar{P}_\mathrm{obs}/P_\mathrm{mod}$. We make several bins of $N(\mathrm{H_2})$ and for each bin, we estimate the $\phi$ values using the above relation. Figure \ref{Figure:phi_NH2_model_1_2} shows the variations of $\phi$ with $N(\mathrm{H_2})$ for each of the above models. We see that $\phi$ is nearly 1 at the outermost region and decreases with the increase in the gas density for the model of $a_\mathrm{max}=1$ $\mu$m with axial ratio = 1.4. For the model of $a_\mathrm{max}=0.45$ $\mu$m with axial ratio = 2, the value of $\phi$ is also nearly 1 at the outermost region and then decreases with the increase in the gas density, but shows some spread at the higher gas density regions. The increase in the value of $\phi$ at the more denser regions could be due to the steeper of the model polarization slope and more closely matching with the observational slope (see Figure \ref{Figure:P_NH2_model_1_2}). The value of $\phi$ would depend on the model parameters, however the variation trends may be similar. We see a nearly similar trend in the variation of $\phi$ with $N(\mathrm{H_2})$ in each of the models. The value of $\phi$ decreases overall from the outermost region with value of nearly 1 to the inner denser regions with values less than 1 in each of the models, which gives an insight into the evolution of the 3D magnetic fields over the filament. For a more detailed discussion, please refer to Section \ref{section: Evidence for varying 3D magnetic field and its effect}.

\section{Discussions} \label{section:Discussions}
In this section, we will discuss the results of the various analyses made in the previous sections from both observational and theoretical modeling perspectives, to explain the grain alignment mechanisms and the grain properties in the filament.

\subsection{Evidence for grain alignment via RAT-A mechanism} \label{section:Evidence for grain alignment via RAT-A mechanism}

\subsubsection{Anti-correlation of polarization fraction with total intensity and gas density: Polarization hole} 
\label{section:Decrease of polarization fraction with total intensity and gas density}

An anti-correlation of the polarization fraction $P$ with the total emission intensity and the gas density, known as polarization hole, is an expectation of the RAT theory (e.g., see \citealt{2020ApJ...896...44L, 2021ApJ...908..218H}), and is observed in various studies in dense star-forming regions (e.g., \citealt{2019FrASS...6...15P, 2021ApJ...908..218H, 2023ApJ...953...66N, 2024ApJ...974..118N, 2025ApJ...981..128P}). The exact origin of the polarization hole is not yet clear. However, the possible explanations are (1) the decrease in grain alignment in denser regions of high gas density, (2) magnetic field tangling along the line of sight, and (3) a combination of both. The main causes of magnetic field fluctuations are the turbulence in the molecular clouds (\citealt{1989ApJ...346..728J, 1992ApJ...389..602J, 2008ApJ...679..537F}).


In our study of the G16 filament, we also observe the polarization hole phenomenon (see Figure \ref{Figure:P_I_NH2}). We investigate whether there is any significant role of magnetic field tangling in causing the observed polarization hole or not by analyzing the variations of $P$ with $S$ and $P \times S$ with $I$, $N(\mathrm{H_2})$ and $T_\mathrm{d}$ (see Figure \ref{Figure:P_S}). We show that the effect of magnetic field tangling to cause the polarization hole is minor and not significant. The G16 filament is a quiescent filament that is in the early stage of star formation and lacks bright embedded sources inside. The denser regions of higher $N(\mathrm{H_2})$ and $I$ values are associated with lower dust temperatures or equivalently lower radiation field strength (see Figures \ref{Figure:NH2_Td_map}, \ref{Figure:Intensity_map}). In these denser regions of weak radiation field strength, a large fraction of grains will not be able to achieve suprathermal rotation and would be easily randomized due to gas-grain collisions, thereby resulting in a decrease in the fraction of aligned grains and hence a decrease in the net polarization fraction, as expected by RAT-A mechanism. Hence, the decrease in the polarization fraction in the denser regions of the filament is majorly due to the decrease in the RAT alignment efficiency of grains in these denser regions, providing evidence for the RAT-A mechanism of grain alignment.

\subsubsection{Correlation of polarization fraction with dust temperature} \label{section:Correlation of polarization fraction with dust temperature}
In the framework of RAT-A theory, another expectation is the increase in polarization fraction with increasing dust temperature or equivalently radiation field strength (see e.g., \citealt{2020ApJ...896...44L, 2021ApJ...908..218H}). The higher the dust temperature or radiation field strength, the fraction of grains rotating suprathermally would be larger, and hence it would result in higher grain alignment efficiency, producing higher polarization fraction. Our study finds that the polarization fraction overall increases with increasing dust temperature (see Figure \ref{Figure:P_Td}). Again, the averaged grain alignment efficiency also increases with increasing dust temperature (see Figure \ref{Figure:P_S}), in a way similar to the variation in Figure \ref{Figure:P_Td}, implying a less significant role of magnetic field tangling. The observed features are in agreement with the expectation of RAT-A theory, further supporting the evidence for the RAT-A mechanism of grain alignment.

\subsubsection{Testing using RAT theory: Increase in $a_{align}$ with I; decrease in P and $P \times S$ with $a_{align}$} \label{section:RAT theory}
We test whether the RAT theory can reproduce our observational results or not by estimating the minimum alignment size of grains using the RAT theory and considering the local physical parameters. In the framework of RAT-A theory, the size distributions of aligned grains range from $a_\mathrm{align}$ to $a_\mathrm{max}$, and this range of grain size distribution determines the polarization fraction. When we consider a fixed value of $a_\mathrm{max}$, an increase in $a_\mathrm{align}$ can result in a narrower size distribution of aligned grains, thereby reducing the polarization fraction while a decrease in $a_\mathrm{align}$ can result in a wider size distribution of aligned grains, thereby increasing the polarization fraction (see Figure 7 in \citealt{2022FrASS...9.3927T}). Hence, an anti-correlation is expected between $P$ and $a_\mathrm{align}$ by RAT theory. Also, RAT theory expects a correlation between $a_\mathrm{align}$ and $I$ in starless clouds.

In G16 filament, which lacks bright embedded sources inside, $a_\mathrm{align}$ increases with increasing $I$ (see Figure \ref{Figure:a_align_I}) and becomes larger in denser regions. Again, the polarization fraction $P$ and the averaged grain alignment efficiency $P \times S$ decrease with the increase in $a_\mathrm{align}$ or equivalently as moving to the denser regions (see Figure \ref{Figure:P_P_S_a_align}). Also, we do not find a significant correlation between $P$ and $S$ (see Figure \ref{Figure:P_S}(a)). The lack of correlation between $P$ and $S$, and the decreases of both $P$ and $P \times S$ with $a_\mathrm{align}$ suggest that the decrease in the polarization fraction is not by magnetic field tangling, but by the reduction in the fraction of aligned grains in the denser regions or equivalently decrease in the RAT alignment efficiency of grains in the denser regions.  
Hence, under the application of the well-established analytical RAT model using the observational data of gas density and dust temperature, and assuming the best values of other parameters suitable for the particular region of our study, we find that the results of the model could reproduce the observational results. The assumed values of the parameters may be subject to change, but this would not affect the results significantly. The consistency between the model and the observation provides an implication that favors the alignment of grains by RATs and helps in further strengthening our explanation of the evidence for the RAT-A mechanism, as we find from the observational analysis.


From the above discussions from different perspectives in Sections \ref{section:Decrease of polarization fraction with total intensity and gas density}, \ref{section:Correlation of polarization fraction with dust temperature} and \ref{section:RAT theory}, we find potential evidence that the RAT-A mechanism could explain the observed depolarization. However, we note that the polarization fraction decreases by an order of magnitude in a limited narrow range of gas column density associated with a factor of 2. The grain alignment efficiency depends not only on the gas density, but also on the dust temperature or equivalently the radiation field intensity and grain properties like size, shape, etc. For the G16 filament, there is no nearby strong radiation source, and the main radiation source is the diffused interstellar radiation field. Since the optical depth is proportional to the gas column density, and the radiation field gets attenuated exponentially with the optical depth, a small change in the gas column density could produce a significant decrease in the radiation field intensity and also changes in other physical conditions.

Quantitatively, we estimate the minimum alignment sizes, $a_\mathrm{align}$, of grains for two cases: one for the outermost region with $T_\mathrm{d}$(outer) $\approx$ 19 K and $n(\mathrm{H_2})$(outer) $\approx$ $1.3 \times 10^4 \, \, \mathrm{cm^{-3}}$, and the other for the innermost region with $T_\mathrm{d}$(inner) $\approx$ 16 K and $n(\mathrm{H_2})$(inner) $\approx$ $9 \times 10^4 \, \, \mathrm{cm^{-3}}$ using equation \ref{equation:a_align} (the outermost and innermost regions mentioned here are indicated with "+" symbols in Figure \ref{Figure:NH2_Td_map}). The grain alignment is more precisely dependent on the gas volume density $n(\mathrm{H_2})$ than the column density $N(\mathrm{H_2})$, and the gas volume density differs by nearly an order of magnitude (see Figure \ref{Figure:NH2_Td_map} (c) ). The $a_\mathrm{align}$ value is found to increase from $\approx$ 0.14 $\mu$m at the outermost region to $\approx$ 0.29 $\mu$m at the innermost region when the gas volume density is increased by a factor of $[n(\mathrm{H_2})$(inner)]/$[n(\mathrm{H_2})$(outer)] $\approx$ 7, and the radiation strength $U$ is decreased by a factor of [$T_\mathrm{d}$(outer)/$T_\mathrm{d}$(inner)]$^6$ $\approx$ 3 (see Figures \ref{Figure:NH2_Td_map} and \ref{Figure:a_align_map_Histogram_a_align}). Therefore, grains with sizes equal to or above 0.14 $\mu$m could be aligned at the outermost region, and grains with sizes equal to or above 0.29 $\mu$m could be aligned and those with sizes below 0.29 $\mu$m could not be aligned at the innermost region. If the maximum grain size or $a_\mathrm{max}$ is small, e.g., 0.3 $\mu$m, the increase in $a_\mathrm{align}$ value from 0.14 $\mu$m to 0.29 $\mu$m would significantly reduce the polarization fraction by a factor of 4, as shown in Figure \ref{Figure:P_NH2_model_1_2}. In our numerical modeling of thermal dust polarization using the RAT paradigm, it is found that to reproduce the observational data an increase in $a_\mathrm{max}$ value, accompanied by an increase in the grain elongation is required. However, the decrease in grain alignment is still found to be insufficient to reproduce the steep decrease in the polarization fraction by an order of magnitude (see Figure \ref{Figure:P_NH2_model_1_2}), and a consideration of the localized variations in the inclination angle of the magnetic field with the line of sight, which introduces a depolarization factor phi (see equation \ref{equation:P_mod}), is important to be taken into account, as inferred by our numerical modeling (see Figure \ref{Figure:phi_NH2_model_1_2} and for a detailed discussion, see Section \ref{section: Evidence for varying 3D magnetic field and its effect}). The combination of the observational data and the modeling results help identify the factors contributing to the steep decrease in the polarization fraction within a narrow range of gas column density, and is attributed to two main effects: a decrease in the grain alignment efficiency by RATs in the denser regions, and changes in the magnetic field's inclination angle relative to the line of sight as one moves from the outer to the inner regions.

\subsection{Significance of magnetic relaxation strength on RAT alignment: Implication for M-RAT alignment mechanism}
The magnetic properties of dust grains are crucial for the alignment of the grains with the external magnetic field. A strong magnetic relaxation strength can help in achieving efficient alignment of grains. Super-paramagnetic grains, which are expected in denser regions due to grain evolution can acquire stronger magnetic relaxation strength in an environment of strong external magnetic fields. A stronger magnetic relaxation strength combined with suprathermal rotation of the super-paramagnetic grains induced by RATs can result in higher grain alignment efficiency, producing higher polarization fraction \citep{2016ApJ...831..159H, 2022AJ....164..248H}. Super-paramagnetic grains can achieve perfect alignment by the M-RAT mechanism when $\delta_\mathrm{mag,sp} > 10$ \citep{2016ApJ...831..159H}. \cite{2024arXiv240710079C} found that observed high polarization fractions in the envelopes of protostellar cores could be produced by the thermal emission of aligned super-paramagnetic grains with $N_\mathrm{cl} \approx 10^2-10^3$, which can produce high intrinsic polarization fraction. The significance of the combined effect of magnetic relaxation strength and suprathermal rotation of grains by RATs to explain the observation of high polarization fractions was studied and quantified in previous studies in G11.11-0.12 filament by \cite{2023ApJ...953...66N} and in G34.43+0.24 filament by \cite{2025ApJ...981..128P}.

In our study for the G16 filament, we observe high polarization fractions of above 10\% up to around 23\%, especially in the outer regions of the filament. We investigate the role of magnetic relaxation strength on the RAT alignment efficiency of grains to explain these observed high $P$ values. We estimate the $\delta_\mathrm{mag,sp}$ values using each of the scaled $B_\mathrm{tot}$ values derived from ADF, DCF, and the average of ADF and DCF methods. We find that the outer less dense regions with $P \geq 8$\% and reaching 23\% are associated with $\delta_\mathrm{mag,sp} > 10$ in each case ($\delta_\mathrm{mag,sp} \gg 10$ in case of DCF method) (see Figure \ref{Figure:mr_ADF_DCF}). These outer, less dense regions have lower $a_\mathrm{align}$ values and stronger RATs. Also, we find that the magnetic relaxation is effective in these regions. An effective magnetic relaxation strength can increase the RAT alignment efficiency of grains. Hence, we get an implication that the observed high polarization fraction of $8-23$\% in the outer regions of the filament could be potentially due to the combined effect of both the effective magnetic relaxation strength and the stronger RAT alignment efficiency of grains, supporting the M-RAT mechanism. Again, we see a significant increase in $P$ in the denser regions for $a_\mathrm{align} > 0.21$ $\mu$m and this increase in $P$ is associated with higher magnetic relaxation strengths (see Figure \ref{Figure:P_a_align_mr}). In these denser regions with higher $a_\mathrm{align}$ values, we expect an increase in maximum grain size and grain elongation to produce the observed high $P$ values. However, the large and more elongated grains of $a \gg a_\mathrm{align}$ need to be perfectly aligned. The perfect alignment of these grains could be produced by the combined effect of both the RAT alignment and the stronger magnetic relaxation strength in these regions, implying the M-RAT mechanism. Hence, we find implications for the alignment of grains by M-RAT mechanism over all the filaments.


Also, we observe that the G16 filament is in the CD regime and hence grains can achieve perfect slow alignment by M-RATs as shown in \cite{2021ApJ...908...12L} and \cite{2025arXiv250502157H}. Our numerical modeling of dust polarization that considers the RAT alignment model with $f_\mathrm{max}=1$ and composite \textsc{Astrodust} model could successfully reproduce the observational data, implying the perfect alignment of large magnetic dust grains of $a \gg a_\mathrm{align}$. Such a perfect alignment of dust grains could not be achieved by the classical RAT theory alone, and it requires the M-RAT mechanism. As mentioned earlier, super-paramagnetic grains can achieve perfect alignment by the M-RAT mechanism when $\delta_\mathrm{mag,sp} > 10$. Our estimation of $\delta_\mathrm{mag,sp}$ taking $\phi_\mathrm{sp}=0.1$ (about 30\% of iron abundance as clusters) results in the value of $\delta_\mathrm{mag,sp} > 10$. The chosen 30\% iron abundance is well within the observational constraint of $\approx$ 95\% of iron abundance as clusters in dust grains. We also estimated $\delta_\mathrm{mag,sp}$ taking a lower value of $\phi_\mathrm{sp}=0.01$ (about 3\% of iron abundance as clusters) and found that the value of $\delta_\mathrm{mag,sp} < 5$ which could not produce the perfect alignment of grains with $f_\mathrm{max}=1$ as required by our modeling. Therefore, our analysis and modeling also suggest a high fraction of iron abundance as clusters inside the grains. 

If we consider that the classical RAT theory is sufficient and M-RAT is not required, by taking a lower value of $f_\mathrm{max} < 1$, then this consideration can be allowed if there is an increase in the grain elongation to have the axial ratio greater than 3. This may be possible because our modeling shows that the grain growth is accompanied by an increase in the grain elongation. However, such extremely elongated grains are less expected. 
Again, the consideration of the classical RAT theory can also be allowed if there is a more intense radiation field near or inside the filament to produce a high polarization fraction. However, there is no high radiation source present near or inside the filament.

\subsection{Implication for anisotropic grain growth of aligned grains} \label{section: Implication for anisotropic grain growth of aligned grains}
The upper limit of the grain size distribution in the interstellar medium is $\approx$ 0.25 $\mu$m as suggested by observational studies (\citealt{1977ApJ...217..425M}; known as MRN distribution).
However, dense starless molecular clouds could acquire the process of grain growth, thereby increasing the maximum grain size, because of gas accretion and grain-grain collisions in these dense regions (e.g, \citealt{2013MNRAS.434L..70H}). Dust polarization can be used as a tracer for grain growth in dense molecular clouds.

From the maps of $a_\mathrm{align}$ (Figure \ref{Figure:a_align_map_Histogram_a_align}) and polarization fraction $P$ (Figure \ref{Figure:Intensity_P_map_Histogram_P}), we constrain the lower limit for the maximum grain size $a_\mathrm{max}$ which is required to reproduce the slope value of $-0.71$ in the variation of polarization fraction with total intensity (Figure \ref{Figure:P_I_NH2} (left)). High polarization fractions are observed in the outer regions and therefore, $a_\mathrm{max}$ needs to be much larger than the typical values of $a_\mathrm{align} \approx 0.14-0.16$ $\mu$m in these outer regions. This can be satisfied with the ISM value of $a_\mathrm{max} \approx 0.25$ $\mu$m from the MRN distribution. In particular, in the densest regions of the filament's spine, the slope value of $|-0.71| < |-1|$ implies that there still exist aligned grains that can emit polarized radiation. This can be satisfied when grain growth occurs within the filament, which increases the maximum grain size beyond the $a_\mathrm{align}$ values of the filament's spine of 0.28 $\mu$m, i.e., $a_\mathrm{max} > a_\mathrm{align} \approx 0.28$ $\mu$m. We also note another possible scenario that may explain the observed slope value without the need for grain growth if we consider a temperature gradient along the line of sight for our filament that assumes a cylindrical model. The outer region of the filament along a particular line of sight is associated with higher dust temperature and lower $a_\mathrm{align}$ values, whereas the inner denser region along this particular line of sight is associated with lower dust temperature and higher $a_\mathrm{align}$ value. This means that the emission of polarized radiation could also be from the outer aligned grains, while the inner region can be highly depolarized due to higher $a_\mathrm{align}$ values. However, the best-fit dust temperature $T_\mathrm{d}$ value for each pixel is derived from the spectral energy distribution (SED) fitting of fluxes at multi-wavelengths, which trace the different temperature regions of the dust layers along the line of sight. From the $T_\mathrm{d}$ map shown in Figure \ref{Figure:NH2_Td_map}(d), we see that $T_\mathrm{d}$ decreases in the denser regions. Along a particular line of sight towards the central denser region, the single effective best-fit $T_\mathrm{d}$ has a lower value which is more weighted to the inner denser region having higher intensity than the outer region, and hence the estimated single $a_\mathrm{align}$ value based on the single $T_\mathrm{d}$ value along the line of sight is more weighted to the inner denser region. In this dense region, we find $a_\mathrm{align}$ value of $\approx 0.28$ $\mu$m, which could imply the presence of large aligned grains with sizes $> a_\mathrm{align} \approx 0.28$ $\mu$m that emit polarized radiations.

The unknown accurate 3D structure of the filament and the assumption of a single averaged dust temperature value along a particular line of sight stand to be a limitation in the exact explanation. However, it is likely expected for a possible grain growth in the inner densest regions of the cold and massive filament having a typical dust temperature of around 16 K and $n(\mathrm{H_2})$ of the order of $\approx 10^5$ $\mathrm{cm^{-3}}$, that could explain the observed slope. Further, we perform numerical modeling of thermal dust polarization using M-RAT theory to constrain the $a_\mathrm{max}$ value and our modeling could reproduce the observational data for $a_\mathrm{max}$ values in the range of $\mbox{$0.4-0.5$ $\mu$m}$, accompanied by a higher grain elongation of axial ratio 2 (see Figures \ref{Figure:P_I_model_2_3}, \ref{Figure:P_NH2_model_1_2} and \ref{Figure:P_Td_model_1_2}). Hence, we get an implication that the observed slope could be possibly explained by a moderate grain growth in the inner very dense regions of the filament.



Although grain growth is expected in dense molecular clouds, a detailed study of the physics of grain growth in these dense regions is still required. The grain growth is assumed to be isotropic in the case of non-aligned grains due to the random motions of gas and grains, and it could result in the increase in grain size isotropically from all directions and can evolve the grains towards spherical shapes. However, recently \cite{2022ApJ...928..102H} showed that grain growth can become anisotropic for the grains aligned with the magnetic field, due to anisotropic gas accretion to the aligned grains and it could result in the increase of not only the grain size but also the elongation of the aligned grains, with the grains of larger radii or larger grains showing higher elongation. The grain elongation determines the polarization cross-section efficiency or the intrinsic polarization. The higher the grain elongation in the denser regions, would increase the polarization efficiency of the grains towards denser regions as long as the grains are still aligned.

We see data points showing higher polarization fractions in the denser regions that correspond to $a_\mathrm{align} > 0.21$ $\mu$m (see Figures \ref{Figure:P_I_NH2}, \ref{Figure:P_S}(b) and (c), \ref{Figure:P_P_S_a_align}), instead of an expected significant decrease. This may imply the insignificant effect of magnetic field tangling in these denser regions and the presence of organized components of magnetic field. Because if the magnetic fields were more disorganized or non-uniform, the polarizations would have shown lower values due to polarization cancellations, which is not the case. Therefore, we try to find out the possible factors that could explain the observed higher polarization fractions in these denser regions. These data points are associated with higher magnetic relaxation strengths (see Figure \ref{Figure:P_a_align_mr}). The observed significant increase in $P$ values in these denser regions could possibly be produced by the perfect alignment, by M-RATs, of those aligned grains that undergo significant grain growth with $a_\mathrm{align} > 0.21$ $\mu$m, accompanied by an increase in the elongation of those grains. A significant increase in the elongation of the large aligned grains in the denser regions, which can increase the intrinsic polarization efficiencies of each of the individual aligned grains, could be caused by the anisotropic grain growth due to anisotropic gas accretion to the aligned grains. Our numerical modeling of thermal dust polarization using M-RAT theory implies the requirement of grain growth accompanied by an increase in the grain elongation to reproduce the observational data (see Figures \ref{Figure:P_I_model_2_3}, \ref{Figure:P_NH2_model_1_2}, \ref{Figure:P_Td_model_1_2}), providing observational evidence that supports the anisotropic grain growth model of aligned grains as predicted in \cite{2022ApJ...928..102H}. 

The first observational evidence of the anisotropic grain growth model was provided in the study of the isolated starless core 109 in the Pipe Nebula (Pipe-109) in \cite{2025arXiv250116079T}. The grain elongation could also be caused due to the effect of the centrifugal force that arises from the fast rotation of the grains by RATs \citep{2024A&A...692A..60R}. However, the dust temperature or equivalently the radiation field strength, is rather weak in the dense regions of the G16 filament, which does not have bright embedded sources and can not significantly make the grains more elongated in the denser core regions through the effect of centrifugal force.

\subsection{Evidence for varying 3D magnetic field and its effect} \label{section: Evidence for varying 3D magnetic field and its effect}
Our numerical modeling assumes that the magnetic fields lie on the plane-of-sky with the inclination angle $\psi=90^\circ$ or the depolarization factor $\phi=1$. However, in a realistic situation, $\psi$ may not be $90^\circ$ and could vary from region to region over the filament. In this situation, the polarization fraction will be reduced by a factor of $\phi < 1$ or $f_\mathrm{max}\mathrm{sin}^2\psi$. 

The polarization slopes from our modeling are found to be shallower than the observed data (see Figures \ref{Figure:P_I_model_2_3} and \ref{Figure:P_NH2_model_1_2}). This can be due to our assumption of a constant value of $\phi=1$ or a constant magnetic field's inclination angle $\psi=90^\circ$ with the line of sight, over the whole filament.
For our model to better reproduce the observational slope, the factor $\phi$ should decrease from the outer to the inner regions with the gas density, instead of a constant value all over the filament.
Hence, our modeling infers the requirement of variation in the inclination angle from the outer to the inner regions of the filament, which is also evident from the observation that the magnetic fields appear to be curved towards the filament bone in the main filament in Figure \ref{Figure:Intensity_P_map_Histogram_P}. Numerical simulations for the formation of cores by \cite{2017ApJ...838...40M} show that the magnetic field inclination angle varies from the outer to the inner region for the case of moderate and strong magnetic fields. \cite{2018A&A...614A.100T} found the bending of magnetic fields towards several filaments. The curved magnetic fields are also observed in the filament bones of the massive filament G11 (see \citealt{2023ApJ...953...66N}) and G34 (see \citealt{2019ApJ...883...95S}). Also, \cite{2024A&A...689A.290T} showed a bow-shaped magnetic field around OMC-1. The bending of the magnetic fields towards the bone of the G11 filament is also inferred through the fitting of the starlight polarization with the modeling (Truong et al., under preparation). Hence, different studies in different environments from both theoretical and observational perspectives show that the magnetic field's inclination angle is found to vary from the outer to the inner regions in star-forming regions or filaments. Therefore, our assumption of a constant $\phi$ value or constant magnetic field's inclination angle over the whole filament limits the accuracy of our modeling to compare and better reproduce the observational data. We need to consider the localised variation in the value of $\psi$ from region to region for accurate modeling, which requires accurate information on the 3D magnetic field geometry. There is a lack of information on the precise 3D magnetic field geometry for the filament of our study. However, we derive quantitatively the magnetic field's inclination angle $\psi$ or the factor $\phi$ using the technique as described in \cite{2024ApJ...965..183H}. The value of $\phi$ decreases overall from the outermost, less dense regions with a value nearly 1 to the inner denser regions with values less than 1 (see Figure \ref{Figure:phi_NH2_model_1_2}), implying the varying nature of the magnetic field's inclination angle.  


Due to the variation in the value of $\psi$, the polarization fraction from our model, that assumes $\psi=90^\circ$, would get decreased by a factor of $\phi < 1$ and hence the model with a fixed grain elongation of 1.4 with $a_\mathrm{max}=0.45$ $\mu$m could not fully reproduce the observational data. This implies that the grain elongation should be increased to reproduce the observational data.  

In future, a consistent model that incorporates the precise 3D magnetic field geometry could help in an accurate modeling, and it could better reproduce the observational data and constrain precisely the grain alignment mechanism and grain properties.

\subsection{Comparison with other studies and future directions}
Several studies in dense star-forming regions use the parameters $S$ and $P \times S$ to investigate whether observed depolarization in the denser regions is due to decrease in grain alignment efficiency or
magnetic field fluctuations. For example, in the study of the dense star-forming regions G11 by \cite{2023ApJ...953...66N}, Musca by \cite{2024ApJ...974..118N}, G34 by \cite{2025ApJ...981..128P} and Cocoon Nebula by \cite{2025ApJ...990...40P}, using the dust polarization observations from SOFIA/HAWC+ and JCMT/POL-2 instruments, they found no significant correlation between $P$ and $S$, and the $P \times S$ decreases in the denser regions, almost similar to the decrease of $P$ in the denser regions, implying that the observed depolarization is majorly due to the decrease in grain alignment efficiency in the denser regions and the effect of magnetic field tangling is minor. However, the above studies are associated with data of around $18''$ and $14''$ resolutions, which could not fully resolve the dense core regions. In our present study with $14''$ resolution, we get the similar conclusions to the above studies. However, for the dense core regions, the depolarization may be caused by less uniform orientations of magnetic fields due to gravity, turbulence, outflows, etc.

In several studies in dense star-forming regions using higher angular resolution data with SMA, CARMA, ALMA, etc., it is found that the morphology of magnetic field gets changed from the large scale towards the small scales (see e.g., \citealt{2014ApJS..213...13H, 2017ApJ...838..121C, 2017ApJ...842L...9H, 2018ApJ...855...39K, 2026AJ....171...50H}). 
Several of these studies show that the magnetic fields in the dense core regions show different morphologies compared to the outer less dense regions. It is not always the case that the magnetic field becomes random in the dense core regions. When the magnetic fields become more random, some of the polarizations may get cancelled, resulting in net decrease in the polarization fraction. There could be organized or ordered component of magnetic field also and the polarization observation could result from those grains aligned with this ordered component. There is still a lack of proper investigation to explain the depolarization, particularly in the dense cold cores using high resolution data.

However, some studies on protostellar cores have been done that use the parameters $P$ and $P \times S$ using high resolution data from ALMA. For example, \cite{2020A&A...644A..11L} performed a statistical analysis of thermal dust polarization towards 11 class 0 protostellar cores, observed by ALMA at wavelengths ranging from 870 $\mu$m to 3 mm and the study found a significant anti-correlation between $P$ and $S$, which could imply the significant contribution of magnetic field tangling to cause the depolarization. The study also found that the parameter $P \times S$ shows a constant profile as a function of gas column density in each of the cores. Also, \cite{2018ApJ...855...39K} studied the polarization properties and magnetic field structures on protostellar cores in the high-mass star-forming region W51 using high resolution ALMA dust polarization observations. They studied the relation between $P$ and $S$ for each of the dense cores and found that the lowest $P$ value is associated with maximum $S$ values, which may imply that the decrease in $P$ with higher $S$ can be due to the cancellation of some polarizations because of rapid changes in the magnetic field orientations. Hence, the explanation for the depolarization could be favored towards the changing of the magnetic field morphology, in case of fully resolved dense cores. However, an extensive study is still required on different types of dense cores (starless, prestellar and protostellar cores). In future, employing high resolution data and fully resolving the dense cores would help in revealing the detailed physical mechanisms occurring there.

\subsection{Limitations of our analysis and modeling}
In our numerical modeling of thermal dust polarization using \textsc{DustPOL\_py}, we first compute the model polarization for an ideal condition of uniform magnetic field on the plane-of-sky. Then, we compute the model polarization taking into account the depolarization effects resulting from the magnetic field tangling along the line of sight and within the beam by incorporating the factor $S^{-\eta}$, but considering the plane-of-sky magnetic field. Our polarization model is based on the assumption of optically thin emission, which is successfully tested with synthetic observations of MHD simulations using POLARIS in \cite{2024ApJ...965..183H}. Although our modeling could reproduce the observational data, there are limitations also.


One limitation of our modeling is the assumption that the maximum grain size is the same in different lines of sight (pixels), which is expected to vary with the local physical parameters like density.
Accurate models that consistently constrain the maximum grain size as a function of local physical conditions could help better compare with the observations.
Another limitation is that we assume the Astrodust grains over the whole filament, which may be subject to uncertainty if separate populations of silicate and carbonaceous grains are found at some regions of the filament. However, we expect that the overall results of our modeling would not be significantly affected, considering the environment of our study.

Again, in this study, our analysis of grain alignment mechanisms and modeling of dust polarization are based on previously derived estimates of the magnetic field strength, which are subject to uncertainty. Our estimation of the gas volume density based on the assumption of a cylindrical model of the overall filament with a spherical model for the embedded cores could be subject to uncertainty since there is a lack of information on the precise 3D structure of the filament. These uncertainties can influence the derived grain alignment size and magnetic relaxation, and polarization degree from our modeling. Nevertheless, we expect that our overall conclusions will remain valid when more accurate measurements of the gas density and magnetic field strength become available.

\section{Conclusions} \label{section:Conclusions}
In this work, we investigate the grain alignment mechanisms in a massive, quiescent and filamentary star-forming IRDC G16.96+0.27 or G16 using thermal dust polarization observations towards this filament with JCMT/POL-2 at 850 $\mu$m. We perform a comprehensive testing of the unified grain alignment theory (M-RAT mechanism) and constrain the grain properties in this filament. Our main results are summarized as follows:

1. We find that the observed polarization fraction $P$ decreases with the increases in the total emission intensity $I$ and the gas column density $N(\mathrm{H_2})$, termed as polarization hole or depolarization, and increases with the increase in the dust temperature $T_\mathrm{d}$.


2. We investigate for any role of magnetic field tangling to cause the observed polarization hole by estimating the polarization angle dispersion function $S$ and the averaged grain alignment efficiency $P \times S$ and then analyzing the variations of $P$ with $S$ and $P \times S$ with $I$, $N(\mathrm{H_2})$ and $T_\mathrm{d}$. We find that the effect of magnetic field tangling on the observed polarization hole is minor and not significant and the observed polarization hole is mainly due to the decrease in RAT alignment efficiency of grains in the denser regions.

3. We further test whether the analytical RAT-A theory can reproduce the observational results by estimating the minimum alignment size of grains, $a_\mathrm{align}$ using RAT-A theory and analyzing the variations of $a_\mathrm{align}$ with $I$, and $P$ and $P \times S$ with $a_\mathrm{align}$. Also, we perform detailed numerical modeling of thermal dust polarization using the physics of grain alignment by RATs and our modeling could reproduce the observational results, further strengthening the evidence for RAT-A mechanism.

4. Using the magnetic field strengths estimated over the filament with ADF, DCF and the average of ADF and DCF methods, we study the significance of magnetic relaxation strength on the RAT alignment for each case. We find that the higher polarization fractions of above 8\% up to around 23\% observed in the outer regions of the filament are associated with effective magnetic relaxation strengths with $\delta_\mathrm{mag,sp} > 10$ and stronger RATs. The observed high polarization fractions could be due to the combined effect of both the stronger RAT alignment efficiency of grains and the effective magnetic relaxation strengths, providing an implication for the alignment of grains by M-RAT mechanism. Again, the observation of a significant increase in the polarization fractions in the denser regions for $a_\mathrm{align} > 0.21$ $\mu$m, associated with higher magnetic relaxation strength, implies the significant role of the stronger magnetic relaxation strengths on the RAT alignment efficiency of those large grains with $a \gg a_\mathrm{align}$, supporting the M-RAT mechanism. Our numerical modeling of dust polarization that considers the perfect alignment of grains with $f_\mathrm{max}=1$, could successfully reproduce the observational data and hence it further supports the M-RAT mechanism.

5. We also find an implication for a possible grain growth in the densest regions of the filament's spine, based on the slope value of the variation of polarization fraction with the total intensity and the RAT alignment theory. In the outer regions, the maximum grain size $a_\mathrm{max}$ values are larger than the typical $a_\mathrm{align}$ value of $\approx 0.15$ $\mu$m. In the densest regions of the filament's spine, $a_\mathrm{max}$ needs to be larger than the $a_\mathrm{align}$ value of $\approx 0.28$ $\mu$m to reproduce the slope of $-0.71$. The observed slope could also be due to the emission of polarized radiations from the aligned grains in the outer region along the line of sight without the need for grain growth in the inner denser region, if we consider a temperature gradient along the line of sight for our filament that assumes a cylindrical model. Our further study with numerical modeling of thermal dust polarization constrains the $a_\mathrm{max}$ value of around $0.4-0.5$ $\mu$m that could reproduce the observational data, providing implications of moderate grain growth in the filament.

6. Apart from the overall implication of grain growth in the densest regions, our numerical modeling of thermal dust polarization requires an increase in the grain elongation of the large aligned grains to reproduce the observational data, providing evidence for the anisotropic grain growth model of aligned grains.

7. Our numerical modeling of thermal dust polarization implies the requirement of variations in the magnetic field's inclination angle with the line of sight, such that it should decrease from the outer to the inner regions of the filament, to reproduce the observational data, providing an insight into the evolution of the 3D magnetic fields over the filament. 
This could potentially explain the steep decrease in the polarization fraction nearly by an order of magnitude within a narrow range of gas column density associated with a factor of 2, which is insufficient to be explained alone by the decrease in the grain alignment efficiency by RATs in the denser regions.  

8. The success of our modeling to reproduce the observational data opens the way forward for testing grain alignment mechanisms and constraining the grain properties in other dense star-forming regions.

\begin{acknowledgments}
This research has made use of observational data from James Clerk Maxwell Telescope (JCMT) POL-2 instrument. JCMT is operated by the East Asian Observatory on behalf of the National Astronomical Observatory of Japan; Academia Sinica Institute of Astronomy and Astrophysics; the Korea Astronomy and Space Science Institute; the Operation, Maintenance and Upgrading Fund for Astronomical Telescopes and Facility Instruments, budgeted from the Ministry of Finance of China and administrated by the Chinese Academy of Sciences and, the National Key R and D Program of China.
N.B.N and P.N.D acknowledge the support from the Vietnam National Foundation for Science and Technology Development (NAFOSTED) under grant number 103.99-2024.36.
T.H. acknowledges the support from the major research project (No. 2025186902) from Korea Astronomy and Space Science Institute (KASI) funded by the Ministry of Science and ICT (MSIT).

\end{acknowledgments}

%
\facilities{JCMT, Herschel Space Observatory}

\software{Astropy \citep{2013A&A...558A..33A, 2018AJ....156..123A}, Scipy \citep{2020NatMe..17..261V}}




\bibliography{G16_references}{}

@ARTICLE{1988QJRAS..29..327H,
       author = {{Hildebrand}, Roger H.},
        title = "{Magnetic fields and stardust}",
      journal = {\qjras},
         year = 1988,
        month = sep,
       volume = {29},
        pages = {327-351},
       adsurl = {https://ui.adsabs.harvard.edu/abs/1988QJRAS..29..327H}
}

@ARTICLE{1949Sci...109..166H,
       author = {{Hall}, John S.},
        title = "{Observations of the Polarized Light from Stars}",
      journal = {Science},
         year = 1949,
        month = feb,
       volume = {109},
       number = {2825},
        pages = {166-167},
          doi = {10.1126/science.109.2825.166},
       adsurl = {https://ui.adsabs.harvard.edu/abs/1949Sci...109..166H}
}

@ARTICLE{1949ApJ...109..471H,
       author = {{Hiltner}, W.~A.},
        title = "{On the Presence of Polarization in the Continuous Radiation of Stars. II.}",
      journal = {\apj},
         year = 1949,
        month = may,
       volume = {109},
        pages = {471},
          doi = {10.1086/145151},
       adsurl = {https://ui.adsabs.harvard.edu/abs/1949ApJ...109..471H}
}

@ARTICLE{2012ARA&A..50...29C,
       author = {{Crutcher}, Richard M.},
        title = "{Magnetic Fields in Molecular Clouds}",
      journal = {\araa},
         year = 2012,
        month = sep,
       volume = {50},
        pages = {29-63},
          doi = {10.1146/annurev-astro-081811-125514},
       adsurl = {https://ui.adsabs.harvard.edu/abs/2012ARA&A..50...29C}
}

@ARTICLE{2019FrASS...6...15P,
       author = {{Pattle}, Kate and {Fissel}, Laura},
        title = "{Submillimeter and Far-infrared Polarimetric Observations of Magnetic Fields in Star-Forming Regions}",
      journal = {Frontiers in Astronomy and Space Sciences},
         year = 2019,
        month = apr,
       volume = {6},
          eid = {15},
        pages = {15},
          doi = {10.3389/fspas.2019.00015},
archivePrefix = {arXiv},
       eprint = {1904.04826},
 primaryClass = {astro-ph.GA},
       adsurl = {https://ui.adsabs.harvard.edu/abs/2019FrASS...6...15P}
}

@ARTICLE{2021ApJ...919...65D,
       author = {{Draine}, B.~T. and {Hensley}, Brandon S.},
        title = "{Using the Starlight Polarization Efficiency Integral to Constrain Shapes and Porosities of Interstellar Grains}",
      journal = {\apj},
         year = 2021,
        month = sep,
       volume = {919},
       number = {1},
          eid = {65},
        pages = {65},
          doi = {10.3847/1538-4357/ac0050},
archivePrefix = {arXiv},
       eprint = {2101.07277},
 primaryClass = {astro-ph.GA},
       adsurl = {https://ui.adsabs.harvard.edu/abs/2021ApJ...919...65D}
}

@ARTICLE{2007MNRAS.378..910L,
       author = {{Lazarian}, A. and {Hoang}, Thiem},
        title = "{Radiative torques: analytical model and basic properties}",
      journal = {\mnras},
         year = 2007,
        month = jul,
       volume = {378},
       number = {3},
        pages = {910-946},
          doi = {10.1111/j.1365-2966.2007.11817.x},
archivePrefix = {arXiv},
       eprint = {0707.0886},
 primaryClass = {astro-ph},
       adsurl = {https://ui.adsabs.harvard.edu/abs/2007MNRAS.378..910L}
}

@ARTICLE{1976Ap&SS..43..291D,
       author = {{Dolginov}, A.~Z. and {Mitrofanov}, I.~G.},
        title = "{Orientation of Cosmic Dust Grains}",
      journal = {\apss},
         year = 1976,
        month = sep,
       volume = {43},
       number = {2},
        pages = {291-317},
          doi = {10.1007/BF00640010},
       adsurl = {https://ui.adsabs.harvard.edu/abs/1976Ap&SS..43..291D}
}

@ARTICLE{1997ApJ...480..633D,
       author = {{Draine}, B.~T. and {Weingartner}, Joseph C.},
        title = "{Radiative Torques on Interstellar Grains. II. Grain Alignment}",
      journal = {\apj},
         year = 1997,
        month = may,
       volume = {480},
       number = {2},
        pages = {633-646},
          doi = {10.1086/304008},
archivePrefix = {arXiv},
       eprint = {astro-ph/9611149},
 primaryClass = {astro-ph},
       adsurl = {https://ui.adsabs.harvard.edu/abs/1997ApJ...480..633D}
}

@ARTICLE{2021ApJ...908..218H,
       author = {{Hoang}, Thiem and {Tram}, Le Ngoc and {Lee}, Hyeseung and {Diep}, Pham Ngoc and {Ngoc}, Nguyen Bich},
        title = "{Grain Alignment and Disruption by Radiative Torques in Dense Molecular Clouds and Implication for Polarization Holes}",
      journal = {\apj},
         year = 2021,
        month = feb,
       volume = {908},
       number = {2},
          eid = {218},
        pages = {218},
          doi = {10.3847/1538-4357/abd54f},
archivePrefix = {arXiv},
       eprint = {2010.07742},
 primaryClass = {astro-ph.GA},
       adsurl = {https://ui.adsabs.harvard.edu/abs/2021ApJ...908..218H}
}

@ARTICLE{2015ARA&A..53..501A,
       author = {{Andersson}, B. -G. and {Lazarian}, A. and {Vaillancourt}, John E.},
        title = "{Interstellar Dust Grain Alignment}",
      journal = {\araa},
         year = 2015,
        month = aug,
       volume = {53},
        pages = {501-539},
          doi = {10.1146/annurev-astro-082214-122414},
       adsurl = {https://ui.adsabs.harvard.edu/abs/2015ARA&A..53..501A}
}

@ARTICLE{1989ApJ...346..728J,
       author = {{Jones}, Terry Jay},
        title = "{Infrared Polarimetry and the Interstellar Magnetic Field}",
      journal = {\apj},
         year = 1989,
        month = nov,
       volume = {346},
        pages = {728},
          doi = {10.1086/168054},
       adsurl = {https://ui.adsabs.harvard.edu/abs/1989ApJ...346..728J}
}

@ARTICLE{1992ApJ...389..602J,
       author = {{Jones}, Terry J. and {Klebe}, Dimitri and {Dickey}, John M.},
        title = "{Infrared Polarimetry and the Galactic Magnetic Field. II. Improved Models}",
      journal = {\apj},
         year = 1992,
        month = apr,
       volume = {389},
        pages = {602},
          doi = {10.1086/171233},
       adsurl = {https://ui.adsabs.harvard.edu/abs/1992ApJ...389..602J}
}

@ARTICLE{2008ApJ...679..537F,
       author = {{Falceta-Gon{\c{c}}alves}, Diego and {Lazarian}, Alex and {Kowal}, Grzegorz},
        title = "{Studies of Regular and Random Magnetic Fields in the ISM: Statistics of Polarization Vectors and the Chandrasekhar-Fermi Technique}",
      journal = {\apj},
         year = 2008,
        month = may,
       volume = {679},
       number = {1},
        pages = {537-551},
          doi = {10.1086/587479},
archivePrefix = {arXiv},
       eprint = {0801.0279},
 primaryClass = {astro-ph},
       adsurl = {https://ui.adsabs.harvard.edu/abs/2008ApJ...679..537F}
}

@ARTICLE{2020A&A...641A..12P,
       author = {{Planck Collaboration} and {Aghanim}, N. and {Akrami}, Y. and {Alves}, M.~I.~R. and {Ashdown}, M. and {Aumont}, J. and {Baccigalupi}, C. and {Ballardini}, M. and {Banday}, A.~J. and {Barreiro}, R.~B. and {Bartolo}, N. and {Basak}, S. and {Benabed}, K. and {Bernard}, J. -P. and {Bersanelli}, M. and {Bielewicz}, P. and {Bock}, J.~J. and {Bond}, J.~R. and {Borrill}, J. and {Bouchet}, F.~R. and {Boulanger}, F. and {Bracco}, A. and {Bucher}, M. and {Burigana}, C. and {Calabrese}, E. and {Cardoso}, J. -F. and {Carron}, J. and {Chary}, R. -R. and {Chiang}, H.~C. and {Colombo}, L.~P.~L. and {Combet}, C. and {Crill}, B.~P. and {Cuttaia}, F. and {de Bernardis}, P. and {de Zotti}, G. and {Delabrouille}, J. and {Delouis}, J. -M. and {Di Valentino}, E. and {Dickinson}, C. and {Diego}, J.~M. and {Dor{\'e}}, O. and {Douspis}, M. and {Ducout}, A. and {Dupac}, X. and {Efstathiou}, G. and {Elsner}, F. and {En{\ss}lin}, T.~A. and {Eriksen}, H.~K. and {Falgarone}, E. and {Fantaye}, Y. and {Fernandez-Cobos}, R. and {Ferri{\`e}re}, K. and {Finelli}, F. and {Forastieri}, F. and {Frailis}, M. and {Fraisse}, A.~A. and {Franceschi}, E. and {Frolov}, A. and {Galeotta}, S. and {Galli}, S. and {Ganga}, K. and {G{\'e}nova-Santos}, R.~T. and {Gerbino}, M. and {Ghosh}, T. and {Gonz{\'a}lez-Nuevo}, J. and {G{\'o}rski}, K.~M. and {Gratton}, S. and {Green}, G. and {Gruppuso}, A. and {Gudmundsson}, J.~E. and {Guillet}, V. and {Handley}, W. and {Hansen}, F.~K. and {Helou}, G. and {Herranz}, D. and {Hivon}, E. and {Huang}, Z. and {Jaffe}, A.~H. and {Jones}, W.~C. and {Keih{\"a}nen}, E. and {Keskitalo}, R. and {Kiiveri}, K. and {Kim}, J. and {Krachmalnicoff}, N. and {Kunz}, M. and {Kurki-Suonio}, H. and {Lagache}, G. and {Lamarre}, J. -M. and {Lasenby}, A. and {Lattanzi}, M. and {Lawrence}, C.~R. and {Le Jeune}, M. and {Levrier}, F. and {Liguori}, M. and {Lilje}, P.~B. and {Lindholm}, V. and {L{\'o}pez-Caniego}, M. and {Lubin}, P.~M. and {Ma}, Y. -Z. and {Mac{\'\i}as-P{\'e}rez}, J.~F. and {Maggio}, G. and {Maino}, D. and {Mandolesi}, N. and {Mangilli}, A. and {Marcos-Caballero}, A. and {Maris}, M. and {Martin}, P.~G. and {Mart{\'\i}nez-Gonz{\'a}lez}, E. and {Matarrese}, S. and {Mauri}, N. and {McEwen}, J.~D. and {Melchiorri}, A. and {Mennella}, A. and {Migliaccio}, M. and {Miville-Desch{\^e}nes}, M. -A. and {Molinari}, D. and {Moneti}, A. and {Montier}, L. and {Morgante}, G. and {Moss}, A. and {Natoli}, P. and {Pagano}, L. and {Paoletti}, D. and {Patanchon}, G. and {Perrotta}, F. and {Pettorino}, V. and {Piacentini}, F. and {Polastri}, L. and {Polenta}, G. and {Puget}, J. -L. and {Rachen}, J.~P. and {Reinecke}, M. and {Remazeilles}, M. and {Renzi}, A. and {Ristorcelli}, I. and {Rocha}, G. and {Rosset}, C. and {Roudier}, G. and {Rubi{\~n}o-Mart{\'\i}n}, J.~A. and {Ruiz-Granados}, B. and {Salvati}, L. and {Sandri}, M. and {Savelainen}, M. and {Scott}, D. and {Sirignano}, C. and {Sunyaev}, R. and {Suur-Uski}, A. -S. and {Tauber}, J.~A. and {Tavagnacco}, D. and {Tenti}, M. and {Toffolatti}, L. and {Tomasi}, M. and {Trombetti}, T. and {Valiviita}, J. and {Vansyngel}, F. and {Van Tent}, B. and {Vielva}, P. and {Villa}, F. and {Vittorio}, N. and {Wandelt}, B.~D. and {Wehus}, I.~K. and {Zacchei}, A. and {Zonca}, A.},
        title = "{Planck 2018 results. XII. Galactic astrophysics using polarized dust emission}",
      journal = {\aap},
         year = 2020,
        month = sep,
       volume = {641},
          eid = {A12},
        pages = {A12},
          doi = {10.1051/0004-6361/201833885},
archivePrefix = {arXiv},
       eprint = {1807.06212},
 primaryClass = {astro-ph.GA},
       adsurl = {https://ui.adsabs.harvard.edu/abs/2020A&A...641A..12P}
}

@ARTICLE{2020ApJ...896...44L,
       author = {{Lee}, Hyeseung and {Hoang}, Thiem and {Le}, Ngan and {Cho}, Jungyeon},
        title = "{Physical Model of Dust Polarization by Radiative Torque Alignment and Disruption and Implications for Grain Internal Structures}",
      journal = {\apj},
         year = 2020,
        month = jun,
       volume = {896},
       number = {1},
          eid = {44},
        pages = {44},
          doi = {10.3847/1538-4357/ab8e33},
archivePrefix = {arXiv},
       eprint = {1911.00654},
 primaryClass = {astro-ph.GA},
       adsurl = {https://ui.adsabs.harvard.edu/abs/2020ApJ...896...44L}
}

@ARTICLE{2006ApJ...641..389R,
       author = {{Rathborne}, J.~M. and {Jackson}, J.~M. and {Simon}, R.},
        title = "{Infrared Dark Clouds: Precursors to Star Clusters}",
      journal = {\apj},
         year = 2006,
        month = apr,
       volume = {641},
       number = {1},
        pages = {389-405},
          doi = {10.1086/500423},
archivePrefix = {arXiv},
       eprint = {astro-ph/0602246},
 primaryClass = {astro-ph},
       adsurl = {https://ui.adsabs.harvard.edu/abs/2006ApJ...641..389R}
}

@ARTICLE{2019ApJ...883...95S,
       author = {{Soam}, Archana and {Liu}, Tie and {Andersson}, B. -G. and {Lee}, Chang Won and {Liu}, Junhao and {Juvela}, Mika and {Li}, Pak Shing and {Goldsmith}, Paul F. and {Zhang}, Qizhou and {Koch}, Patrick M. and {Kim}, Kee-Tae and {Qiu}, Keping and {Evans}, Neal J., II and {Johnstone}, Doug and {Thompson}, Mark and {Ward-Thompson}, Derek and {Di Francesco}, James and {Tang}, Ya-Wen and {Montillaud}, Julien and {Kim}, Gwanjeong and {Mairs}, Steve and {Sanhueza}, Patricio and {Kim}, Shinyoung and {Berry}, David and {Gordon}, Michael S. and {Tatematsu}, Ken'ichi and {Liu}, Sheng-Yuan and {Pattle}, Kate and {Eden}, David and {McGehee}, Peregrine M. and {Wang}, Ke and {Ristorcelli}, I. and {Graves}, Sarah F. and {Alina}, Dana and {Lacaille}, Kevin M. and {Montier}, Ludovic and {Park}, Geumsook and {Kwon}, Woojin and {Chung}, Eun Jung and {Pelkonen}, Veli-Matti and {Micelotta}, Elisabetta R. and {Saajasto}, Mika and {Fuller}, Gary},
        title = "{Magnetic Fields in the Infrared Dark Cloud G34.43+0.24}",
      journal = {\apj},
         year = 2019,
        month = sep,
       volume = {883},
       number = {1},
          eid = {95},
        pages = {95},
          doi = {10.3847/1538-4357/ab39dd},
archivePrefix = {arXiv},
       eprint = {1908.03624},
 primaryClass = {astro-ph.GA},
       adsurl = {https://ui.adsabs.harvard.edu/abs/2019ApJ...883...95S}
}

@ARTICLE{2021ApJ...906..115T,
       author = {{Tram}, Le Ngoc and {Hoang}, Thiem and {Lee}, Hyeseung and {Santos}, Fabio P. and {Soam}, Archana and {Lesaffre}, Pierre and {Gusdorf}, Antoine and {Reach}, William T.},
        title = "{Understanding Polarized Dust Emission from {\ensuremath{\rho}} Ophiuchi A in Light of Grain Alignment and Disruption by Radiative Torques}",
      journal = {\apj},
         year = 2021,
        month = jan,
       volume = {906},
       number = {2},
          eid = {115},
        pages = {115},
          doi = {10.3847/1538-4357/abc6fe},
archivePrefix = {arXiv},
       eprint = {2007.10621},
 primaryClass = {astro-ph.GA},
       adsurl = {https://ui.adsabs.harvard.edu/abs/2021ApJ...906..115T}
}

@ARTICLE{2016A&A...595A..57A,
       author = {{Alina}, D. and {Montier}, L. and {Ristorcelli}, I. and {Bernard}, J. -P. and {Levrier}, F. and {Abdikamalov}, E.},
        title = "{Polarization measurement analysis. III. Analysis of the polarization angle dispersion function with high precision polarization data}",
      journal = {\aap},
         year = 2016,
        month = oct,
       volume = {595},
          eid = {A57},
        pages = {A57},
          doi = {10.1051/0004-6361/201628809},
archivePrefix = {arXiv},
       eprint = {1608.07105},
 primaryClass = {astro-ph.GA},
       adsurl = {https://ui.adsabs.harvard.edu/abs/2016A&A...595A..57A}
}

@ARTICLE{2008MNRAS.388..117H,
       author = {{Hoang}, Thiem and {Lazarian}, A.},
        title = "{Radiative torque alignment: essential physical processes}",
      journal = {\mnras},
         year = 2008,
        month = jul,
       volume = {388},
       number = {1},
        pages = {117-143},
          doi = {10.1111/j.1365-2966.2008.13249.x},
archivePrefix = {arXiv},
       eprint = {0707.3645},
 primaryClass = {astro-ph},
       adsurl = {https://ui.adsabs.harvard.edu/abs/2008MNRAS.388..117H}
}

@ARTICLE{2016ApJ...831..159H,
       author = {{Hoang}, Thiem and {Lazarian}, A.},
        title = "{A Unified Model of Grain Alignment: Radiative Alignment of Interstellar Grains with Magnetic Inclusions}",
      journal = {\apj},
         year = 2016,
        month = nov,
       volume = {831},
       number = {2},
          eid = {159},
        pages = {159},
          doi = {10.3847/0004-637X/831/2/159},
archivePrefix = {arXiv},
       eprint = {1605.02828},
 primaryClass = {astro-ph.GA},
       adsurl = {https://ui.adsabs.harvard.edu/abs/2016ApJ...831..159H}
}

@ARTICLE{2014MNRAS.438..680H,
       author = {{Hoang}, Thiem and {Lazarian}, A.},
        title = "{Grain alignment by radiative torques in special conditions and implications}",
      journal = {\mnras},
         year = 2014,
        month = feb,
       volume = {438},
       number = {1},
        pages = {680-703},
          doi = {10.1093/mnras/stt2240},
archivePrefix = {arXiv},
       eprint = {1407.8228},
 primaryClass = {astro-ph.GA},
       adsurl = {https://ui.adsabs.harvard.edu/abs/2014MNRAS.438..680H}
}

@ARTICLE{2007ApJ...663.1055B,
       author = {{Bethell}, T.~J. and {Chepurnov}, A. and {Lazarian}, A. and {Kim}, J.},
        title = "{Polarization of Dust Emission in Clumpy Molecular Clouds and Cores}",
      journal = {\apj},
         year = 2007,
        month = jul,
       volume = {663},
       number = {2},
        pages = {1055-1068},
          doi = {10.1086/516622},
archivePrefix = {arXiv},
       eprint = {astro-ph/0611324},
 primaryClass = {astro-ph},
       adsurl = {https://ui.adsabs.harvard.edu/abs/2007ApJ...663.1055B}
}

@BOOK{2011piim.book.....D,
       author = {{Draine}, Bruce T.},
        title = "{Physics of the Interstellar and Intergalactic Medium}",
         year = 2011,
       adsurl = {https://ui.adsabs.harvard.edu/abs/2011piim.book.....D}
}

@ARTICLE{2022AJ....164..248H,
       author = {{Hoang}, Thiem and {Tram}, Le Ngoc and {Minh Phan}, Vo Hong and {Giang}, Nguyen Chau and {Phuong}, Nguyen Thi and {Dieu}, Nguyen Duc},
        title = "{On Internal and External Alignment of Dust Grains in Protostellar Environments}",
      journal = {\aj},
         year = 2022,
        month = dec,
       volume = {164},
       number = {6},
          eid = {248},
        pages = {248},
          doi = {10.3847/1538-3881/ac9af5},
archivePrefix = {arXiv},
       eprint = {2205.02334},
 primaryClass = {astro-ph.GA},
       adsurl = {https://ui.adsabs.harvard.edu/abs/2022AJ....164..248H}
}

@ARTICLE{1951ApJ...114..206D,
       author = {{Davis}, Leverett, Jr. and {Greenstein}, Jesse L.},
        title = "{The Polarization of Starlight by Aligned Dust Grains.}",
      journal = {\apj},
         year = 1951,
        month = sep,
       volume = {114},
        pages = {206},
          doi = {10.1086/145464},
       adsurl = {https://ui.adsabs.harvard.edu/abs/1951ApJ...114..206D}
}

@ARTICLE{2021ApJ...913...63H,
       author = {{Herranen}, Joonas and {Lazarian}, A. and {Hoang}, Thiem},
        title = "{Alignment of Irregular Grains by Radiative Torques: Efficiency Study}",
      journal = {\apj},
         year = 2021,
        month = may,
       volume = {913},
       number = {1},
          eid = {63},
        pages = {63},
          doi = {10.3847/1538-4357/abf096},
archivePrefix = {arXiv},
       eprint = {2006.16563},
 primaryClass = {astro-ph.GA},
       adsurl = {https://ui.adsabs.harvard.edu/abs/2021ApJ...913...63H}
}

@ARTICLE{2004ApJ...600..279C,
       author = {{Crutcher}, Richard M. and {Nutter}, D.~J. and {Ward-Thompson}, D. and {Kirk}, J.~M.},
        title = "{SCUBA Polarization Measurements of the Magnetic Field Strengths in the L183, L1544, and L43 Prestellar Cores}",
      journal = {\apj},
         year = 2004,
        month = jan,
       volume = {600},
       number = {1},
        pages = {279-285},
          doi = {10.1086/379705},
archivePrefix = {arXiv},
       eprint = {astro-ph/0305604},
 primaryClass = {astro-ph},
       adsurl = {https://ui.adsabs.harvard.edu/abs/2004ApJ...600..279C}
}

@ARTICLE{2023ApJ...953...66N,
       author = {{Ngoc}, Nguyen Bich and {Diep}, Pham Ngoc and {Hoang}, Thiem and {Tram}, Le Ngoc and {Giang}, Nguyen Chau and {L{\^e}}, Ng{\^a}n and {Hoang}, Thuong D. and {Phuong}, Nguyen Thi and {Khang}, Nguyen Minh and {Nguyen}, Dieu D. and {Truong}, Bao},
        title = "{B-fields and Dust in Interstellar Filaments Using Dust Polarization (BALLAD-POL). I. The Massive Filament G11.11-0.12 Observed by SOFIA/HAWC+}",
      journal = {\apj},
         year = 2023,
        month = aug,
       volume = {953},
       number = {1},
          eid = {66},
        pages = {66},
          doi = {10.3847/1538-4357/acdb6e},
archivePrefix = {arXiv},
       eprint = {2302.10543},
 primaryClass = {astro-ph.GA},
       adsurl = {https://ui.adsabs.harvard.edu/abs/2023ApJ...953...66N}
}

@ARTICLE{2013A&A...558A..33A,
       author = {{Astropy Collaboration} and {Robitaille}, Thomas P. and {Tollerud}, Erik J. and {Greenfield}, Perry and {Droettboom}, Michael and {Bray}, Erik and {Aldcroft}, Tom and {Davis}, Matt and {Ginsburg}, Adam and {Price-Whelan}, Adrian M. and {Kerzendorf}, Wolfgang E. and {Conley}, Alexander and {Crighton}, Neil and {Barbary}, Kyle and {Muna}, Demitri and {Ferguson}, Henry and {Grollier}, Fr{\'e}d{\'e}ric and {Parikh}, Madhura M. and {Nair}, Prasanth H. and {Unther}, Hans M. and {Deil}, Christoph and {Woillez}, Julien and {Conseil}, Simon and {Kramer}, Roban and {Turner}, James E.~H. and {Singer}, Leo and {Fox}, Ryan and {Weaver}, Benjamin A. and {Zabalza}, Victor and {Edwards}, Zachary I. and {Azalee Bostroem}, K. and {Burke}, D.~J. and {Casey}, Andrew R. and {Crawford}, Steven M. and {Dencheva}, Nadia and {Ely}, Justin and {Jenness}, Tim and {Labrie}, Kathleen and {Lim}, Pey Lian and {Pierfederici}, Francesco and {Pontzen}, Andrew and {Ptak}, Andy and {Refsdal}, Brian and {Servillat}, Mathieu and {Streicher}, Ole},
        title = "{Astropy: A community Python package for astronomy}",
      journal = {\aap},
         year = 2013,
        month = oct,
       volume = {558},
          eid = {A33},
        pages = {A33},
          doi = {10.1051/0004-6361/201322068},
archivePrefix = {arXiv},
       eprint = {1307.6212},
 primaryClass = {astro-ph.IM},
       adsurl = {https://ui.adsabs.harvard.edu/abs/2013A&A...558A..33A}
}

@ARTICLE{2018AJ....156..123A,
       author = {{Astropy Collaboration} and {Price-Whelan}, A.~M. and {Sip{\H{o}}cz}, B.~M. and {G{\"u}nther}, H.~M. and {Lim}, P.~L. and {Crawford}, S.~M. and {Conseil}, S. and {Shupe}, D.~L. and {Craig}, M.~W. and {Dencheva}, N. and {Ginsburg}, A. and {VanderPlas}, J.~T. and {Bradley}, L.~D. and {P{\'e}rez-Su{\'a}rez}, D. and {de Val-Borro}, M. and {Aldcroft}, T.~L. and {Cruz}, K.~L. and {Robitaille}, T.~P. and {Tollerud}, E.~J. and {Ardelean}, C. and {Babej}, T. and {Bach}, Y.~P. and {Bachetti}, M. and {Bakanov}, A.~V. and {Bamford}, S.~P. and {Barentsen}, G. and {Barmby}, P. and {Baumbach}, A. and {Berry}, K.~L. and {Biscani}, F. and {Boquien}, M. and {Bostroem}, K.~A. and {Bouma}, L.~G. and {Brammer}, G.~B. and {Bray}, E.~M. and {Breytenbach}, H. and {Buddelmeijer}, H. and {Burke}, D.~J. and {Calderone}, G. and {Cano Rodr{\'\i}guez}, J.~L. and {Cara}, M. and {Cardoso}, J.~V.~M. and {Cheedella}, S. and {Copin}, Y. and {Corrales}, L. and {Crichton}, D. and {D'Avella}, D. and {Deil}, C. and {Depagne}, {\'E}. and {Dietrich}, J.~P. and {Donath}, A. and {Droettboom}, M. and {Earl}, N. and {Erben}, T. and {Fabbro}, S. and {Ferreira}, L.~A. and {Finethy}, T. and {Fox}, R.~T. and {Garrison}, L.~H. and {Gibbons}, S.~L.~J. and {Goldstein}, D.~A. and {Gommers}, R. and {Greco}, J.~P. and {Greenfield}, P. and {Groener}, A.~M. and {Grollier}, F. and {Hagen}, A. and {Hirst}, P. and {Homeier}, D. and {Horton}, A.~J. and {Hosseinzadeh}, G. and {Hu}, L. and {Hunkeler}, J.~S. and {Ivezi{\'c}}, {\v{Z}}. and {Jain}, A. and {Jenness}, T. and {Kanarek}, G. and {Kendrew}, S. and {Kern}, N.~S. and {Kerzendorf}, W.~E. and {Khvalko}, A. and {King}, J. and {Kirkby}, D. and {Kulkarni}, A.~M. and {Kumar}, A. and {Lee}, A. and {Lenz}, D. and {Littlefair}, S.~P. and {Ma}, Z. and {Macleod}, D.~M. and {Mastropietro}, M. and {McCully}, C. and {Montagnac}, S. and {Morris}, B.~M. and {Mueller}, M. and {Mumford}, S.~J. and {Muna}, D. and {Murphy}, N.~A. and {Nelson}, S. and {Nguyen}, G.~H. and {Ninan}, J.~P. and {N{\"o}the}, M. and {Ogaz}, S. and {Oh}, S. and {Parejko}, J.~K. and {Parley}, N. and {Pascual}, S. and {Patil}, R. and {Patil}, A.~A. and {Plunkett}, A.~L. and {Prochaska}, J.~X. and {Rastogi}, T. and {Reddy Janga}, V. and {Sabater}, J. and {Sakurikar}, P. and {Seifert}, M. and {Sherbert}, L.~E. and {Sherwood-Taylor}, H. and {Shih}, A.~Y. and {Sick}, J. and {Silbiger}, M.~T. and {Singanamalla}, S. and {Singer}, L.~P. and {Sladen}, P.~H. and {Sooley}, K.~A. and {Sornarajah}, S. and {Streicher}, O. and {Teuben}, P. and {Thomas}, S.~W. and {Tremblay}, G.~R. and {Turner}, J.~E.~H. and {Terr{\'o}n}, V. and {van Kerkwijk}, M.~H. and {de la Vega}, A. and {Watkins}, L.~L. and {Weaver}, B.~A. and {Whitmore}, J.~B. and {Woillez}, J. and {Zabalza}, V. and {Astropy Contributors}},
        title = "{The Astropy Project: Building an Open-science Project and Status of the v2.0 Core Package}",
      journal = {\aj},
         year = 2018,
        month = sep,
       volume = {156},
       number = {3},
          eid = {123},
        pages = {123},
          doi = {10.3847/1538-3881/aabc4f},
archivePrefix = {arXiv},
       eprint = {1801.02634},
 primaryClass = {astro-ph.IM},
       adsurl = {https://ui.adsabs.harvard.edu/abs/2018AJ....156..123A}
}

@ARTICLE{2024ApJ...965..183H,
       author = {{Hoang}, Thiem and {Truong}, Bao},
        title = "{Probing 3D Magnetic Fields Using Thermal Dust Polarization and Grain Alignment Theory}",
      journal = {\apj},
         year = 2024,
        month = apr,
       volume = {965},
       number = {2},
          eid = {183},
        pages = {183},
          doi = {10.3847/1538-4357/ad2a56},
archivePrefix = {arXiv},
       eprint = {2310.17048},
 primaryClass = {astro-ph.GA},
       adsurl = {https://ui.adsabs.harvard.edu/abs/2024ApJ...965..183H}
}

@ARTICLE{2024arXiv240714896T,
       author = {{Truong}, Bao and {Hoang}, Thiem},
        title = "{Probing 3D magnetic fields using starlight polarization and grain alignment theory}",
      journal = {arXiv e-prints},
         year = 2024,
        month = jul,
          eid = {arXiv:2407.14896},
        pages = {arXiv:2407.14896},
          doi = {10.48550/arXiv.2407.14896},
archivePrefix = {arXiv},
       eprint = {2407.14896},
 primaryClass = {astro-ph.GA},
       adsurl = {https://ui.adsabs.harvard.edu/abs/2024arXiv240714896T}
}

@ARTICLE{2007JQSRT.106..225L,
       author = {{Lazarian}, A.},
        title = "{Tracing magnetic fields with aligned grains}",
      journal = {\jqsrt},
         year = 2007,
        month = jul,
       volume = {106},
        pages = {225-256},
          doi = {10.1016/j.jqsrt.2007.01.038},
archivePrefix = {arXiv},
       eprint = {0707.0858},
 primaryClass = {astro-ph},
       adsurl = {https://ui.adsabs.harvard.edu/abs/2007JQSRT.106..225L}
}

@INCOLLECTION{2015psps.book...81L,
       author = {{Lazarian}, Alexander and {Andersson}, B. -G. and {Hoang}, Thiem},
        title = "{Grain alignment: Role of radiative torques and paramagnetic relaxation}",
    booktitle = {Polarimetry of Stars and Planetary Systems},
         year = 2015,
       editor = {{Kolokolova}, Ludmilla and {Hough}, James and {Levasseur-Regourd}, Anny-Chantal},
        pages = {81},
          doi = {10.48550/arXiv.1511.03696},
       adsurl = {https://ui.adsabs.harvard.edu/abs/2015psps.book...81L}
}

@ARTICLE{2022FrASS...9.3927T,
       author = {{Tram}, Le Ngoc and {Hoang}, Thiem},
        title = "{Recent progress in theory and observational study of dust grain alignment and rotational disruption in star-forming regions}",
      journal = {Frontiers in Astronomy and Space Sciences},
         year = 2022,
        month = oct,
       volume = {9},
          eid = {923927},
        pages = {923927},
          doi = {10.3389/fspas.2022.923927},
archivePrefix = {arXiv},
       eprint = {2208.13195},
 primaryClass = {astro-ph.GA},
       adsurl = {https://ui.adsabs.harvard.edu/abs/2022FrASS...9.3927T}
}

@ARTICLE{2020NatMe..17..261V,
       author = {{Virtanen}, Pauli and {Gommers}, Ralf and {Oliphant}, Travis E. and {Haberland}, Matt and {Reddy}, Tyler and {Cournapeau}, David and {Burovski}, Evgeni and {Peterson}, Pearu and {Weckesser}, Warren and {Bright}, Jonathan and {van der Walt}, St{\'e}fan J. and {Brett}, Matthew and {Wilson}, Joshua and {Millman}, K. Jarrod and {Mayorov}, Nikolay and {Nelson}, Andrew R.~J. and {Jones}, Eric and {Kern}, Robert and {Larson}, Eric and {Carey}, C.~J. and {Polat}, {\.I}lhan and {Feng}, Yu and {Moore}, Eric W. and {VanderPlas}, Jake and {Laxalde}, Denis and {Perktold}, Josef and {Cimrman}, Robert and {Henriksen}, Ian and {Quintero}, E.~A. and {Harris}, Charles R. and {Archibald}, Anne M. and {Ribeiro}, Ant{\^o}nio H. and {Pedregosa}, Fabian and {van Mulbregt}, Paul and {SciPy 1. 0 Contributors}},
        title = "{SciPy 1.0: fundamental algorithms for scientific computing in Python}",
      journal = {Nature Methods},
         year = 2020,
        month = feb,
       volume = {17},
        pages = {261-272},
          doi = {10.1038/s41592-019-0686-2},
archivePrefix = {arXiv},
       eprint = {1907.10121},
 primaryClass = {cs.MS},
       adsurl = {https://ui.adsabs.harvard.edu/abs/2020NatMe..17..261V}
}

@ARTICLE{1977ApJ...217..425M,
       author = {{Mathis}, J.~S. and {Rumpl}, W. and {Nordsieck}, K.~H.},
        title = "{The size distribution of interstellar grains.}",
      journal = {\apj},
         year = 1977,
        month = oct,
       volume = {217},
        pages = {425-433},
          doi = {10.1086/155591},
       adsurl = {https://ui.adsabs.harvard.edu/abs/1977ApJ...217..425M}
}

@ARTICLE{2024ApJ...974..118N,
       author = {{Ngoc}, Nguyen Bich and {Hoang}, Thiem and {Diep}, Pham Ngoc and {Tram}, Le Ngoc},
        title = "{B-fields and Dust in Interstellar Filaments Using Dust Polarization (BALLAD-POL). II. Testing the Radiative Torque Paradigm in Musca and OMC-1}",
      journal = {\apj},
         year = 2024,
        month = oct,
       volume = {974},
       number = {1},
          eid = {118},
        pages = {118},
          doi = {10.3847/1538-4357/ad6a5e},
archivePrefix = {arXiv},
       eprint = {2403.16857},
 primaryClass = {astro-ph.GA},
       adsurl = {https://ui.adsabs.harvard.edu/abs/2024ApJ...974..118N}
}

@ARTICLE{2003ARA&A..41..241D,
       author = {{Draine}, B.~T.},
        title = "{Interstellar Dust Grains}",
      journal = {\araa},
         year = 2003,
        month = jan,
       volume = {41},
        pages = {241-289},
          doi = {10.1146/annurev.astro.41.011802.094840},
archivePrefix = {arXiv},
       eprint = {astro-ph/0304489},
 primaryClass = {astro-ph},
       adsurl = {https://ui.adsabs.harvard.edu/abs/2003ARA&A..41..241D}
}

@INPROCEEDINGS{2016SPIE.9914E..03F,
       author = {{Friberg}, Per and {Bastien}, Pierre and {Berry}, David and {Savini}, Giorgio and {Graves}, Sarah F. and {Pattle}, Kate},
        title = "{POL-2: a polarimeter for the James-Clerk-Maxwell telescope}",
    booktitle = {Millimeter, Submillimeter, and Far-Infrared Detectors and Instrumentation for Astronomy VIII},
         year = 2016,
       editor = {{Holland}, Wayne S. and {Zmuidzinas}, Jonas},
       series = {Society of Photo-Optical Instrumentation Engineers (SPIE) Conference Series},
       volume = {9914},
        month = jul,
          eid = {991403},
        pages = {991403},
          doi = {10.1117/12.2231943},
       adsurl = {https://ui.adsabs.harvard.edu/abs/2016SPIE.9914E..03F}
}

@ARTICLE{2022A&A...657L..13P,
       author = {{Panopoulou}, G.~V. and {Clark}, S.~E. and {Hacar}, A. and {Heitsch}, F. and {Kainulainen}, J. and {Ntormousi}, E. and {Seifried}, D. and {Smith}, R.~J.},
        title = "{The width of Herschel filaments varies with distance}",
      journal = {\aap},
         year = 2022,
        month = jan,
       volume = {657},
          eid = {L13},
        pages = {L13},
          doi = {10.1051/0004-6361/202142281},
archivePrefix = {arXiv},
       eprint = {2111.08125},
 primaryClass = {astro-ph.GA},
       adsurl = {https://ui.adsabs.harvard.edu/abs/2022A&A...657L..13P}
}

@ARTICLE{2014ApJS..213...13H,
       author = {{Hull}, Charles L.~H. and {Plambeck}, Richard L. and {Kwon}, Woojin and {Bower}, Geoffrey C. and {Carpenter}, John M. and {Crutcher}, Richard M. and {Fiege}, Jason D. and {Franzmann}, Erica and {Hakobian}, Nicholas S. and {Heiles}, Carl and {Houde}, Martin and {Hughes}, A. Meredith and {Lamb}, James W. and {Looney}, Leslie W. and {Marrone}, Daniel P. and {Matthews}, Brenda C. and {Pillai}, Thushara and {Pound}, Marc W. and {Rahman}, Nurur and {Sandell}, G{\"o}ran and {Stephens}, Ian W. and {Tobin}, John J. and {Vaillancourt}, John E. and {Volgenau}, N.~H. and {Wright}, Melvyn C.~H.},
        title = "{TADPOL: A 1.3 mm Survey of Dust Polarization in Star-forming Cores and Regions}",
      journal = {\apjs},
         year = 2014,
        month = jul,
       volume = {213},
       number = {1},
          eid = {13},
        pages = {13},
          doi = {10.1088/0067-0049/213/1/13},
archivePrefix = {arXiv},
       eprint = {1310.6653},
 primaryClass = {astro-ph.SR},
       adsurl = {https://ui.adsabs.harvard.edu/abs/2014ApJS..213...13H}
}

@ARTICLE{2020A&A...644A..11L,
       author = {{Le Gouellec}, V.~J.~M. and {Maury}, A.~J. and {Guillet}, V. and {Hull}, C.~L.~H. and {Girart}, J.~M. and {Verliat}, A. and {Mignon-Risse}, R. and {Valdivia}, V. and {Hennebelle}, P. and {Gonz{\'a}lez}, M. and {Louvet}, F.},
        title = "{A statistical analysis of dust polarization properties in ALMA observations of Class 0 protostellar cores}",
      journal = {\aap},
         year = 2020,
        month = dec,
       volume = {644},
          eid = {A11},
        pages = {A11},
          doi = {10.1051/0004-6361/202038404},
archivePrefix = {arXiv},
       eprint = {2009.07186},
 primaryClass = {astro-ph.GA},
       adsurl = {https://ui.adsabs.harvard.edu/abs/2020A&A...644A..11L}
}

@ARTICLE{2024arXiv240710079C,
       author = {{Chau Giang}, Nguyen and {Le Gouellec}, V.~J.~M. and {Hoang}, Thiem and {Maury}, A.~J. and {Hennebelle}, P.},
        title = "{Synthetic Modelling of Polarized Dust Emission in Intermediate-Mass YSOs: I: Constraining the Role of Iron Inclusions and Inelastic Relaxation on Grain Alignment with ALMA Polarization}",
      journal = {arXiv e-prints},
         year = 2024,
        month = jul,
          eid = {arXiv:2407.10079},
        pages = {arXiv:2407.10079},
          doi = {10.48550/arXiv.2407.10079},
archivePrefix = {arXiv},
       eprint = {2407.10079},
 primaryClass = {astro-ph.GA},
       adsurl = {https://ui.adsabs.harvard.edu/abs/2024arXiv240710079C}
}

@ARTICLE{2015A&C....10...22B,
       author = {{Berry}, D.~S.},
        title = "{FellWalker-A clump identification algorithm}",
      journal = {Astronomy and Computing},
         year = 2015,
        month = apr,
       volume = {10},
        pages = {22-31},
          doi = {10.1016/j.ascom.2014.11.004},
archivePrefix = {arXiv},
       eprint = {1411.6267},
 primaryClass = {astro-ph.IM},
       adsurl = {https://ui.adsabs.harvard.edu/abs/2015A&C....10...22B}
}

@ARTICLE{2024ApJ...976..249G,
       author = {{Gu}, Qi-Lao and {Liu}, Tie and {Shen}, Zhi-Qiang and {Jiao}, Sihan and {Montillaud}, Julien and {Juvela}, Mika and {Lu}, Xing and {Lee}, Chang Won and {Liu}, Junhao and {Li}, Pak Shing and {Liu}, Xunchuan and {Johnstone}, Doug and {Kwon}, Woojin and {Kim}, Kee-Tae and {Tatematsu}, Ken'ichi and {Sanhueza}, Patricio and {Ristorcelli}, Isabelle and {Koch}, Patrick and {Zhang}, Qizhou and {Pattle}, Kate and {Hirano}, Naomi and {Alina}, Dana and {Di Francesco}, James},
        title = "{The Magnetic Field in Quiescent Star-forming Filament G16.96+0.27}",
      journal = {\apj},
         year = 2024,
        month = dec,
       volume = {976},
       number = {2},
          eid = {249},
        pages = {249},
          doi = {10.3847/1538-4357/ad8912},
archivePrefix = {arXiv},
       eprint = {2410.15913},
 primaryClass = {astro-ph.GA},
       adsurl = {https://ui.adsabs.harvard.edu/abs/2024ApJ...976..249G}
}

@ARTICLE{2013MNRAS.430.2513H,
       author = {{Holland}, W.~S. and {Bintley}, D. and {Chapin}, E.~L. and {Chrysostomou}, A. and {Davis}, G.~R. and {Dempsey}, J.~T. and {Duncan}, W.~D. and {Fich}, M. and {Friberg}, P. and {Halpern}, M. and {Irwin}, K.~D. and {Jenness}, T. and {Kelly}, B.~D. and {MacIntosh}, M.~J. and {Robson}, E.~I. and {Scott}, D. and {Ade}, P.~A.~R. and {Atad-Ettedgui}, E. and {Berry}, D.~S. and {Craig}, S.~C. and {Gao}, X. and {Gibb}, A.~G. and {Hilton}, G.~C. and {Hollister}, M.~I. and {Kycia}, J.~B. and {Lunney}, D.~W. and {McGregor}, H. and {Montgomery}, D. and {Parkes}, W. and {Tilanus}, R.~P.~J. and {Ullom}, J.~N. and {Walther}, C.~A. and {Walton}, A.~J. and {Woodcraft}, A.~L. and {Amiri}, M. and {Atkinson}, D. and {Burger}, B. and {Chuter}, T. and {Coulson}, I.~M. and {Doriese}, W.~B. and {Dunare}, C. and {Economou}, F. and {Niemack}, M.~D. and {Parsons}, H.~A.~L. and {Reintsema}, C.~D. and {Sibthorpe}, B. and {Smail}, I. and {Sudiwala}, R. and {Thomas}, H.~S.},
        title = "{SCUBA-2: the 10 000 pixel bolometer camera on the James Clerk Maxwell Telescope}",
      journal = {\mnras},
         year = 2013,
        month = apr,
       volume = {430},
       number = {4},
        pages = {2513-2533},
          doi = {10.1093/mnras/sts612},
archivePrefix = {arXiv},
       eprint = {1301.3650},
 primaryClass = {astro-ph.IM},
       adsurl = {https://ui.adsabs.harvard.edu/abs/2013MNRAS.430.2513H}
}

@INPROCEEDINGS{2018SPIE10708E..3MF,
       author = {{Friberg}, Per and {Berry}, David and {Savini}, Giorgio and {Bintley}, Dan and {Dempsey}, Jessica and {Graves}, Sarah and {Parsons}, Harriet},
        title = "{Characterizing and reducing the POL-2 instrumental polarization}",
    booktitle = {Millimeter, Submillimeter, and Far-Infrared Detectors and Instrumentation for Astronomy IX},
         year = 2018,
       editor = {{Zmuidzinas}, Jonas and {Gao}, Jian-Rong},
       series = {Society of Photo-Optical Instrumentation Engineers (SPIE) Conference Series},
       volume = {10708},
        month = jul,
          eid = {107083M},
        pages = {107083M},
          doi = {10.1117/12.2314345},
       adsurl = {https://ui.adsabs.harvard.edu/abs/2018SPIE10708E..3MF}
}

@ARTICLE{2021AJ....162..191M,
       author = {{Mairs}, Steve and {Dempsey}, Jessica T. and {Bell}, Graham S. and {Parsons}, Harriet and {Currie}, Malcolm J. and {Friberg}, Per and {Jiang}, Xue-Jian and {Tetarenko}, Alexandra J. and {Bintley}, Dan and {Cookson}, Jamie and {Li}, Shaoliang and {Rawlings}, Mark G. and {Wouterloot}, Jan and {Berry}, David and {Graves}, Sarah and {Mizuno}, Izumi and {Acohido}, Alexis Ann and {Clark}, Alyssa and {Cox}, Jeff and {Fuchs}, Miriam and {Hoge}, James and {Kemp}, Johnathon and {Lee}, E'lisa and {Matulonis}, Callie and {Montgomerie}, William and {Silva}, Kevin and {Smith}, Patrice},
        title = "{A Decade of SCUBA-2: A Comprehensive Guide to Calibrating 450 {\ensuremath{\mu}}m and 850 {\ensuremath{\mu}}m Continuum Data at the JCMT}",
      journal = {\aj},
         year = 2021,
        month = nov,
       volume = {162},
       number = {5},
          eid = {191},
        pages = {191},
          doi = {10.3847/1538-3881/ac18bf},
archivePrefix = {arXiv},
       eprint = {2107.13558},
 primaryClass = {astro-ph.IM},
       adsurl = {https://ui.adsabs.harvard.edu/abs/2021AJ....162..191M}
}

@ARTICLE{2013MNRAS.430.2545C,
       author = {{Chapin}, Edward L. and {Berry}, David S. and {Gibb}, Andrew G. and {Jenness}, Tim and {Scott}, Douglas and {Tilanus}, Remo P.~J. and {Economou}, Frossie and {Holland}, Wayne S.},
        title = "{SCUBA-2: iterative map-making with the Sub-Millimetre User Reduction Facility}",
      journal = {\mnras},
         year = 2013,
        month = apr,
       volume = {430},
       number = {4},
        pages = {2545-2573},
          doi = {10.1093/mnras/stt052},
archivePrefix = {arXiv},
       eprint = {1301.3652},
 primaryClass = {astro-ph.IM},
       adsurl = {https://ui.adsabs.harvard.edu/abs/2013MNRAS.430.2545C}
}

@INPROCEEDINGS{2014ASPC..485..391C,
       author = {{Currie}, M.~J. and {Berry}, D.~S. and {Jenness}, T. and {Gibb}, A.~G. and {Bell}, G.~S. and {Draper}, P.~W.},
        title = "{Starlink Software in 2013}",
    booktitle = {Astronomical Data Analysis Software and Systems XXIII},
         year = 2014,
       editor = {{Manset}, N. and {Forshay}, P.},
       series = {Astronomical Society of the Pacific Conference Series},
       volume = {485},
        month = may,
        pages = {391},
       adsurl = {https://ui.adsabs.harvard.edu/abs/2014ASPC..485..391C}
}

@ARTICLE{2022SCPMA..6599511J,
       author = {{Jiao}, Sihan and {Lin}, Yuxin and {Shui}, Xiangyu and {Wu}, Jingwen and {Ren}, Zhiyuan and {Li}, Di},
        title = "{J-comb: An image fusion algorithm to combine observations covering different spatial frequency ranges}",
      journal = {Science China Physics, Mechanics, and Astronomy},
         year = 2022,
        month = sep,
       volume = {65},
       number = {9},
          eid = {299511},
        pages = {299511},
          doi = {10.1007/s11433-021-1902-3},
archivePrefix = {arXiv},
       eprint = {2208.00588},
 primaryClass = {astro-ph.IM},
       adsurl = {https://ui.adsabs.harvard.edu/abs/2022SCPMA..6599511J}
}

@ARTICLE{2018ApJS..234...28L,
       author = {{Liu}, Tie and {Kim}, Kee-Tae and {Juvela}, Mika and {Wang}, Ke and {Tatematsu}, Ken'ichi and {Di Francesco}, James and {Liu}, Sheng-Yuan and {Wu}, Yuefang and {Thompson}, Mark and {Fuller}, Gary and {Eden}, David and {Li}, Di and {Ristorcelli}, I. and {Kang}, Sung-ju and {Lin}, Yuxin and {Johnstone}, D. and {He}, J.~H. and {Koch}, P.~M. and {Sanhueza}, Patricio and {Qin}, Sheng-Li and {Zhang}, Q. and {Hirano}, N. and {Goldsmith}, Paul F. and {Evans}, II, Neal J. and {White}, Glenn J. and {Choi}, Minho and {Lee}, Chang Won and {Toth}, L.~V. and {Mairs}, Steve and {Yi}, H. -W. and {Tang}, Mengyao and {Soam}, Archana and {Peretto}, N. and {Samal}, Manash R. and {Fich}, Michel and {Parsons}, Harriet and {Yuan}, Jinghua and {Zhang}, Chuan-Peng and {Malinen}, Johanna and {Bendo}, George J. and {Rivera-Ingraham}, A. and {Liu}, Hong-Li and {Wouterloot}, Jan and {Li}, Pak Shing and {Qian}, Lei and {Rawlings}, Jonathan and {Rawlings}, Mark G. and {Feng}, Siyi and {Aikawa}, Yuri and {Akhter}, S. and {Alina}, Dana and {Bell}, Graham and {Bernard}, J. -P. and {Blain}, Andrew and {B{\H{o}}gner}, Rebeka and {Bronfman}, L. and {Byun}, D. -Y. and {Chapman}, Scott and {Chen}, Huei-Ru and {Chen}, M. and {Chen}, Wen-Ping and {Chen}, X. and {Chen}, Xuepeng and {Chrysostomou}, A. and {Cosentino}, Giuliana and {Cunningham}, M.~R. and {Demyk}, K. and {Drabek-Maunder}, Emily and {Doi}, Yasuo and {Eswaraiah}, C. and {Falgarone}, Edith and {Feh{\'e}r}, O. and {Fraser}, Helen and {Friberg}, Per and {Garay}, G. and {Ge}, J.~X. and {Gear}, W.~K. and {Greaves}, Jane and {Guan}, X. and {Harvey-Smith}, Lisa and {HASEGAWA}, Tetsuo and {Hatchell}, J. and {He}, Yuxin and {Henkel}, C. and {Hirota}, T. and {Holland}, W. and {Hughes}, A. and {Jarken}, E. and {Ji}, Tae-Geun and {Jimenez-Serra}, Izaskun and {Kang}, Miju and {Kawabata}, Koji S. and {Kim}, Gwanjeong and {Kim}, Jungha and {Kim}, Jongsoo and {Kim}, Shinyoung and {Koo}, B. -C. and {Kwon}, Woojin and {Kuan}, Yi-Jehng and {Lacaille}, K.~M. and {Lai}, Shih-Ping and {Lee}, C.~F. and {Lee}, J. -E. and {Lee}, Y. -U. and {Li}, Dalei and {Li}, Hua-bai and {Lo}, N. and {Lopez}, John A.~P. and {Lu}, Xing and {Lyo}, A. -Ran and {Mardones}, D. and {Marston}, A. and {McGehee}, P. and {Meng}, F. and {Montier}, L. and {Montillaud}, Julien and {Moore}, T. and {Morata}, O. and {Moriarty-Schieven}, Gerald H. and {Ohashi}, S. and {Pak}, Soojong and {Park}, Geumsook and {Paladini}, R. and {Pattle}, Kate M. and {Pech}, Gerardo and {Pelkonen}, V. -M. and {Qiu}, K. and {Ren}, Zhi-Yuan and {Richer}, John and {Saito}, M. and {Sakai}, Takeshi and {Shang}, H. and {Shinnaga}, Hiroko and {Stamatellos}, Dimitris and {Tang}, Y. -W. and {Traficante}, Alessio and {Vastel}, Charlotte and {Viti}, S. and {Walsh}, Andrew and {Wang}, Bingru and {Wang}, Hongchi and {Wang}, Junzhi and {Ward-Thompson}, D. and {Whitworth}, Anthony and {Xu}, Ye and {Yang}, J. and {Yang}, Yao-Lun and {Yuan}, Lixia and {Zavagno}, A. and {Zhang}, Guoyin and {Zhang}, H. -W. and {Zhou}, Chenlin and {Zhou}, Jianjun and {Zhu}, Lei and {Zuo}, Pei and {Zhang}, Chao},
        title = "{The TOP-SCOPE Survey of Planck Galactic Cold Clumps: Survey Overview and Results of an Exemplar Source, PGCC G26.53+0.17}",
      journal = {\apjs},
         year = 2018,
        month = feb,
       volume = {234},
       number = {2},
          eid = {28},
        pages = {28},
          doi = {10.3847/1538-4365/aaa3dd},
archivePrefix = {arXiv},
       eprint = {1711.04382},
 primaryClass = {astro-ph.GA},
       adsurl = {https://ui.adsabs.harvard.edu/abs/2018ApJS..234...28L}
}

@ARTICLE{2020ApJS..249...33K,
       author = {{Kim}, Gwanjeong and {Tatematsu}, Ken'ichi and {Liu}, Tie and {Yi}, Hee-Weon and {He}, Jinhua and {Hirano}, Naomi and {Liu}, Sheng-Yuan and {Choi}, Minho and {Sanhueza}, Patricio and {T{\'o}th}, L. Viktor and {Evans}, II, Neal J. and {Feng}, Siyi and {Juvela}, Mika and {Kim}, Kee-Tae and {Vastel}, Charlotte and {Lee}, Jeong-Eun and {Nguyễn Lu'o'ng}, Quang and {Kang}, Miju and {Ristorcelli}, Isabelle and {Feh{\'e}r}, Orsolya and {Wu}, Yuefang and {Ohashi}, Satoshi and {Wang}, Ke and {Kandori}, Ryo and {Hirota}, Tomoya and {Sakai}, Takeshi and {Lu}, Xing and {Thompson}, Mark A. and {Fuller}, Gary A. and {Li}, Di and {Shinnaga}, Hiroko and {Kim}, Jungha},
        title = "{Molecular Cloud Cores with a High Deuterium Fraction: Nobeyama Single-pointing Survey}",
      journal = {\apjs},
         year = 2020,
        month = aug,
       volume = {249},
       number = {2},
          eid = {33},
        pages = {33},
          doi = {10.3847/1538-4365/aba746},
archivePrefix = {arXiv},
       eprint = {2007.12319},
 primaryClass = {astro-ph.GA},
       adsurl = {https://ui.adsabs.harvard.edu/abs/2020ApJS..249...33K}
}

@ARTICLE{2021A&A...654A.123M,
       author = {{Mannfors}, E. and {Juvela}, M. and {Bronfman}, L. and {Eden}, D.~J. and {He}, J. and {Kim}, G. and {Kim}, K. -T. and {Kirppu}, H. and {Liu}, T. and {Montillaud}, J. and {Parsons}, H. and {Sanhueza}, P. and {Shang}, H. and {Soam}, A. and {Tatematsu}, K. and {Traficante}, A. and {V{\"a}is{\"a}l{\"a}}, M.~S. and {Lee}, C.~W.},
        title = "{Characterization of dense Planck clumps observed with Herschel and SCUBA-2}",
      journal = {\aap},
         year = 2021,
        month = oct,
       volume = {654},
          eid = {A123},
        pages = {A123},
          doi = {10.1051/0004-6361/202037791},
archivePrefix = {arXiv},
       eprint = {2106.10114},
 primaryClass = {astro-ph.GA},
       adsurl = {https://ui.adsabs.harvard.edu/abs/2021A&A...654A.123M}
}

@ARTICLE{2021ApJS..256...25T,
       author = {{Tatematsu}, Ken'ichi and {Kim}, Gwanjeong and {Liu}, Tie and {Evans}, II, Neal J. and {Yi}, Hee-Weon and {Lee}, Jeong-Eun and {Wu}, Yuefang and {Hirano}, Naomi and {Liu}, Sheng-Yuan and {Dutta}, Somnath and {Sahu}, Dipen and {Sanhueza}, Patricio and {Kim}, Kee-Tae and {Juvela}, Mika and {T{\'o}th}, L. Viktor and {Feh{\'e}r}, Orsolya and {He}, Jinhua and {Ge}, Jixing and {Feng}, Siyi and {Choi}, Minho and {Kang}, Miju and {Thompson}, Mark A. and {Fuller}, Gary A. and {Li}, Di and {Ristorcelli}, Isabelle and {Wang}, Ke and {di Francesco}, James and {Eden}, David and {Ohashi}, Satoshi and {Kandori}, Ryo and {Vastel}, Charlotte and {Hirota}, Tomoya and {Sakai}, Takeshi and {Lu}, Xing and {Nguy{\^e}n Lu'O'Ng}, Quang and {Shinnaga}, Hiroko and {Kim}, Jungha and {Scope Collaboration} and {Jcmt Large Program}},
        title = "{Molecular Cloud Cores with High Deuterium Fractions: Nobeyama Mapping Survey}",
      journal = {\apjs},
         year = 2021,
        month = oct,
       volume = {256},
       number = {2},
          eid = {25},
        pages = {25},
          doi = {10.3847/1538-4365/ac0978},
archivePrefix = {arXiv},
       eprint = {2106.04052},
 primaryClass = {astro-ph.GA},
       adsurl = {https://ui.adsabs.harvard.edu/abs/2021ApJS..256...25T}
}

@ARTICLE{2013MNRAS.434L..70H,
       author = {{Hirashita}, H. and {Li}, Z. -Y.},
        title = "{Condition for the formation of micron-sized dust grains in dense  molecular cloud cores.}",
      journal = {\mnras},
         year = 2013,
        month = jul,
       volume = {434},
        pages = {L70-L74},
          doi = {10.1093/mnrasl/slt081},
archivePrefix = {arXiv},
       eprint = {1306.5575},
 primaryClass = {astro-ph.GA},
       adsurl = {https://ui.adsabs.harvard.edu/abs/2013MNRAS.434L..70H}
}

@ARTICLE{2022ApJ...928..102H,
       author = {{Hoang}, Thiem},
        title = "{Effects of Grain Alignment with Magnetic Fields on Grain Growth and the Structure of Dust Aggregates}",
      journal = {\apj},
         year = 2022,
        month = apr,
       volume = {928},
       number = {2},
          eid = {102},
        pages = {102},
          doi = {10.3847/1538-4357/ac5408},
archivePrefix = {arXiv},
       eprint = {2109.07669},
 primaryClass = {astro-ph.GA},
       adsurl = {https://ui.adsabs.harvard.edu/abs/2022ApJ...928..102H}
}

@ARTICLE{2003JQSRT..79..881L,
       author = {{Lazarian}, A.},
        title = "{Magnetic Fields via Polarimetry: Progress of Grain Alignment Theory}",
      journal = {\jqsrt},
         year = 2003,
        month = jan,
       volume = {79-80},
        pages = {881},
          doi = {10.1016/S0022-4073(02)00326-6},
archivePrefix = {arXiv},
       eprint = {astro-ph/0208487},
 primaryClass = {astro-ph},
       adsurl = {https://ui.adsabs.harvard.edu/abs/2003JQSRT..79..881L}
}

@ARTICLE{2024A&A...692A..60R,
       author = {{Reissl}, Stefan and {Nguyen}, Philipp and {Jordan}, Lucas M. and {Klessen}, Ralf S.},
        title = "{The rotational disruption of porous dust aggregates from ab initio kinematic calculations}",
      journal = {\aap},
         year = 2024,
        month = dec,
       volume = {692},
          eid = {A60},
        pages = {A60},
          doi = {10.1051/0004-6361/202346068},
archivePrefix = {arXiv},
       eprint = {2301.12889},
 primaryClass = {astro-ph.GA},
       adsurl = {https://ui.adsabs.harvard.edu/abs/2024A&A...692A..60R}
}

@ARTICLE{2025ApJ...981..128P,
       author = {{Pravash}, Saikhom and {Soam}, Archana and {Diep}, Pham Ngoc and {Hoang}, Thiem and {Ngoc}, Nguyen Bich and {Tram}, Le Ngoc},
        title = "{B-fields and Dust in Interstellar Filaments Using Dust Polarization (BALLAD-POL). III. Grain Alignment and Disruption Mechanisms in G34.43+0.24 Using Polarization Observations from JCMT/POL-2}",
      journal = {\apj},
         year = 2025,
        month = mar,
       volume = {981},
       number = {2},
          eid = {128},
        pages = {128},
          doi = {10.3847/1538-4357/adae06},
archivePrefix = {arXiv},
       eprint = {2501.11634},
 primaryClass = {astro-ph.GA},
       adsurl = {https://ui.adsabs.harvard.edu/abs/2025ApJ...981..128P}
}

@ARTICLE{2025arXiv250502157H,
       author = {{Hoang}, Thiem},
        title = "{Toward A General Theory of Grain Alignment and Disruption by Radiative Torques and Magnetic Relaxation}",
      journal = {arXiv e-prints},
         year = 2025,
        month = may,
          eid = {arXiv:2505.02157},
        pages = {arXiv:2505.02157},
          doi = {10.48550/arXiv.2505.02157},
archivePrefix = {arXiv},
       eprint = {2505.02157},
 primaryClass = {astro-ph.GA},
       adsurl = {https://ui.adsabs.harvard.edu/abs/2025arXiv250502157H}
}

@ARTICLE{2021ApJ...908...12L,
       author = {{Lazarian}, A. and {Hoang}, Thiem},
        title = "{Alignment and Rotational Disruption of Dust}",
      journal = {\apj},
         year = 2021,
        month = feb,
       volume = {908},
       number = {1},
          eid = {12},
        pages = {12},
          doi = {10.3847/1538-4357/abd02c},
archivePrefix = {arXiv},
       eprint = {2010.15301},
 primaryClass = {astro-ph.GA},
       adsurl = {https://ui.adsabs.harvard.edu/abs/2021ApJ...908...12L}
}

@ARTICLE{2024A&A...689A.290T,
       author = {{Tram}, Le Ngoc and {Hoang}, Thiem and {Wiesemeyer}, Helmut and {Ristorcelli}, Isabelle and {Menten}, Karl M. and {Ngoc}, Nguyen Bich and {Diep}, Pham Ngoc},
        title = "{Understanding the multi-wavelength thermal dust polarisation from the Orion molecular cloud in light of the radiative torque paradigm}",
      journal = {\aap},
         year = 2024,
        month = sep,
       volume = {689},
          eid = {A290},
        pages = {A290},
          doi = {10.1051/0004-6361/202450127},
archivePrefix = {arXiv},
       eprint = {2403.17088},
 primaryClass = {astro-ph.GA},
       adsurl = {https://ui.adsabs.harvard.edu/abs/2024A&A...689A.290T}
}

@ARTICLE{2025arXiv250116079T,
       author = {{Tram}, Le Ngoc and {Hoang}, Thiem and {Lazarian}, Alex and {Seifried}, Daniel and {Andersson}, B-G and {Pillai}, Thushara G.~S. and {Truong}, Bao and {Diep}, Pham Ngoc and {Fanciullo}, Lapo},
        title = "{Grain Alignment and Dust Evolution Physics with Polarisation (GRADE-POL). I. Dust Polarisation Modelling for Isolated Starless Cores}",
      journal = {arXiv e-prints},
         year = 2025,
        month = jan,
          eid = {arXiv:2501.16079},
        pages = {arXiv:2501.16079},
          doi = {10.48550/arXiv.2501.16079},
archivePrefix = {arXiv},
       eprint = {2501.16079},
 primaryClass = {astro-ph.GA},
       adsurl = {https://ui.adsabs.harvard.edu/abs/2025arXiv250116079T}
}

@ARTICLE{2023ApJ...948...55H,
       author = {{Hensley}, Brandon S. and {Draine}, B.~T.},
        title = "{The Astrodust+PAH Model: A Unified Description of the Extinction, Emission, and Polarization from Dust in the Diffuse Interstellar Medium}",
      journal = {\apj},
         year = 2023,
        month = may,
       volume = {948},
       number = {1},
          eid = {55},
        pages = {55},
          doi = {10.3847/1538-4357/acc4c2},
archivePrefix = {arXiv},
       eprint = {2208.12365},
 primaryClass = {astro-ph.GA},
       adsurl = {https://ui.adsabs.harvard.edu/abs/2023ApJ...948...55H}
}

@ARTICLE{2013A&A...558A..62J,
       author = {{Jones}, A.~P. and {Fanciullo}, L. and {K{\"o}hler}, M. and {Verstraete}, L. and {Guillet}, V. and {Bocchio}, M. and {Ysard}, N.},
        title = "{The evolution of amorphous hydrocarbons in the ISM: dust modelling from a new vantage point}",
      journal = {\aap},
         year = 2013,
        month = oct,
       volume = {558},
          eid = {A62},
        pages = {A62},
          doi = {10.1051/0004-6361/201321686},
archivePrefix = {arXiv},
       eprint = {1411.6293},
 primaryClass = {astro-ph.GA},
       adsurl = {https://ui.adsabs.harvard.edu/abs/2013A&A...558A..62J}
}

@ARTICLE{2018A&A...614A.100T,
       author = {{Tahani}, M. and {Plume}, R. and {Brown}, J.~C. and {Kainulainen}, J.},
        title = "{Helical magnetic fields in molecular clouds?. A new method to determine the line-of-sight magnetic field structure in molecular clouds}",
      journal = {\aap},
         year = 2018,
        month = jun,
       volume = {614},
          eid = {A100},
        pages = {A100},
          doi = {10.1051/0004-6361/201732219},
archivePrefix = {arXiv},
       eprint = {1802.07831},
 primaryClass = {astro-ph.GA},
       adsurl = {https://ui.adsabs.harvard.edu/abs/2018A&A...614A.100T}
}

@ARTICLE{2017ApJ...838...40M,
       author = {{Mocz}, Philip and {Burkhart}, Blakesley and {Hernquist}, Lars and {McKee}, Christopher F. and {Springel}, Volker},
        title = "{Moving-mesh Simulations of Star-forming Cores in Magneto-gravo-turbulence}",
      journal = {\apj},
         year = 2017,
        month = mar,
       volume = {838},
       number = {1},
          eid = {40},
        pages = {40},
          doi = {10.3847/1538-4357/aa6475},
archivePrefix = {arXiv},
       eprint = {1702.06133},
 primaryClass = {astro-ph.GA},
       adsurl = {https://ui.adsabs.harvard.edu/abs/2017ApJ...838...40M}
}

@ARTICLE{2010ApJ...725..466C,
       author = {{Crutcher}, Richard M. and {Wandelt}, Benjamin and {Heiles}, Carl and {Falgarone}, Edith and {Troland}, Thomas H.},
        title = "{Magnetic Fields in Interstellar Clouds from Zeeman Observations: Inference of Total Field Strengths by Bayesian Analysis}",
      journal = {\apj},
         year = 2010,
        month = dec,
       volume = {725},
       number = {1},
        pages = {466-479},
          doi = {10.1088/0004-637X/725/1/466},
       adsurl = {https://ui.adsabs.harvard.edu/abs/2010ApJ...725..466C}
}

@ARTICLE{2015A&A...576A.104P,
       author = {{Planck Collaboration} and {Ade}, P.~A.~R. and {Aghanim}, N. and {Alina}, D. and {Alves}, M.~I.~R. and {Armitage-Caplan}, C. and {Arnaud}, M. and {Arzoumanian}, D. and {Ashdown}, M. and {Atrio-Barandela}, F. and {Aumont}, J. and {Baccigalupi}, C. and {Banday}, A.~J. and {Barreiro}, R.~B. and {Battaner}, E. and {Benabed}, K. and {Benoit-L{\'e}vy}, A. and {Bernard}, J.-P. and {Bersanelli}, M. and {Bielewicz}, P. and {Bock}, J.~J. and {Bond}, J.~R. and {Borrill}, J. and {Bouchet}, F.~R. and {Boulanger}, F. and {Bracco}, A. and {Burigana}, C. and {Butler}, R.~C. and {Cardoso}, J.-F. and {Catalano}, A. and {Chamballu}, A. and {Chary}, R.-R. and {Chiang}, H.~C. and {Christensen}, P.~R. and {Colombi}, S. and {Colombo}, L.~P.~L. and {Combet}, C. and {Couchot}, F. and {Coulais}, A. and {Crill}, B.~P. and {Curto}, A. and {Cuttaia}, F. and {Danese}, L. and {Davies}, R.~D. and {Davis}, R.~J. and {de Bernardis}, P. and {de Gouveia Dal Pino}, E.~M. and {de Rosa}, A. and {de Zotti}, G. and {Delabrouille}, J. and {D{\'e}sert}, F.-X. and {Dickinson}, C. and {Diego}, J.~M. and {Donzelli}, S. and {Dor{\'e}}, O. and {Douspis}, M. and {Dunkley}, J. and {Dupac}, X. and {Efstathiou}, G. and {En{\ss}lin}, T.~A. and {Eriksen}, H.~K. and {Falgarone}, E. and {Ferri{\`e}re}, K. and {Finelli}, F. and {Forni}, O. and {Frailis}, M. and {Fraisse}, A.~A. and {Franceschi}, E. and {Galeotta}, S. and {Ganga}, K. and {Ghosh}, T. and {Giard}, M. and {Giraud-H{\'e}raud}, Y. and {Gonz{\'a}lez-Nuevo}, J. and {G{\'o}rski}, K.~M. and {Gregorio}, A. and {Gruppuso}, A. and {Guillet}, V. and {Hansen}, F.~K. and {Harrison}, D.~L. and {Helou}, G. and {Hern{\'a}ndez-Monteagudo}, C. and {Hildebrandt}, S.~R. and {Hivon}, E. and {Hobson}, M. and {Holmes}, W.~A. and {Hornstrup}, A. and {Huffenberger}, K.~M. and {Jaffe}, A.~H. and {Jaffe}, T.~R. and {Jones}, W.~C. and {Juvela}, M. and {Keih{\"a}nen}, E. and {Keskitalo}, R. and {Kisner}, T.~S. and {Kneissl}, R. and {Knoche}, J. and {Kunz}, M. and {Kurki-Suonio}, H. and {Lagache}, G. and {L{\"a}hteenm{\"a}ki}, A. and {Lamarre}, J.-M. and {Lasenby}, A. and {Lawrence}, C.~R. and {Leahy}, J.~P. and {Leonardi}, R. and {Levrier}, F. and {Liguori}, M. and {Lilje}, P.~B. and {Linden-V{\o}rnle}, M. and {L{\'o}pez-Caniego}, M. and {Lubin}, P.~M. and {Mac{\'\i}as-P{\'e}rez}, J.~F. and {Maffei}, B. and {Magalh{\~a}es}, A.~M. and {Maino}, D. and {Mandolesi}, N. and {Maris}, M. and {Marshall}, D.~J. and {Martin}, P.~G. and {Mart{\'\i}nez-Gonz{\'a}lez}, E. and {Masi}, S. and {Matarrese}, S. and {Mazzotta}, P. and {Melchiorri}, A. and {Mendes}, L. and {Mennella}, A. and {Migliaccio}, M. and {Miville-Desch{\^e}nes}, M.-A. and {Moneti}, A. and {Montier}, L. and {Morgante}, G. and {Mortlock}, D. and {Munshi}, D. and {Murphy}, J.~A. and {Naselsky}, P. and {Nati}, F. and {Natoli}, P. and {Netterfield}, C.~B. and {Noviello}, F. and {Novikov}, D. and {Novikov}, I. and {Oxborrow}, C.~A. and {Pagano}, L. and {Pajot}, F. and {Paladini}, R. and {Paoletti}, D. and {Pasian}, F. and {Pearson}, T.~J. and {Perdereau}, O. and {Perotto}, L. and {Perrotta}, F. and {Piacentini}, F. and {Piat}, M. and {Pietrobon}, D. and {Plaszczynski}, S. and {Poidevin}, F. and {Pointecouteau}, E. and {Polenta}, G. and {Popa}, L. and {Pratt}, G.~W. and {Prunet}, S. and {Puget}, J.-L. and {Rachen}, J.~P. and {Reach}, W.~T. and {Rebolo}, R. and {Reinecke}, M. and {Remazeilles}, M. and {Renault}, C. and {Ricciardi}, S. and {Riller}, T. and {Ristorcelli}, I. and {Rocha}, G. and {Rosset}, C. and {Roudier}, G. and {Rubi{\~n}o-Mart{\'\i}n}, J.~A. and {Rusholme}, B. and {Sandri}, M. and {Savini}, G. and {Scott}, D. and {Spencer}, L.~D. and {Stolyarov}, V. and {Stompor}, R. and {Sudiwala}, R. and {Sutton}, D. and {Suur-Uski}, A.-S. and {Sygnet}, J.-F. and {Tauber}, J.~A. and {Terenzi}, L. and {Toffolatti}, L. and {Tomasi}, M. and {Tristram}, M. and {Tucci}, M. and {Umana}, G. and {Valenziano}, L. and {Valiviita}, J. and {Van Tent}, B. and {Vielva}, P. and {Villa}, F. and {Wade}, L.~A.},
        title = "{Planck intermediate results. XIX. An overview of the polarized thermal emission from Galactic dust}",
      journal = {\aap},
         year = 2015,
        month = apr,
       volume = {576},
          eid = {A104},
        pages = {A104},
          doi = {10.1051/0004-6361/201424082},
archivePrefix = {arXiv},
       eprint = {1405.0871},
 primaryClass = {astro-ph.GA},
       adsurl = {https://ui.adsabs.harvard.edu/abs/2015A&A...576A.104P}
}

@ARTICLE{1984ApJ...285...89D,
       author = {{Draine}, B.~T. and {Lee}, H.~M.},
        title = "{Optical Properties of Interstellar Graphite and Silicate Grains}",
      journal = {\apj},
         year = 1984,
        month = oct,
       volume = {285},
        pages = {89},
          doi = {10.1086/162480},
       adsurl = {https://ui.adsabs.harvard.edu/abs/1984ApJ...285...89D}
}

@ARTICLE{2001ApJ...548..296W,
       author = {{Weingartner}, Joseph C. and {Draine}, B.~T.},
        title = "{Dust Grain-Size Distributions and Extinction in the Milky Way, Large Magellanic Cloud, and Small Magellanic Cloud}",
      journal = {\apj},
         year = 2001,
        month = feb,
       volume = {548},
       number = {1},
        pages = {296-309},
          doi = {10.1086/318651},
archivePrefix = {arXiv},
       eprint = {astro-ph/0008146},
 primaryClass = {astro-ph},
       adsurl = {https://ui.adsabs.harvard.edu/abs/2001ApJ...548..296W}
}

@ARTICLE{2021ApJ...909...94D,
       author = {{Draine}, B.~T. and {Hensley}, Brandon S.},
        title = "{The Dielectric Function of ``Astrodust'' and Predictions for Polarization in the 3.4 and 10 {\ensuremath{\mu}}m Features}",
      journal = {\apj},
         year = 2021,
        month = mar,
       volume = {909},
       number = {1},
          eid = {94},
        pages = {94},
          doi = {10.3847/1538-4357/abd6c6},
archivePrefix = {arXiv},
       eprint = {2009.11314},
 primaryClass = {astro-ph.GA},
       adsurl = {https://ui.adsabs.harvard.edu/abs/2021ApJ...909...94D}
}

@INPROCEEDINGS{1990ASPC...12..193D,
       author = {{Draine}, Bruce T.},
        title = "{Evolution of interstellar dust.}",
    booktitle = {The Evolution of the Interstellar Medium},
         year = 1990,
       editor = {{Blitz}, Leo},
       series = {Astronomical Society of the Pacific Conference Series},
       volume = {12},
        month = jan,
        pages = {193-205},
       adsurl = {https://ui.adsabs.harvard.edu/abs/1990ASPC...12..193D}
}

@INPROCEEDINGS{2009ASPC..414..453D,
       author = {{Draine}, B.~T.},
        title = "{Interstellar Dust Models and Evolutionary Implications}",
    booktitle = {Cosmic Dust - Near and Far},
         year = 2009,
       editor = {{Henning}, T. and {Gr{\"u}n}, E. and {Steinacker}, J.},
       series = {Astronomical Society of the Pacific Conference Series},
       volume = {414},
        month = dec,
        pages = {453},
          doi = {10.48550/arXiv.0903.1658},
archivePrefix = {arXiv},
       eprint = {0903.1658},
 primaryClass = {astro-ph.GA},
       adsurl = {https://ui.adsabs.harvard.edu/abs/2009ASPC..414..453D}
}

@ARTICLE{2017ApJ...838..121C,
       author = {{Ching}, Tao-Chung and {Lai}, Shih-Ping and {Zhang}, Qizhou and {Girart}, Josep M. and {Qiu}, Keping and {Liu}, Hauyu B.},
        title = "{Magnetic Fields in the Massive Dense Cores of the DR21 Filament: Weakly Magnetized Cores in a Strongly Magnetized Filament}",
      journal = {\apj},
         year = 2017,
        month = apr,
       volume = {838},
       number = {2},
          eid = {121},
        pages = {121},
          doi = {10.3847/1538-4357/aa65cc},
archivePrefix = {arXiv},
       eprint = {1703.02566},
 primaryClass = {astro-ph.GA},
       adsurl = {https://ui.adsabs.harvard.edu/abs/2017ApJ...838..121C}
}

@ARTICLE{2026AJ....171...50H,
       author = {{Hwang}, Jihye and {Sanhueza}, Patricio and {Girart}, Josep Miquel and {Stephens}, Ian W. and {Beltr{\'a}n}, Maria T. and {Law}, Chi Yan and {Zhang}, Qizhou and {Liu}, Junhao and {Cort{\'e}s}, Paulo and {Olguin}, Fernando A. and {Koch}, Patrick M. and {Nakamura}, Fumitaka and {Saha}, Piyali and {Wang}, Jia-Wei and {Xu}, Fengwei and {Beuther}, Henrik and {Morii}, Kaho and {Fern{\'a}ndez L{\'o}pez}, Manuel and {Jiao}, Wenyu and {Kim}, Kee-Tae and {Li}, Shanghuo and {Zapata}, Luis A. and {Kim}, Jongsoo and {Choudhury}, Spandan and {Cheng}, Yu and {Pattle}, Kate and {Eswaraiah}, Chakali and {Sandhyarani}, Panigrahy and {Dewangan}, L.~K. and {Jadhav}, O.~R.},
        title = "{Magnetic Fields in Massive Star-forming Regions (MagMaR). VI. Magnetic Field Dragging in the Filamentary High-mass Star-forming Region G35.20─0.74N Due to Gravity}",
      journal = {\aj},
         year = 2026,
        month = jan,
       volume = {171},
       number = {1},
          eid = {50},
        pages = {50},
          doi = {10.3847/1538-3881/ae18c9},
archivePrefix = {arXiv},
       eprint = {2510.25078},
 primaryClass = {astro-ph.GA},
       adsurl = {https://ui.adsabs.harvard.edu/abs/2026AJ....171...50H}
}

@ARTICLE{2017ApJ...842L...9H,
       author = {{Hull}, Charles L.~H. and {Mocz}, Philip and {Burkhart}, Blakesley and {Goodman}, Alyssa A. and {Girart}, Josep M. and {Cort{\'e}s}, Paulo C. and {Hernquist}, Lars and {Springel}, Volker and {Li}, Zhi-Yun and {Lai}, Shih-Ping},
        title = "{Unveiling the Role of the Magnetic Field at the Smallest Scales of Star Formation}",
      journal = {\apjl},
         year = 2017,
        month = jun,
       volume = {842},
       number = {2},
          eid = {L9},
        pages = {L9},
          doi = {10.3847/2041-8213/aa71b7},
archivePrefix = {arXiv},
       eprint = {1706.03806},
 primaryClass = {astro-ph.GA},
       adsurl = {https://ui.adsabs.harvard.edu/abs/2017ApJ...842L...9H}
}

@ARTICLE{2018ApJ...855...39K,
       author = {{Koch}, Patrick M. and {Tang}, Ya-Wen and {Ho}, Paul T.~P. and {Yen}, Hsi-Wei and {Su}, Yu-Nung and {Takakuwa}, Shigehisa},
        title = "{Polarization Properties and Magnetic Field Structures in the High-mass Star-forming Region W51 Observed with ALMA}",
      journal = {\apj},
         year = 2018,
        month = mar,
       volume = {855},
       number = {1},
          eid = {39},
        pages = {39},
          doi = {10.3847/1538-4357/aaa4c1},
archivePrefix = {arXiv},
       eprint = {1801.08264},
 primaryClass = {astro-ph.GA},
       adsurl = {https://ui.adsabs.harvard.edu/abs/2018ApJ...855...39K}
}

@ARTICLE{2023ApJ...954..216H,
       author = {{Hoang}, Thiem and {Minh Phan}, Vo Hong and {Tram}, Le Ngoc},
        title = "{Internal and External Alignment of Carbonaceous Grains within the Radiative Torque Paradigm}",
      journal = {\apj},
         year = 2023,
        month = sep,
       volume = {954},
       number = {2},
          eid = {216},
        pages = {216},
          doi = {10.3847/1538-4357/ace788},
archivePrefix = {arXiv},
       eprint = {2301.07832},
 primaryClass = {astro-ph.GA},
       adsurl = {https://ui.adsabs.harvard.edu/abs/2023ApJ...954..216H}
}

@ARTICLE{2025ApJ...990...40P,
       author = {{Pravash}, Saikhom and {Hoang}, Thiem and {Soam}, Archana and {Chung}, Eun Jung and {Pham}, Diep Ngoc and {Ngoc}, Nguyen Bich and {Tram}, Le Ngoc},
        title = "{B-fields and Dust in Interstellar Filaments Using Dust Polarization (BALLAD-POL). IV. Grain Alignment Mechanisms in the Cocoon Nebula (IC 5146) Using Polarization Observations from JCMT/POL-2}",
      journal = {\apj},
         year = 2025,
        month = sep,
       volume = {990},
       number = {1},
          eid = {40},
        pages = {40},
          doi = {10.3847/1538-4357/adef09},
archivePrefix = {arXiv},
       eprint = {2507.07205},
 primaryClass = {astro-ph.GA},
       adsurl = {https://ui.adsabs.harvard.edu/abs/2025ApJ...990...40P}
}
\bibliographystyle{aasjournalv7}



\end{document}